\documentclass[preprint,12pt]{elsarticle}
\pdfoutput=1

\usepackage{amssymb}

\journal{ArXiv}
\usepackage{hyperref}
\usepackage{graphicx}
\usepackage{dcolumn}
\usepackage{bm}

\usepackage{subcaption}
\usepackage{mathptmx}
\usepackage[active]{srcltx}
\usepackage{amssymb}
\usepackage{amsmath}
\usepackage{amsfonts}
\usepackage[margin=1in]{geometry}

\usepackage{color,soul}
\usepackage{xcolor}

\newcommand{\bea}{\begin{eqnarray}}
\newcommand{\eea}{\end{eqnarray}}
\newcommand{\nbea}{\begin{eqnarray*}}
\newcommand{\neea}{\end{eqnarray*}}

\newcommand{\ba}{\begin{align}}

\newcommand{\be}{\begin{equation}}
\newcommand{\ee}{\end{equation}}
\newcommand{\ea}{\end{align}}

\newcommand{\ie}{{\mbox{\it i.e.~}}}

\definecolor{MyGreen}{HTML}{006400}
\definecolor{maroon}{HTML}{800000}

\makeatletter
\renewcommand*\env@matrix[1][*\c@MaxMatrixCols c]{%
  \hskip -\arraycolsep
  \let\@ifnextchar\new@ifnextchar
  \array{#1}}
\makeatother

\makeindex             

\begin{document}

\begin{frontmatter}
\title{Super-resolved shear shock focusing in the human head}

\author[1]{Bharat B Tripathi}
\ead{bharat.tripathi@nuigalway.ie}
\address[1]{School of Mathematics, Statistics and Applied Mathematics, National University of Ireland Galway, University Road Galway H91TK33, Ireland}

\author[2,3]{Sandhya Chandrasekaran}%
\address[2]{Department of Mechanical Engineering, North Carolina State University}
\author[3]{Gianmarco F. Pinton\corref{cor1}}
\ead{gia@email.unc.edu}
\cortext[cor1]{Corresponding Author}
\address[3]{Joint Department of Biomedical Engineering,  University of North Carolina at Chapel Hill and North Carolina State University, 116 Manning Drive, 9212A Mary Ellen Jones, Chapel Hill, North Carolina-27599, USA.}

\begin{abstract}
Shear shocks, which exist in a completely different regime from compressional shocks, were recently observed in the brain. These low phase speed ($\approx$ 2 m/s) high Mach number ($\approx$ 1) waves could be the primary mechanism behind diffuse axonal injury due to a very high local acceleration at the shock front. The extreme nonlinearity of these waves results in unique behaviors that are different from more commonly studied nonlinear compressional waves. Here we show the first observation of super-resolved shear shock wave focusing. Shear shock wave imaging and numerical simulations in a human head phantom over a range of frequencies/amplitudes shows the super-resolution of shock waves in the low strain and high strain-rate regime. These results suggest that even for mild accelerations injuries as small as a grain of rice on the scale of mm$^2$ can be easily created deep inside the brain.
\end{abstract}




\end{frontmatter}

\section{Introduction}
Traumatic Brain Injuries (TBIs) are a major cause of disability and mortality worldwide. Recent estimates indicate that each year in the United States 1.1 million are treated in emergency departments, 235,000 are hospitalized for nonfatal TBI, and 50,000 die~\cite{corrigan2010epidemiology, guskiewicz2000epidemiology}. 
According to a European survey, 51\% of brain injuries are motor-vehicle related, which explains why globally the incidence of TBI's is rising sharply as transportation becomes more widely available \cite{meythaler2001current,tagliaferri2006systematic,maas2008moderate}. One of the most common type of TBI is the 
Diffuse Axonal Injuries (DAI's) have been linked to progressive neurodegenerative diseases like Chronic Traumatic Encephalopathy (CTE), Parksinon's, and Alzheimer's \cite{chen2004long, mckee2009chronic, johnson2010traumatic}. And yet, the relationship between brain motion and injury remains poorly understood. 

Directly measuring local brain dynamics during impact has been a persistently challenging task. If obtainable, such measurements would have a significant impact on our ability to understand, prevent, and treat brain injury by generating accurate relationships between impact, brain deformation, and injury.
Head acceleration, although not the same as brain motion, can be easily measured with accelerometers, which is why in the past 50 years brain injury has been postulated principally in terms of head motion, typically using head, mouth, ear, skin based sensors and at times together with video-graphic data from the impact~\cite{Beckwith2012,Camarillo2013,Salzar2008_Bass,Kim1993}. Current predictors of injury thus rely on measurements of the acceleration/time history of the impact. The motion can be aggregated, as in the head injury criterion (HIC), or the linear and rotational acceleration can be considered separately \cite{rimel1981disability,greenwald2008head}. However these injury metrics are often poor predictors of injury with errors as high as 500\% \cite{Wu2016_Camarillo} and their link to mechanisms of injury has not been conclusively established~\cite{guskiewicz2007measurement}. There is thus a clear motivation to increase the biofidelic accuracy of injury measurements, yet there have only been a few successful attempts to directly measure human brain deformation and its relationship to injury. Previous measurements of brain phantom motion have relied on optically transparent gels with a grid pattern that were filmed with a high speed camera~\cite{margulies1990physical,meaney1995biomechanical}, or with markers implanted {\it ex vivo} that were tracked with high speed biplanar X-ray imaging~\cite{hardy2001investigation}. More recently MRI has been used to non-invasively measure low-level brain motion using fast gradient-echo sequences to maximize the frame rates~\cite{bayly2005deformation,clayton2012transmission}. MR imaging of the head is completely non-invasive and suitable for {\it in vivo} imaging in humans. However, due to fundamental limits imposed by the spin-relaxation time constants, MRI has acquisition of roughly 40-50 frames per second. 

The field of brain injury biomechanics still suffers from ``a dearth of large strain and high rate mechanical properties for brain tissue'' \cite{Macmanus2018}. 
At a cellular level, it has been shown that the tau-proteins break at strain-rate of 44 1/s with just 5\% of strain whereas they remain intact for up to 100\% strain at lower strain-rate of 0.01 1/s process \cite{Ahmadzadeh2014}. 
To the best of our knowledge, the highest strain-rate we have seen in TBI related literature is up to 250 1/s \cite{Ghajari2017}, and consequently  higher strain-rate regime is relatively unexplored. 
However, the high frame-rate (up to 10,000 images/second) 2D ultrasound techniques have recently quantified the formation of shear shock waves in  brain in a wide field of view (4 cm $\times$ 6 cm), with a high displacement sensitivity ($<1~\mu$m)~\cite{Espindola2017} with strain-rate over 400 1/s. 
These experiments showed that smooth (35$g$) shear waves  develop into destructive (320$g$) shear shock waves (not compressional shock waves like in blast-TBI) deep inside the brain. These planar shock waves are governed by cubically nonlinear viscoelastic behavior and they can be simulated using custom methods developed for this purpose~\cite{Tripathi2019_PPM1D_CT,Tripathi2019_PPM2D_CT}. We hypothesize that the violent gradients in these recently discovered shear shock waves are the primary biomechanical origin for neuronal damage deep inside the brain, ranging from  diffuse axonal injuries to chronic traumatic encephalopathy.

Unlike acoustical shocks which have been studied extensively including within the context of traumatic brain injuries, shear shocks are fundamentally different and are relatively unstudied. The soft tissue in the brain has nonlinear shear properties that are several orders of magnitude larger than its compressional properties. A typical Mach number (particle-velocity/wave-speed) for compressional waves in soft tissue is on the order of $10^{-4}$, and for shear waves it is on the order of one \cite{Pinton2010,Espindola2017}. This is due to the very low value of the shear wave velocity (typically 2 m/s) which in the case of a violent impact is the same magnitude as the particle velocity (typically 2 m/s or higher).  Consequently these extremely nonlinear shear waves can generate shock fronts within a single propagation wavelength, i.e., areas with smooth shear waves can be adjacent to areas with violent shear shocks. There are few reports of shear shock wave modeling in soft tissue. 
Models for wave propagation in nonlinear soft solids have been developed based on Landau's description of nonlinear elasticity \cite{Landau1986,Zabolotskaya2004,Destrade2010}, other shear shock  descriptions have also been proposed.\cite{Chockalingam2020,Ziv2019}
There are also models that are specific to soft-tissue  which, in addition to the shear wave nonlinearity, include the non-classical viscous or attenuating behavior soft tissue~\cite{Giammarinaro2016,Tripathi2017,Tripathi2019_PPM1D_CT,Tripathi2019_PPM2D_CT}. Describing the non-classical viscous behavior is of fundamental importance because its effect on the wave dynamics is just as significant as nonlinearity when estimating injury-relevant metrics such as the local acceleration or strain-rate. 

Solutions to these models, especially in configurations that describe injuries in humans, require numerical solvers since injury experiments on humans is not a possibility. 
Experiments with excised cadaveric brains are infeasible for shear shock waves because neural  tissue decomposes and liquefies, within 24  hours,  destroying its ability to support shear stress and obtaining specimens in less than a day is nearly impossible. 
In the absence of direct experimental measurements of brain motion, simulation tools must rely on material properties measured by mechanical testing of brain samples, where even fundamental measurements of linear elastic constants vary by three orders of magnitude depending on the method used~\cite{chatelin2010fifty}. 
Commercial finite element (FE) tools such as LS-DYNA (Livermore, CA) or ABAQUS (Johnston, RI) are widely used in the study of TBI's because they contain well-developed contact algorithms for efficient modeling of impacts ranging from closed skull to controlled cortical impacts \cite{Dixit2017,horgan2003creation,horgan2004influence,Taylor2009,wittek2011algorithms,yang2019modelling}. They are also capable of describing brain anatomy in great detail using unstructured meshes and sophisticated nonlinear viscoelastic material models~\cite{Zienkiewicz2005,Ye2017b}. However, these finite element tools have not been used to model the brain biomechanics of shear shock formation.  

This motivated the development of custom piecewise parabolic finite volume simulation tool designed specifically for simulating the formation and propagation of shear shock wave in relaxing soft solids like brain which was thoroughly validated by direct measurements of shear shock waves in tissue mimicking gelatin phantoms \cite{Tripathi2019_PPM1D_CT, Tripathi2019_PPM2D_CT}. In the next section, we summarize the theoretical model and the numerical method. In Section \ref{Sec:Experiments}, we present experimental validation of shear shock formation in human head phantom filled with tissue mimicking gelatin phantom which further validates the numerical solver. Then the numerical method is used to simulate shear shock waves in the same skull geometry with the brain material properties. This lead to the discovery of three distinct regimes appear 1) at low frequencies shear shock waves develop at the geometric focus of the head 2) at intermediate frequencies shocks form near the brain surface and at the focus 3) at high frequencies shock form only near the brain surface. It is shown that these three regimes arise from the interplay of attenuation and nonlinearity both of which are frequency-dependent and only one of which is amplitude-dependent. Finally it is shown that super-resolution occurs in the focal regime when the highly nonlinear harmonics generate acceleration and strain-rate focal zones that are much smaller than the impact wavelength. Together these simulations and experiments demonstrate the existence and determine the extent of the regimes where this previously unappreciated shear shock wave physics plays the leading-order role in brain biomechanics as discussed in section \ref{Sec:Conclusions}.

\section{Theoretical Model and Numerical Method}\label{Sec:Theory}
The system of equations describing the nonlinear propagation of linearly-polarized shear wave, i.e., particle displacement is confined to the axis orthogonal to the plane of propagation, in a homogeneous, isotropic, relaxing media is given by \cite{Tripathi2019_PPM2D_CT}: 
\bea
\begin{bmatrix}
v\\
r\\
s\\
\bar \xi_{1x}\\
\bar \xi_{2x}\\
\bar \xi_{3x}\\
\bar \xi_{1y}\\
\bar \xi_{2y}\\
\bar \xi_{3y}
\end{bmatrix}_t
+
\begin{bmatrix}
 -\sigma_{zx}/\rho_0 \\
 -v\\
 0\\
 -\omega_{1}v\\
 -\omega_{2}v\\
 -\omega_{3}v\\
 0\\
 0\\
 0
\end{bmatrix}_x
+
\begin{bmatrix}
-\sigma_{zy}/\rho_0\\
0\\
-v\\
0\\
 0\\
 0\\
-\omega_{1}v\\
- \omega_{2}v\\
- \omega_{3}v 
\end{bmatrix}_y
=
\begin{bmatrix}
  0\\
 - \sum_{l=1}^{3} \bar r_{l} \bar \xi_{lx}\\
 - \sum_{l=1}^{3} \bar r_{l} \bar \xi_{ly}\\
 -\omega_1 \bar \xi_{1x}\\
 -\omega_2 \bar \xi_{2x}\\
 -\omega_3 \bar \xi_{3x}\\
 -\omega_1 \bar \xi_{1y}\\
 -\omega_2 \bar \xi_{2y}\\
 -\omega_3 \bar \xi_{3y}
\end{bmatrix},
\label{NLEDHoReM2D} 
\eea
where $v$ is the particle velocity, $r,s$ are strain-like auxiliary variables. Here the first three equations model the lossless propagation with the cubic nonlinear stress terms:
\be
\sigma_{zx} = \mu r +  \frac{2\mu\beta}{3} r(r^2 + s^2)
\ee
and 
\be
\sigma_{zy} = \mu s +  \frac{2\mu\beta}{3} s(r^2 + s^2)
\ee
where $\rho_0$ is the material density, $\mu$ is the unrelaxed shear modulus, and $\beta$ is the coefficient of nonlinearity. On the other hand, the last 6 equations are resulting from a generalized Maxwell body (GMB) consisting of three Maxwell bodies and an elastic element, all connected in parallel.  Each relaxation mechanism (Maxwell body) has a variable associated along each direction, which are: $\bar{\xi}_{lx}, \bar{\xi}_{ly},~l=1,2,3$ corresponding to the three relaxation frequencies $\omega_l,~ l=1, 2, 3$.  The relaxation constants $\bar{r}_{lx}, \bar{r}_{ly},~l=1,2,3$ are determined after fitting a GMB with attenuation law: $\alpha(\omega) = a\omega^b$ along with its dispersion relation given by the Kramers-Kronig causality conditions \cite{Tripathi2019_PPM1D_CT, Tripathi2019_PPM2D_CT}.

The resulting system of equation was solved using a custom piecewise parabolic method, a high-order finite volume method \cite{Colella1984, Miller2002}.  Finite volume methods are the first choice for simulating shock waves. It discretizes the domain into volumes/cells and is designed to conserve the net-flux of the material in and out of the volume. This characteristic is important for shock wave propagation as it ensures that the Rankine-Hugoniot jump conditions are satsified \cite{Smoller2012}. A detailed illustration of the theoretical and the numerical model can be found in the references \cite{Tripathi2019_PPM1D_CT, Tripathi2019_PPM2D_CT}. 

\section{Experimental Validation}\label{Sec:Experiments}

\begin{figure}[t]
    \centering
    \begin{subfigure}{0.47\textwidth}
    \centering
        \caption{}
    \includegraphics[width=\textwidth]{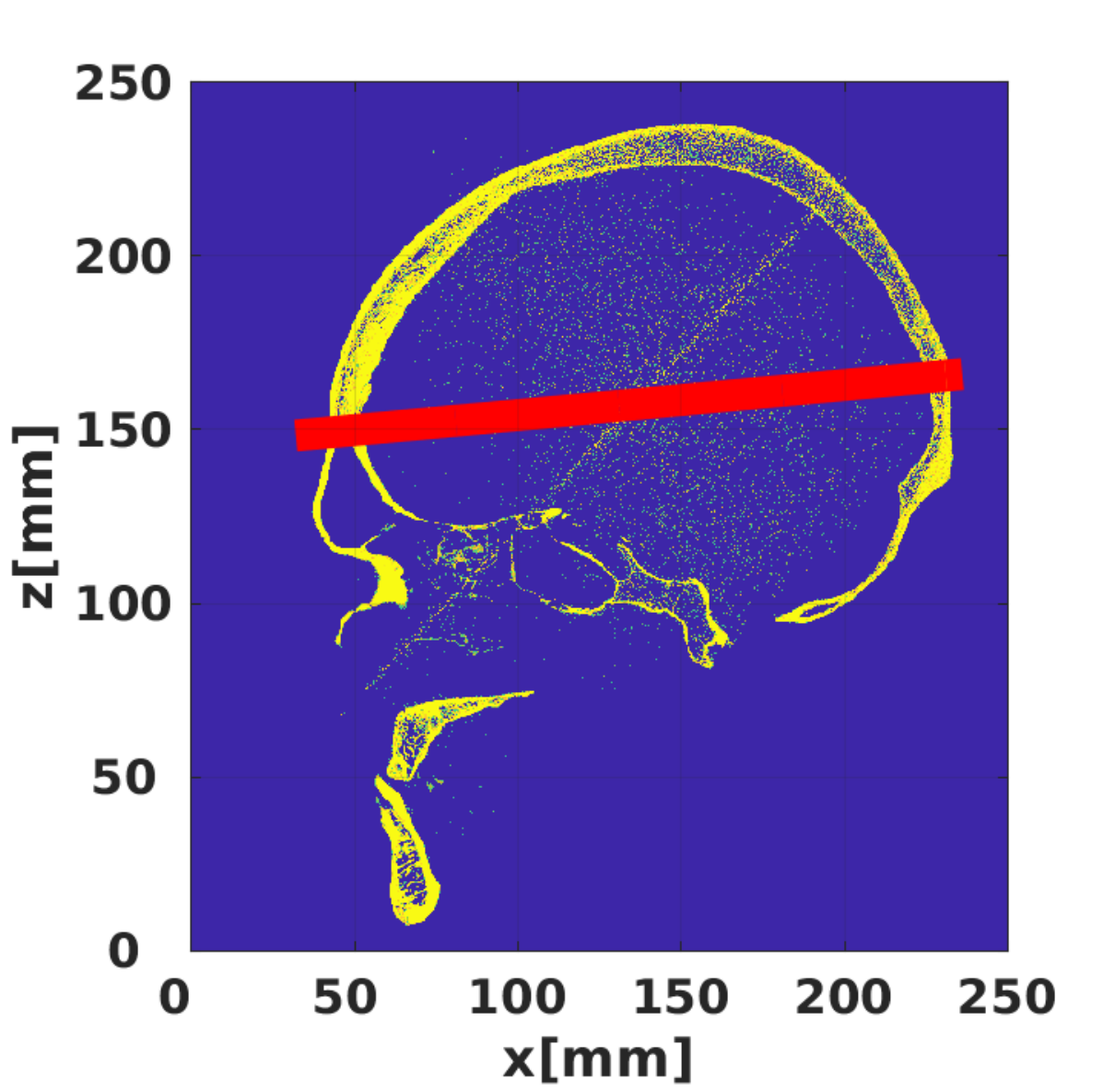}
        \label{Fig:Skull-VS-HS-a}
    \end{subfigure}
    \begin{subfigure}{0.40\textwidth}
    \centering
        \caption{}
        \includegraphics[width=\textwidth]{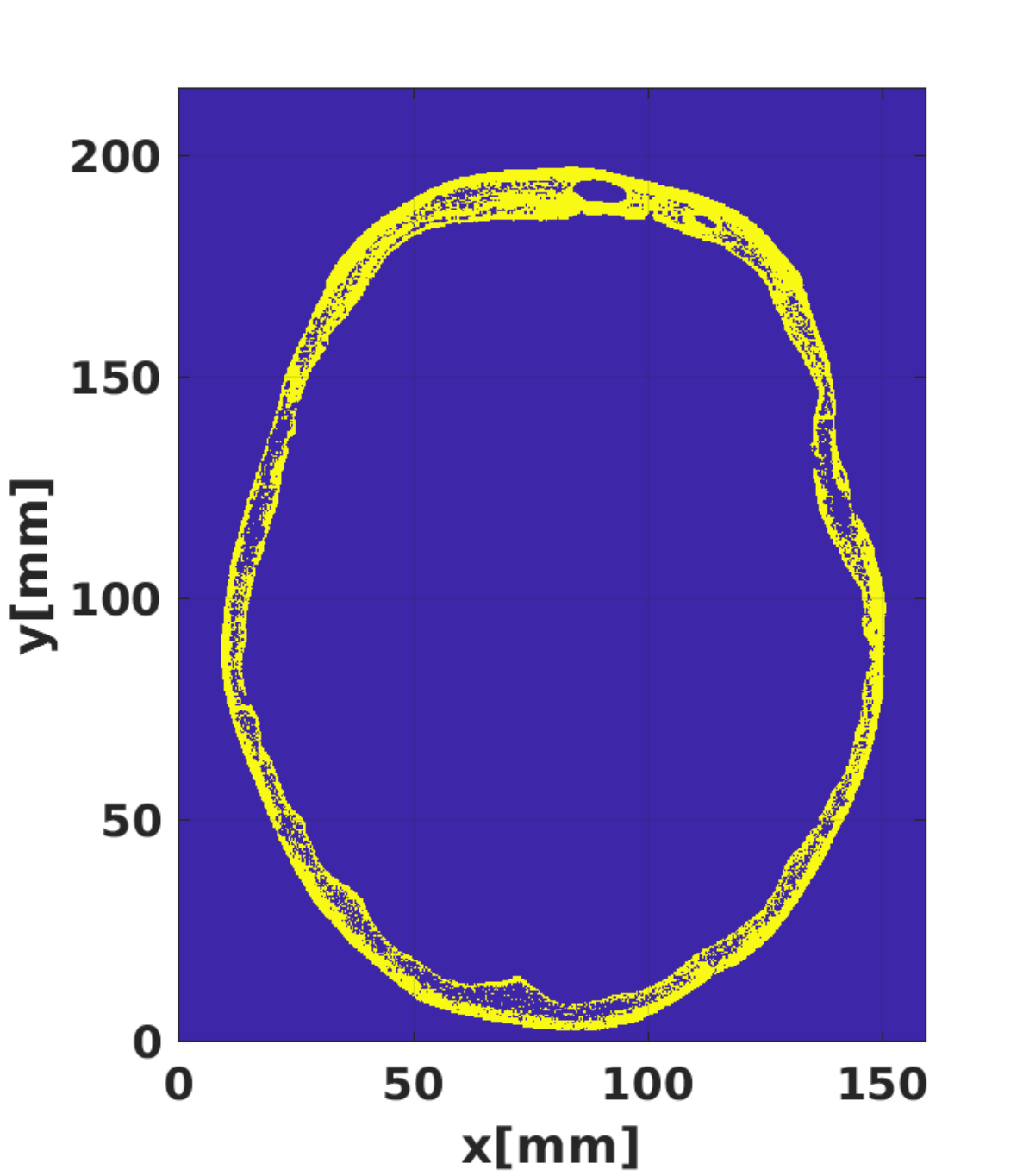} 
        \label{Fig:Skull-VS-HS-b}
    \end{subfigure}
    \caption{Sagittal section (a) of the human head CT scan showing with the red line showing the location of (b) the coronal plane  used in expriments and simulations of shear shock wave generation within the head. This interior surface of this 2D section is used as a source boundary generating linearly-polarized shear waves inside the closed geometry.}
    \label{Fig:Skull-VS-HS}
\end{figure}

\begin{figure}[htbp]
\begin{picture}(500.8,50.75)(-0.05,0.25)
\setlength{\unitlength}{0.48\textwidth}

\put(0.08,-0.38){\includegraphics[trim= 0 0 0 0, clip, height = 0.4\textwidth]{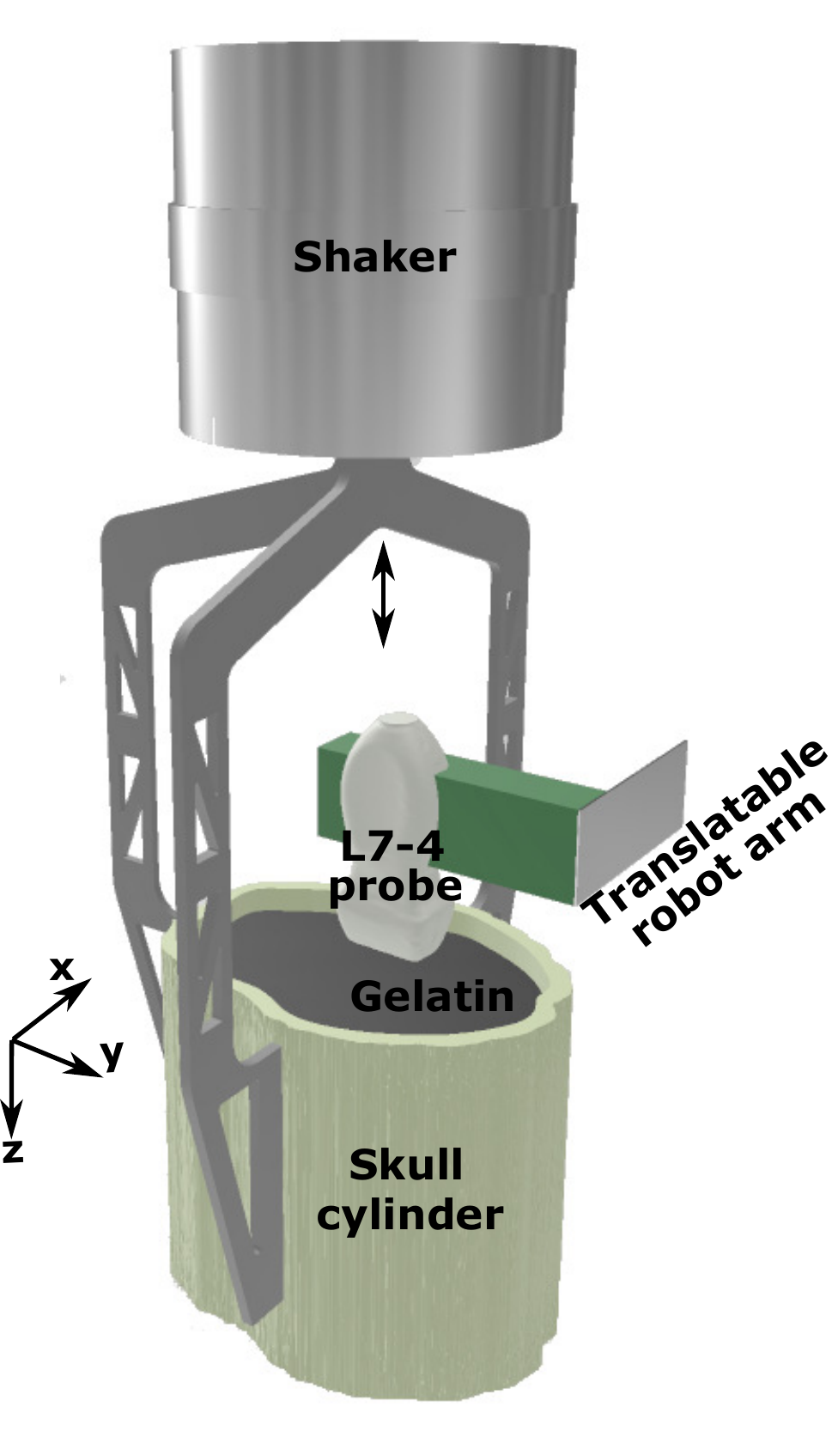}}
\put(0.5,0.30){\bf \color{black}\scriptsize{(a)}}

\put(0.59,-0.35){\includegraphics[trim= 4 1 10 3, clip, width=0.35\textwidth]{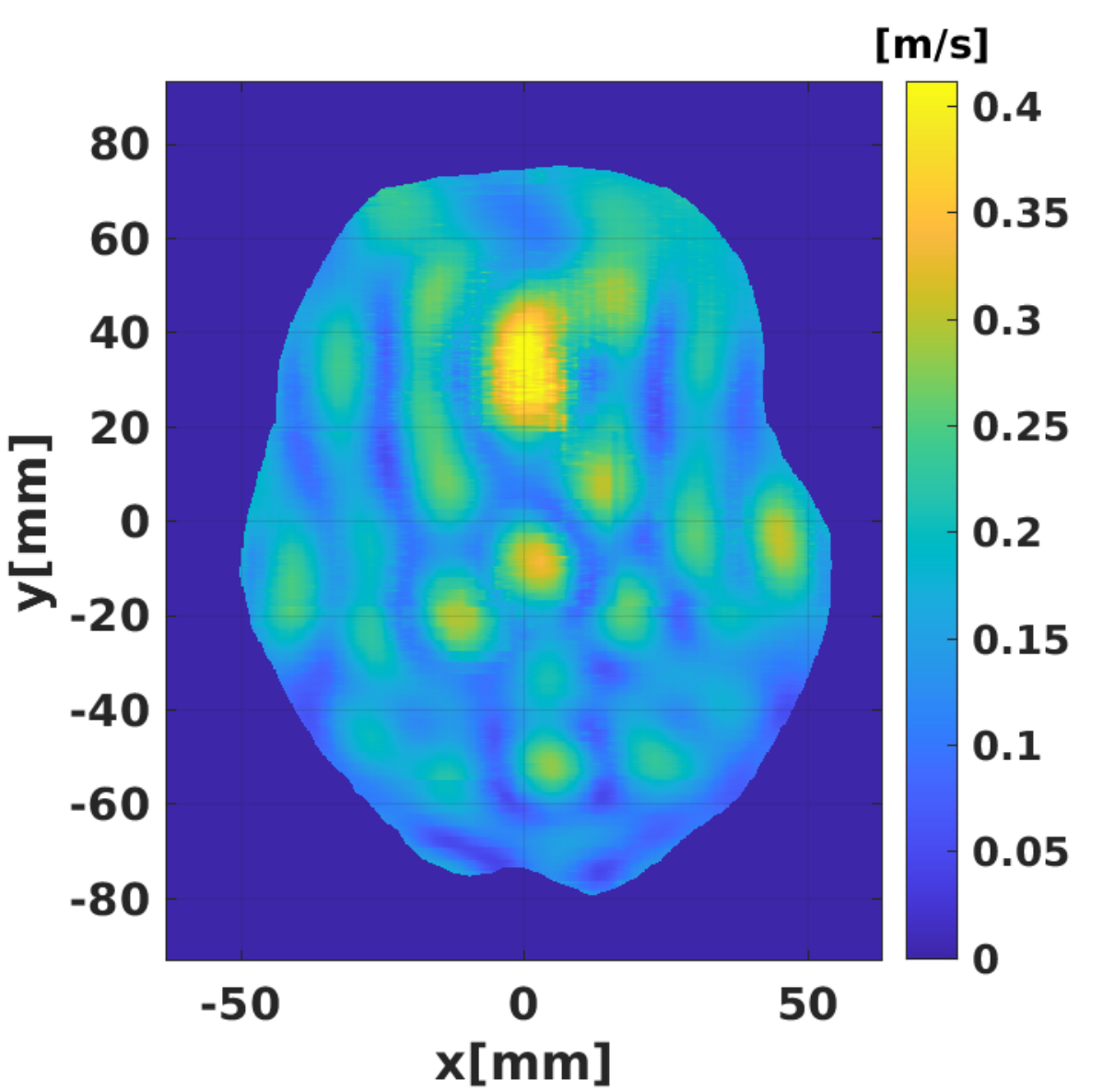}}
\put(1.13,0.30){\bf \color{white}\scriptsize{(b)}}

\put(1.3,-0.35){\includegraphics[trim= 4 1 10 3, clip, width=0.35\textwidth]{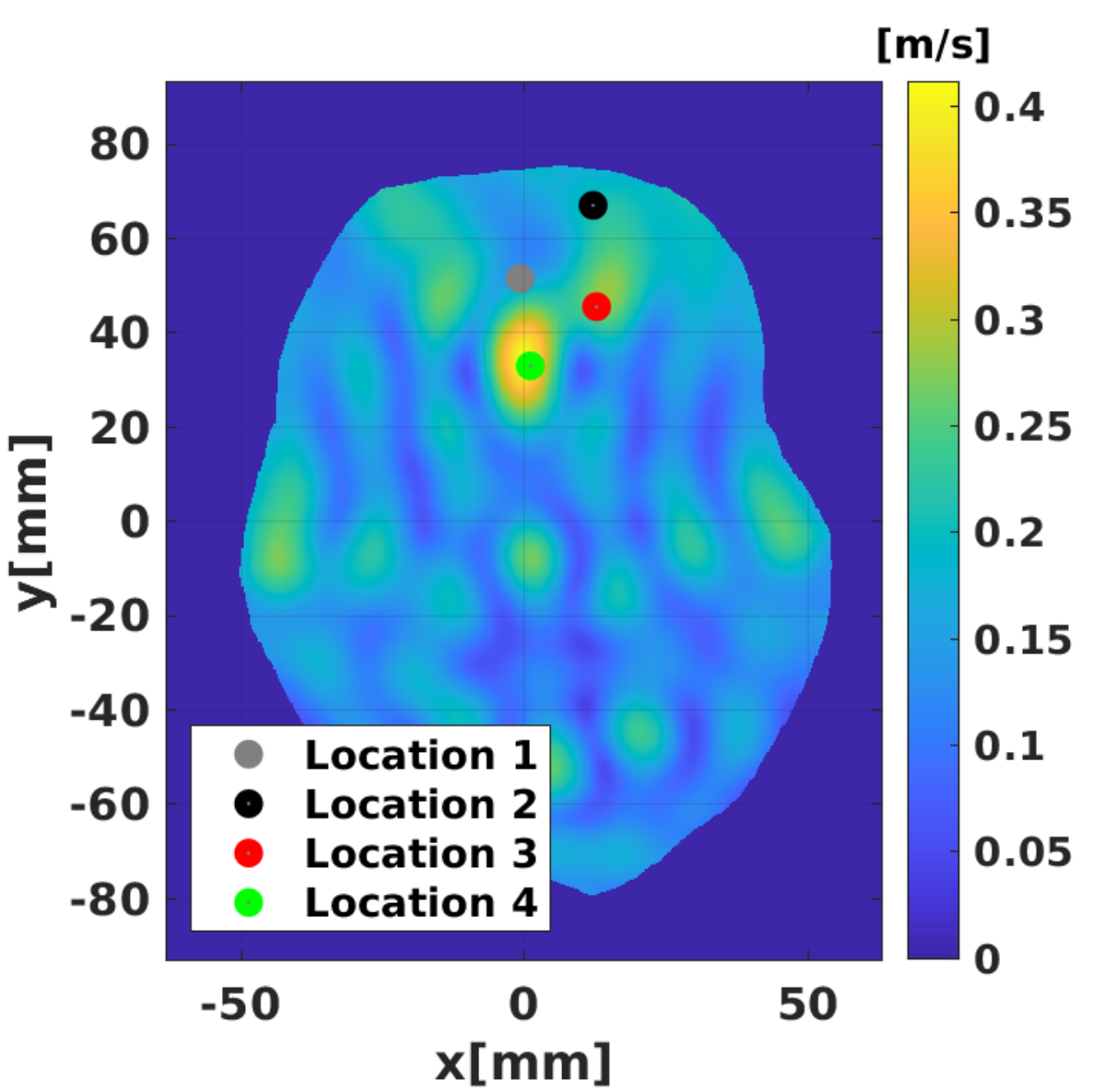}}
\put(1.85,0.30){\bf \color{white}\scriptsize{(c)}}

\end{picture}

\vspace{45mm}
\begin{picture}(500.8,0.75)(-0.05,0.25)
\setlength{\unitlength}{0.45\textwidth}
\put(0,-0.35){\includegraphics[trim= 4 1 15 3, clip, width=0.26\textwidth]{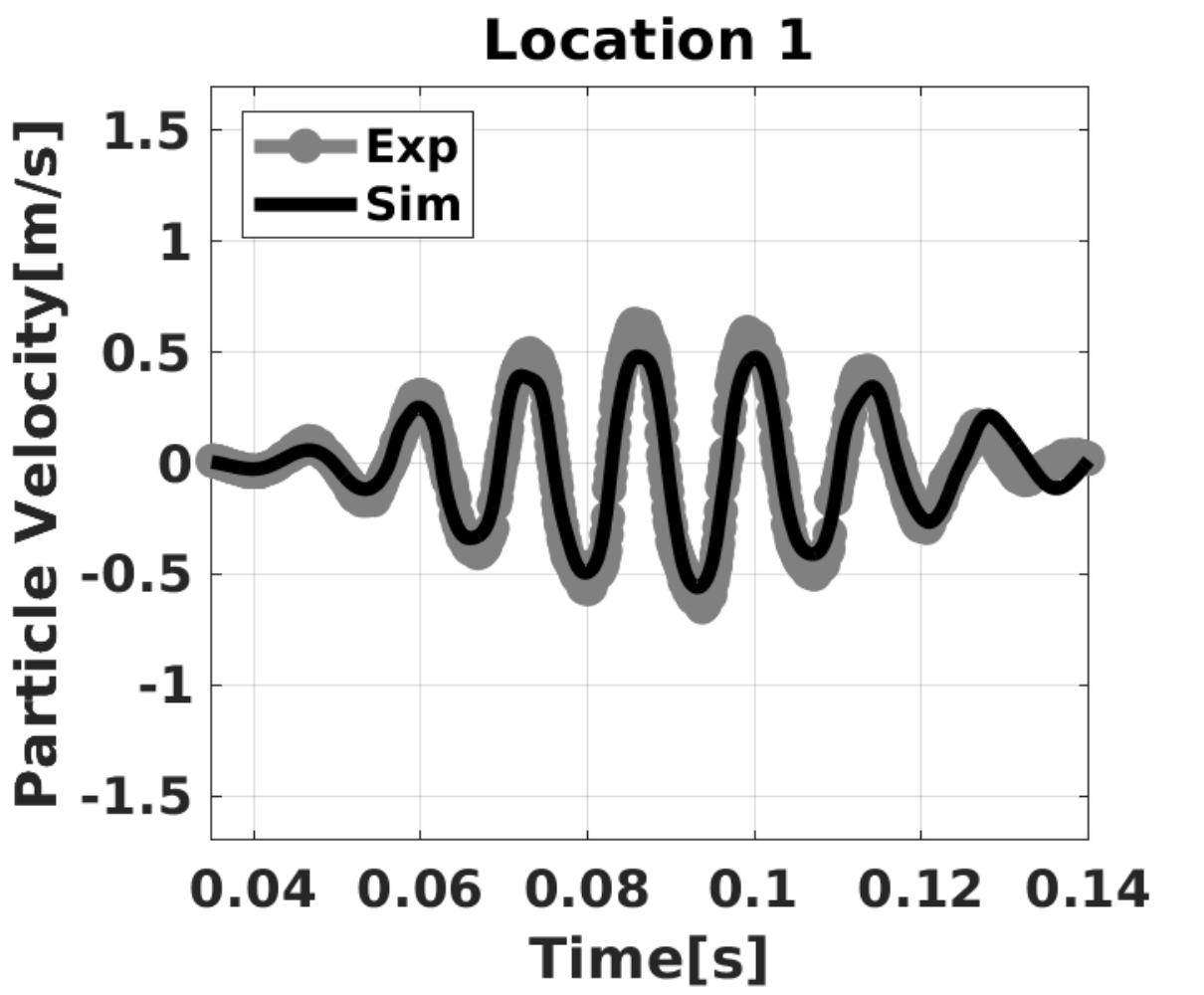}}
\put(0.5,0.08){\bf \color{black}\scriptsize{(d)}}

\put(0.60,-0.35){\includegraphics[trim= 55 1 15 3, clip, width=0.22\textwidth]{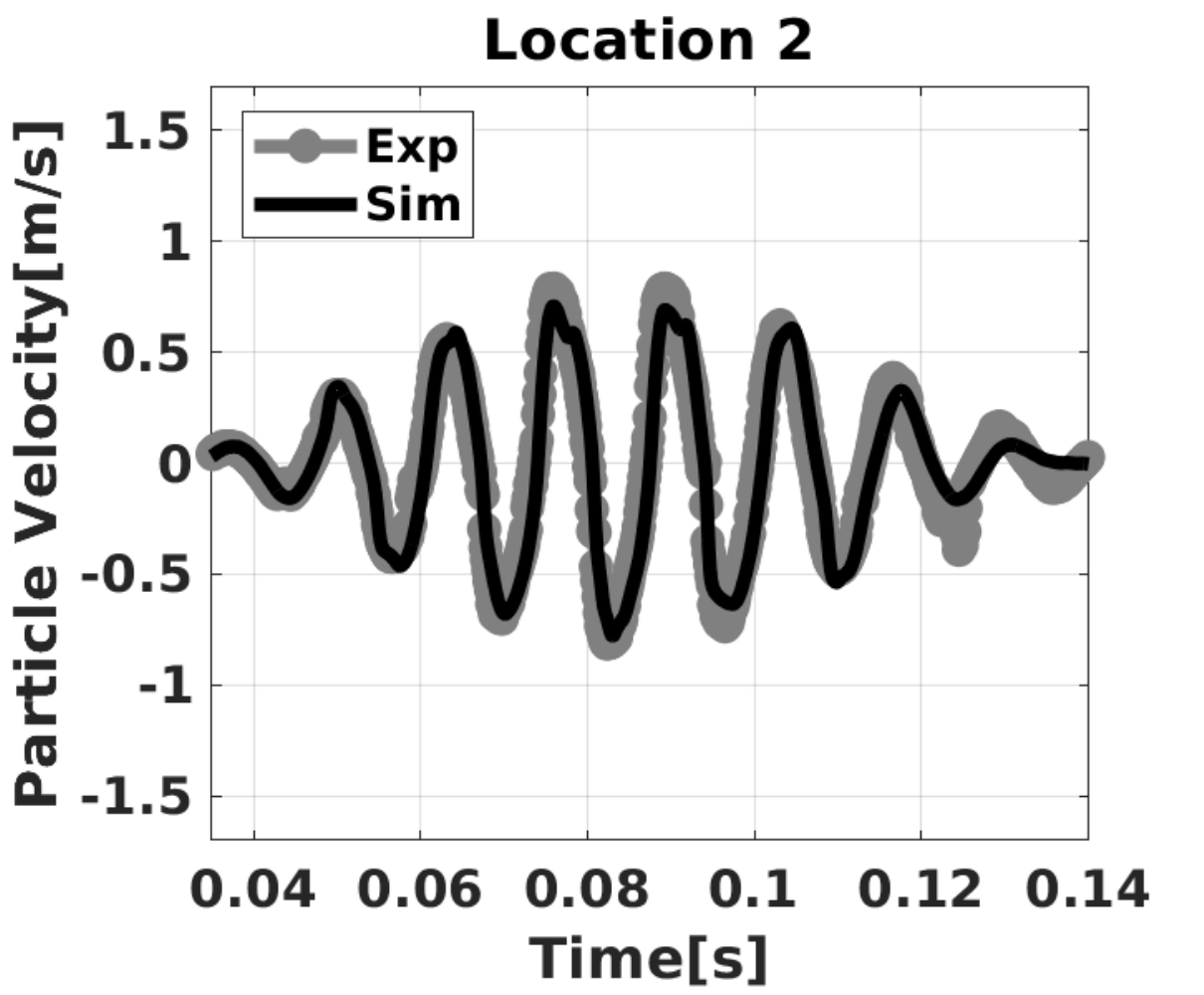}}
\put(1,0.08){\bf \color{black}\scriptsize{(e)}}
\put(1.11,-0.35){\includegraphics[trim= 55 1 15 3, clip, width=0.22\textwidth]{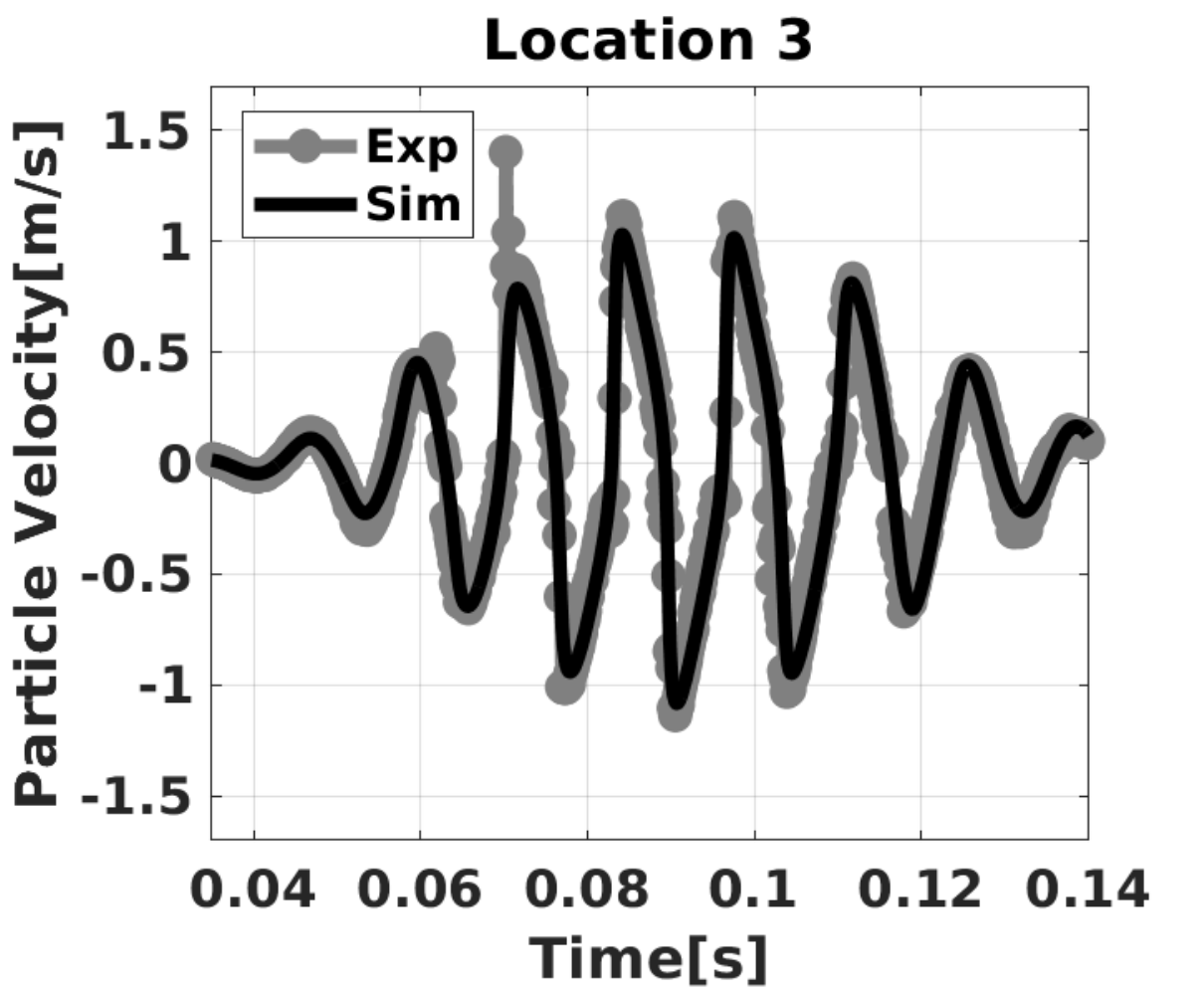}}
\put(1.52,0.08){\bf \color{black}\scriptsize{(f)}}

\put(1.62,-0.35){\includegraphics[trim= 55 1 15 3, clip, width=0.22\textwidth]{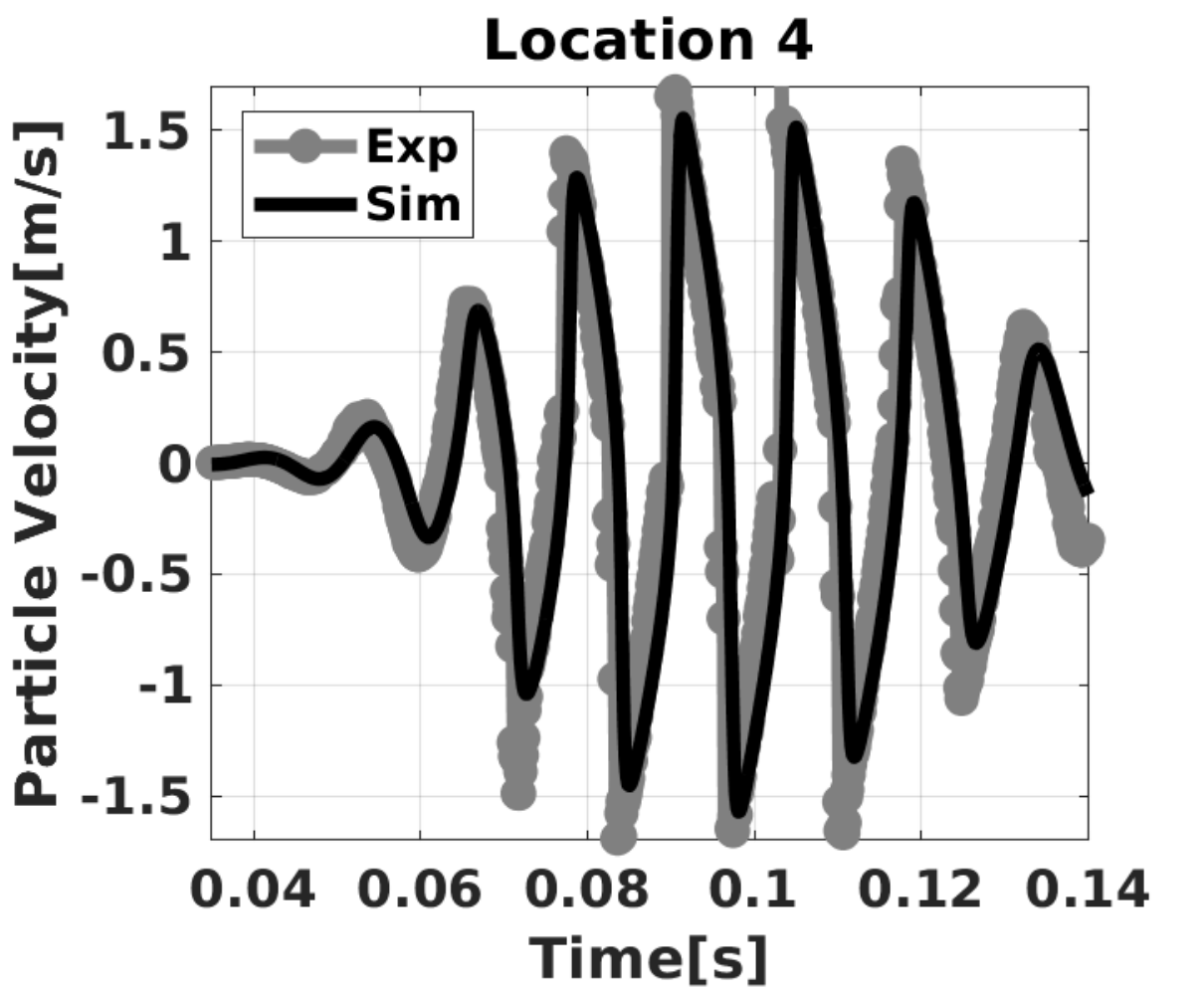}}
\put(2.02,0.08){\bf \color{black}\scriptsize{(g)}}

\end{picture}

\vspace{37mm}
\begin{picture}(500.8,0.75)(-0.05,0.25)
\setlength{\unitlength}{0.45\textwidth}
\put(0,-0.35){\includegraphics[trim= 4 1 15 3, clip, width=0.263\textwidth]{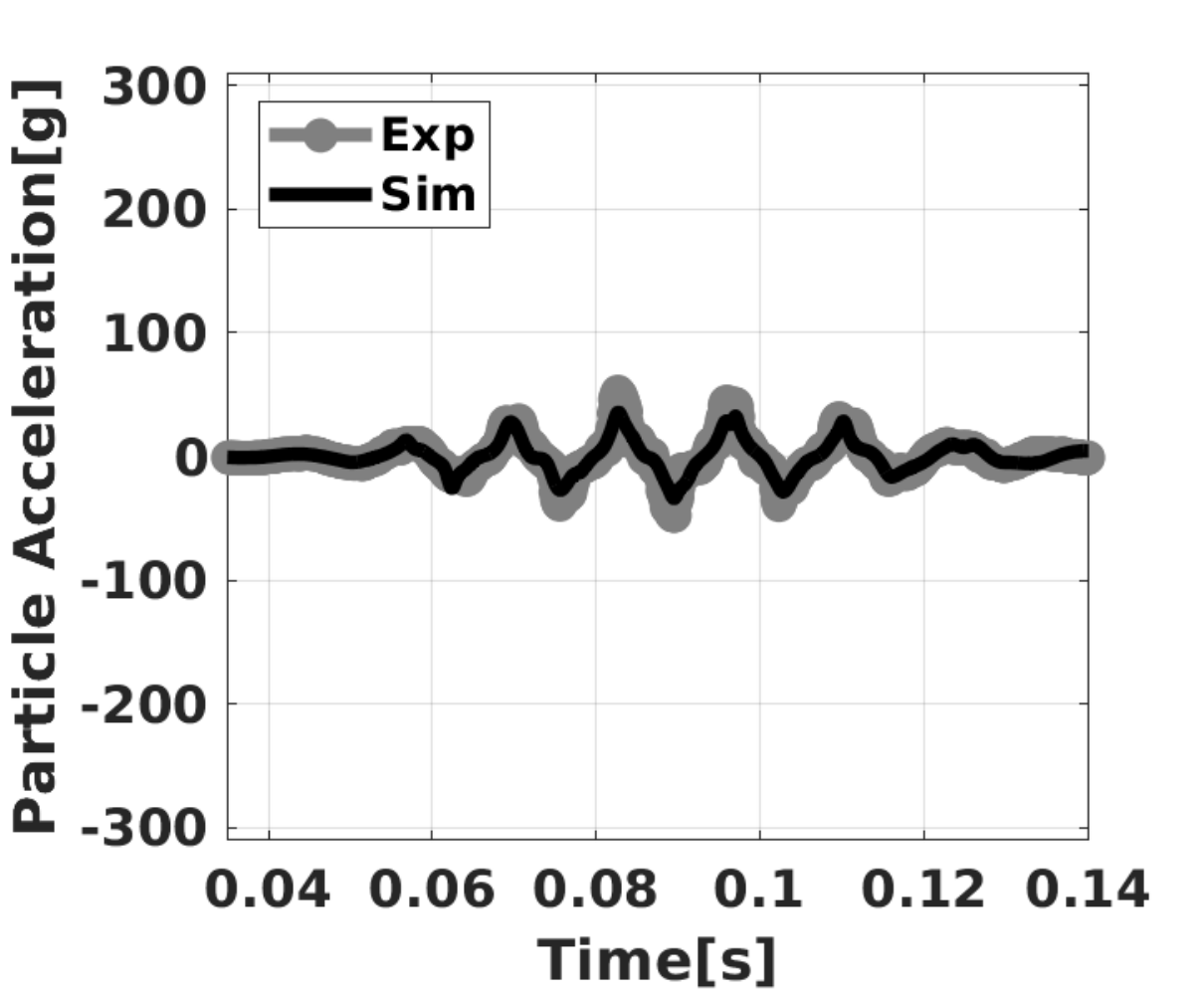}}
\put(0.5,0.08){\bf \color{black}\scriptsize{(h)}}

\put(0.60,-0.35){\includegraphics[trim= 60 1 15 3, clip, width=0.22\textwidth]{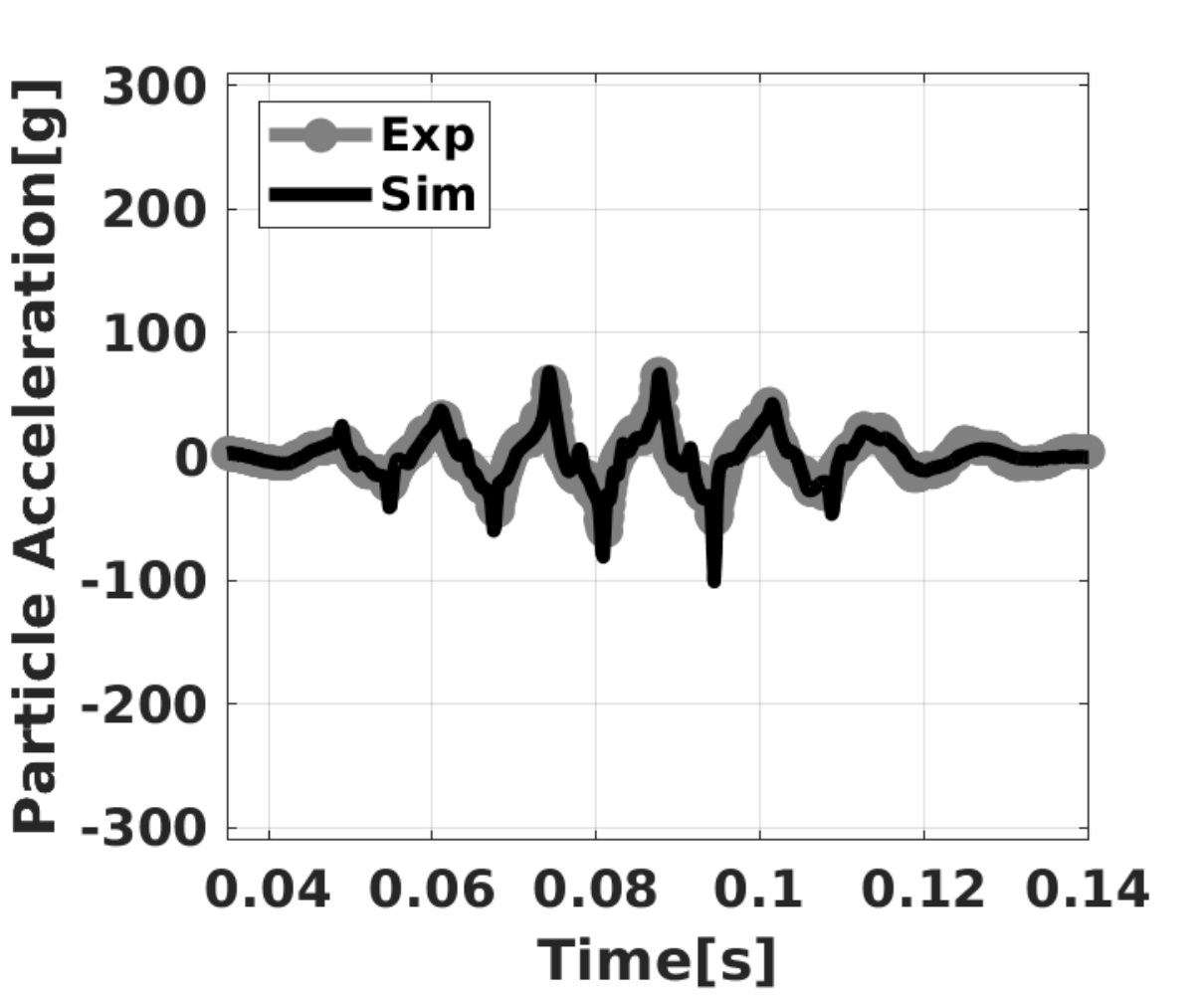}}
\put(1,0.08){\bf \color{black}\scriptsize{(i)}}

\put(1.11,-0.35){\includegraphics[trim= 60 1 15 3, clip, width=0.22\textwidth]{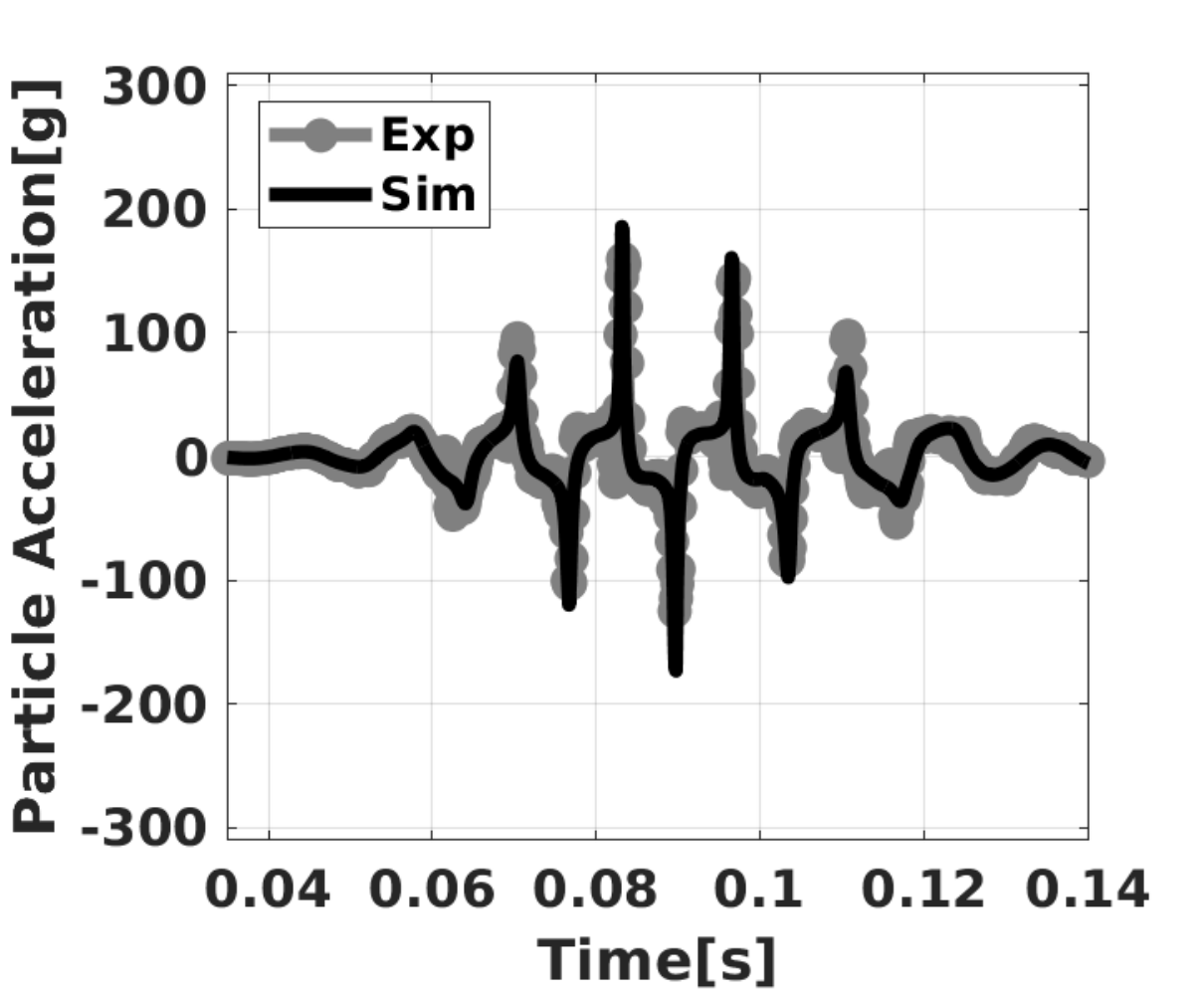}}
\put(1.52,0.08){\bf \color{black}\scriptsize{(j)}}

\put(1.62,-0.35){\includegraphics[trim= 60 1 15 3, clip, width=0.22\textwidth]{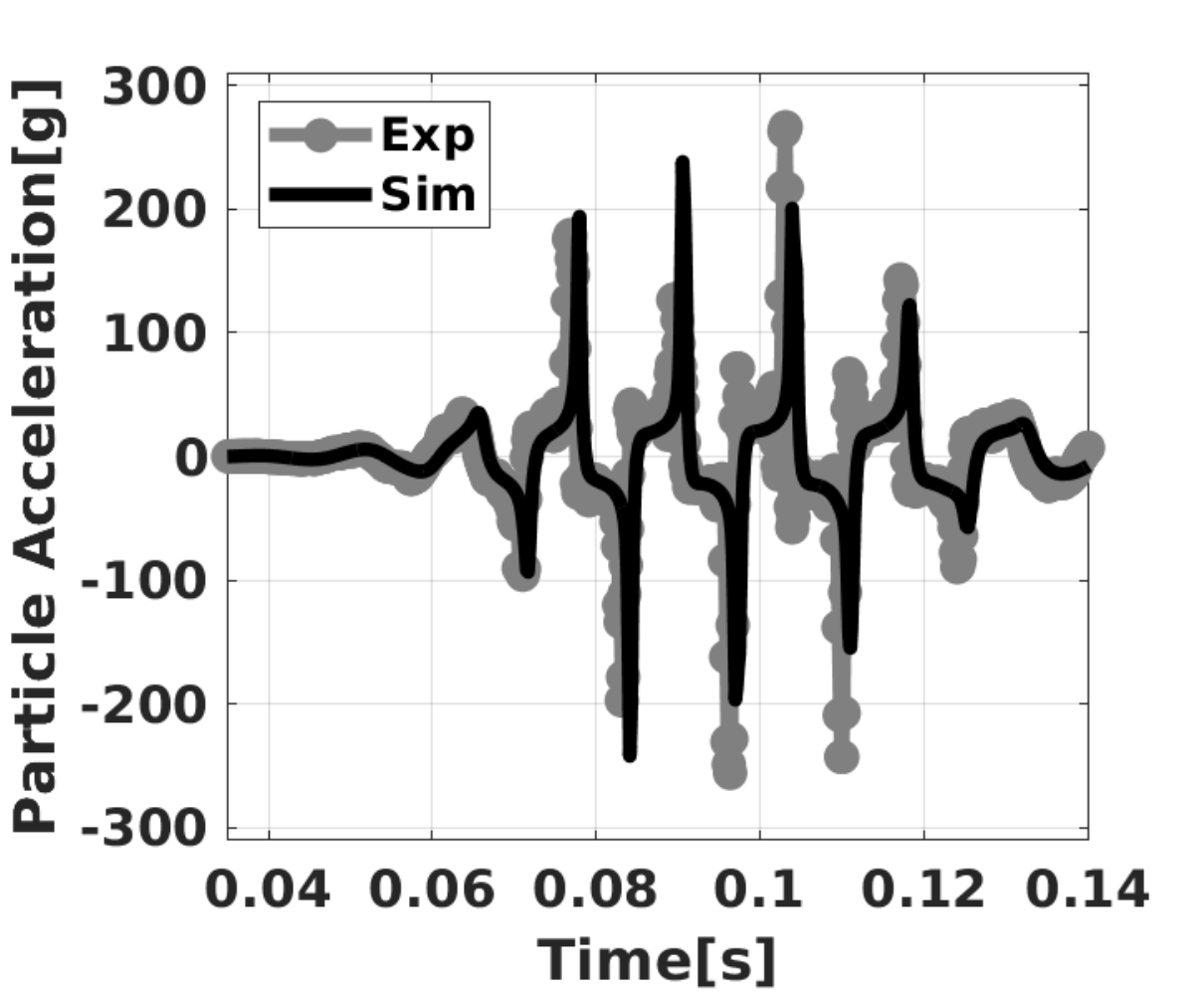}}
\put(2.02,0.08){\bf \color{black}\scriptsize{(k)}}

\end{picture}

\vspace{35mm}
\begin{picture}(500.8,0.75)(-0.05,0.25)
\setlength{\unitlength}{0.45\textwidth}
\put(0,-0.35){\includegraphics[trim= 4 1 15 3, clip, width=0.26\textwidth]{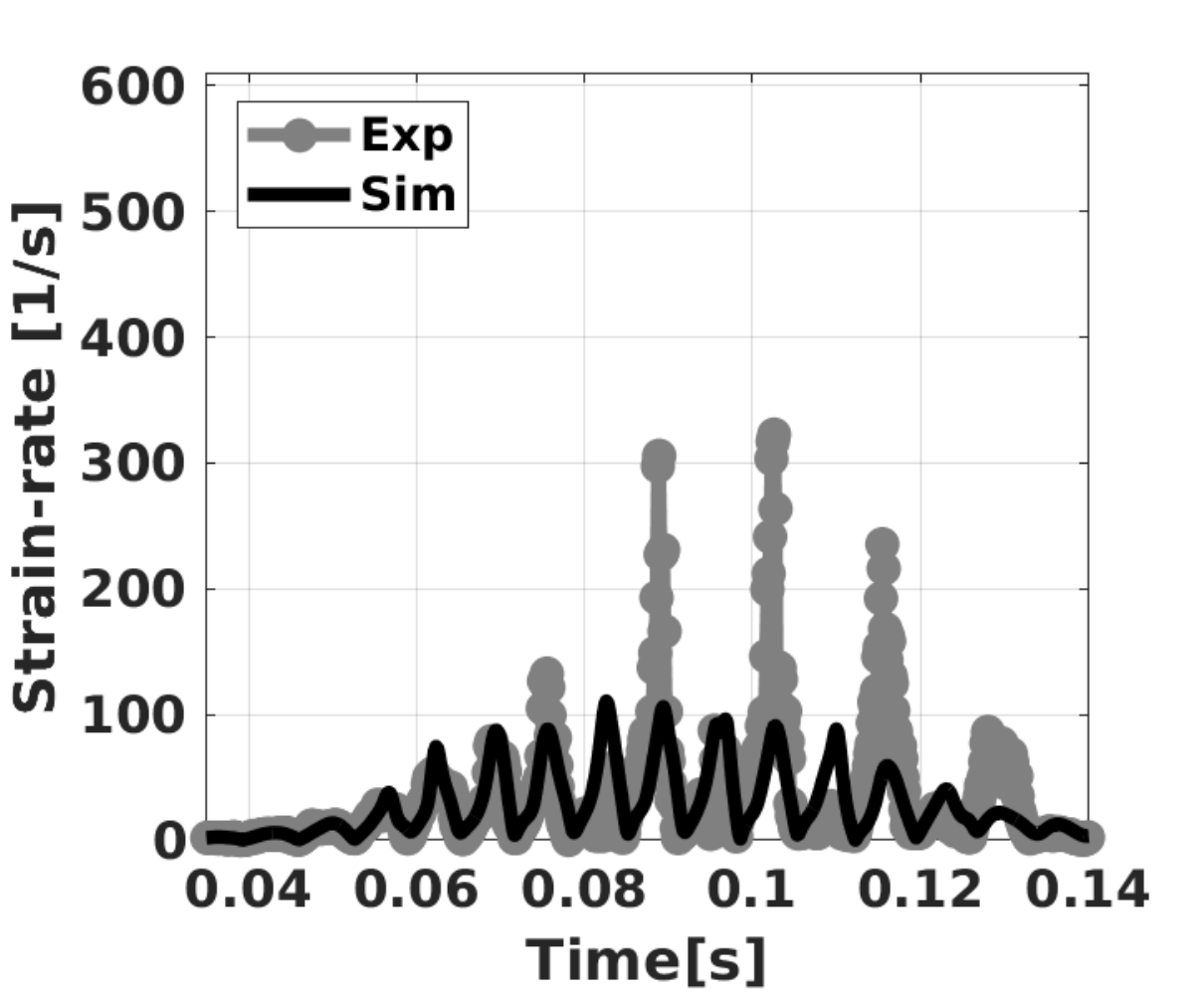}}
\put(0.5,0.08){\bf \color{black}\scriptsize{(l)}}

\put(0.60,-0.35){\includegraphics[trim= 55 1 15 3, clip, width=0.22\textwidth]{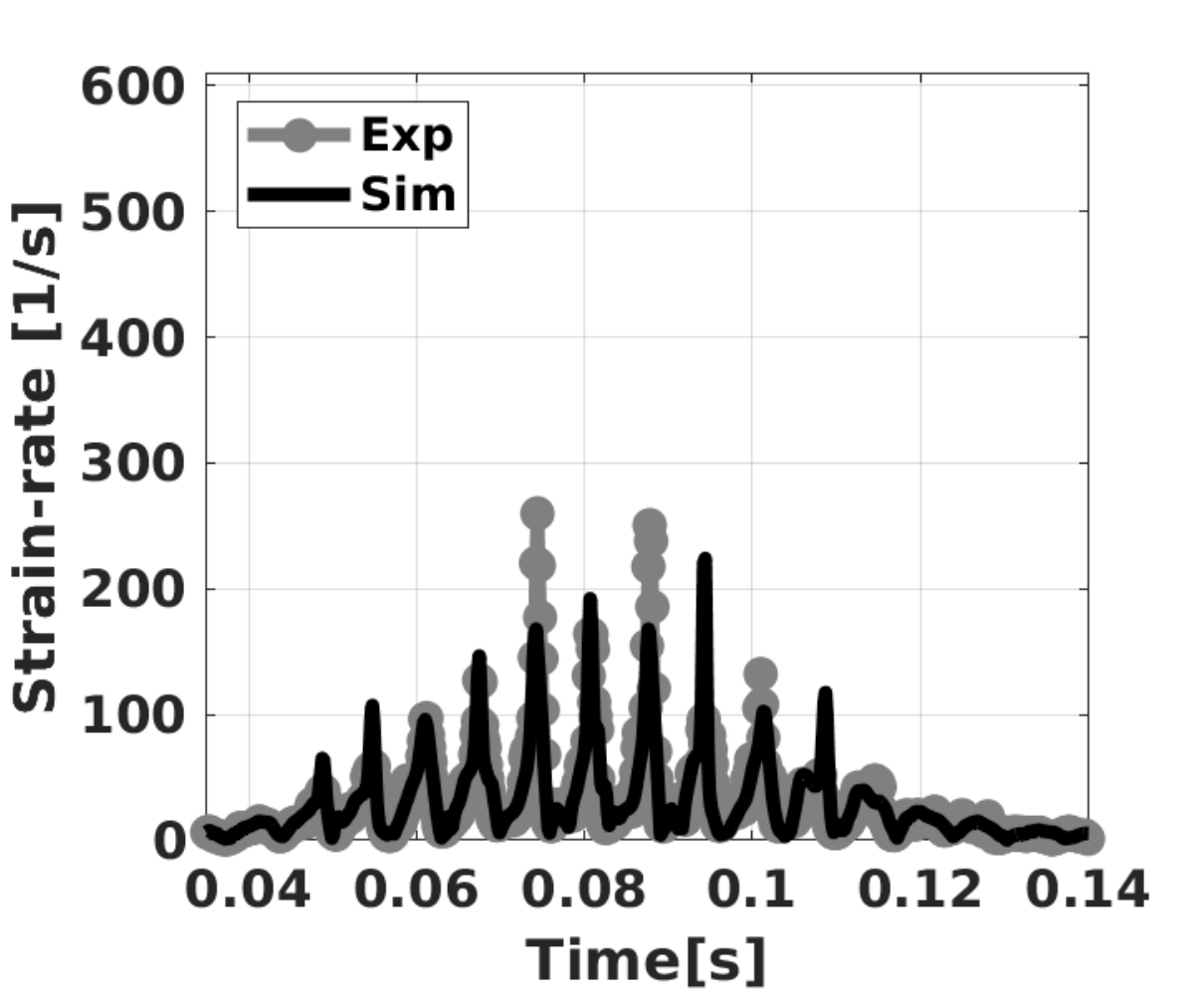}}
\put(1,0.08){\bf \color{black}\scriptsize{(m)}}

\put(1.11,-0.35){\includegraphics[trim= 55 1 15 3, clip, width=0.22\textwidth]{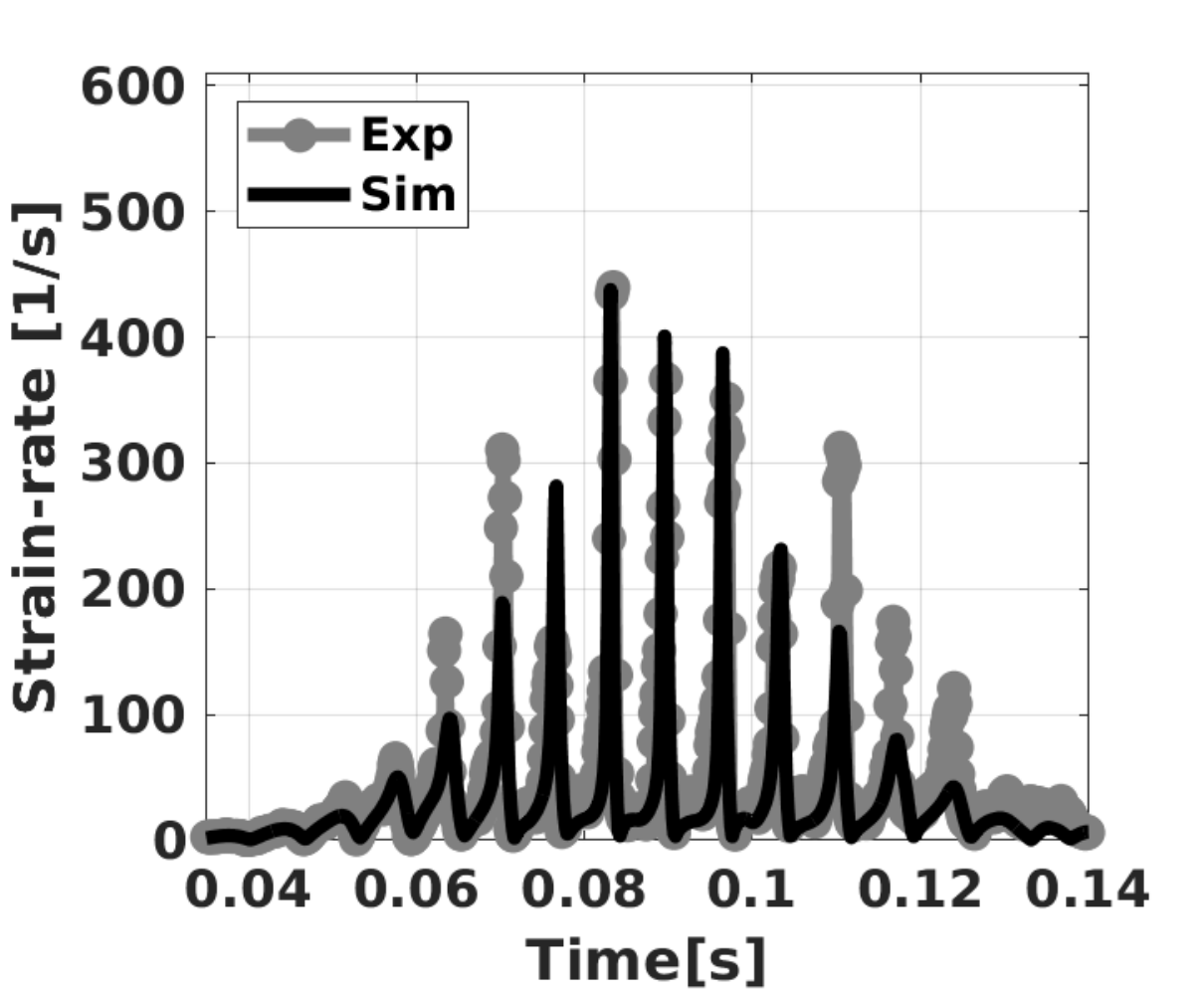}}
\put(1.52,0.08){\bf \color{black}\scriptsize{(n)}}

\put(1.62,-0.35){\includegraphics[trim= 55 1 15 3, clip, width=0.22\textwidth]{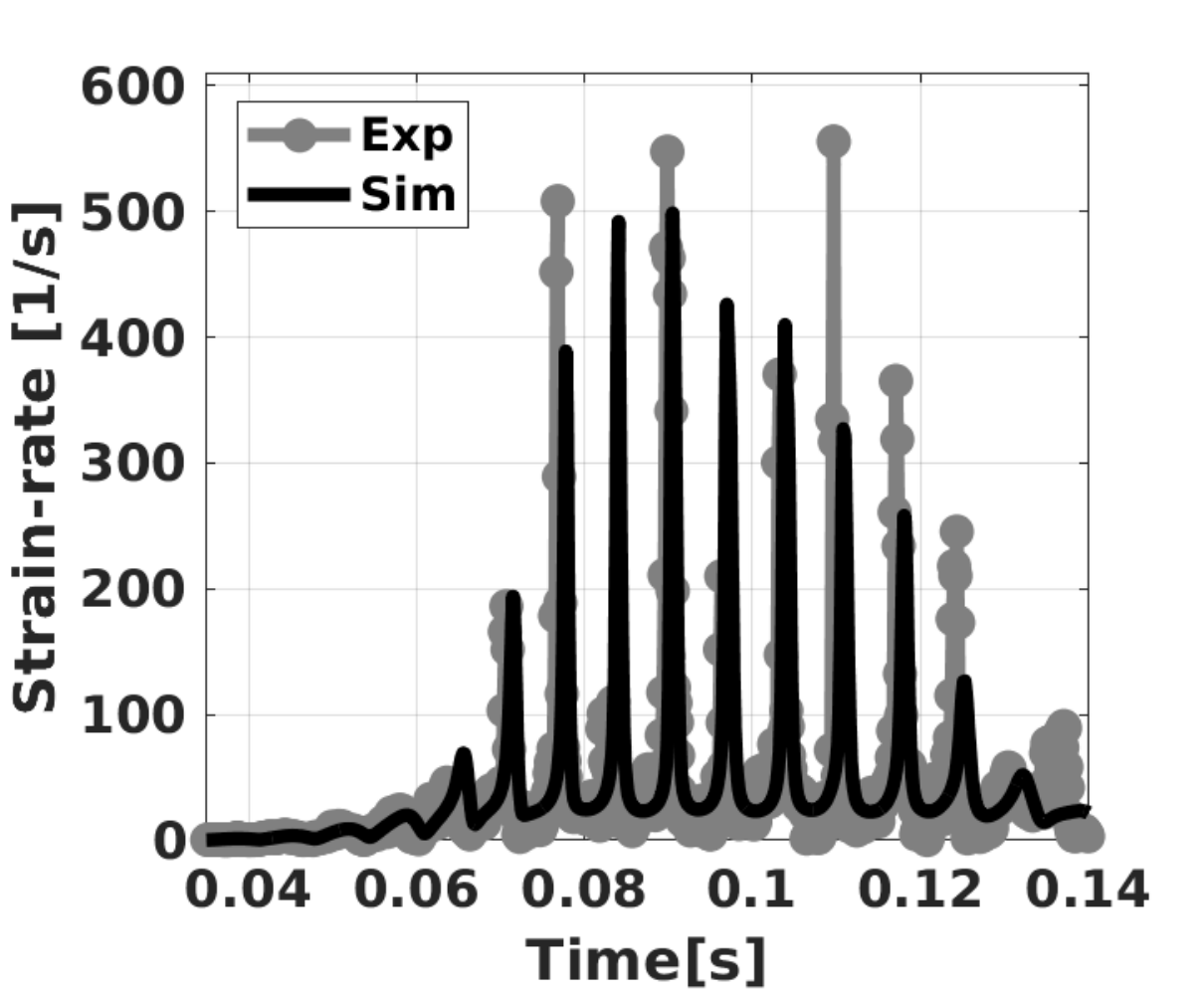}}
\put(2.02,0.08){\bf \color{black}\scriptsize{(o)}}

\end{picture}

\vspace{30mm}
\caption{(a): Snapshot of the experimental setup showing the human head phantom clamped from three sides filled with a gelatin phantom submerged in water with an ultrasound probe over it. (b): the RMS average in time calculated for all space points using the experimental data. (c): the RMS average in time calculated for the simulated data using the physical parameters obtained from the planar experiments: $\beta=3.5$, c(75 Hz) = 1.9 m/s, $\alpha(\omega) = 0.012\omega^{1.12}$.
Second, third, and fourth row show velocity, acceleration, strain-rate at 4 different spatial location (black), shown in the legend in (c) plot, along with the experimental reference (grey). Note the steepening of the waveform with increase in amplitude, and spike in the local acceleration and strain-rate at in the vicinity of the steep profile. Propagation movie for both the experiment and simulation is provided in the supplementary material. 
}
\label{Fig:Skull-Exp}
\end{figure}

To further validate the numerical solver in a realistic morphology of human head an experiments was performed. An  axial section of human skull was extracted along the red line from the CT image of a human head as shown in  Fig. \ref{Fig:Skull-VS-HS}a was extracted. The axial section Fig. \ref{Fig:Skull-VS-HS}b was extruded and 3D printed into a mould for fabricating a fiberglass skull cylinder (Fig. \ref{Fig:Skull-Exp}a). Over the mould surface, two plies of a fiberglass fabric were laid up using an epoxy resin and allowed to cure for 24hrs. The resulting skull-cylinder was subsequently filled with a brain mimicking gelatin mixture (5\% by volume). Based on the linear planar experiments the linear shear wave speed and the attenuation power-law were obtained to be $c(75 {\rm Hz})= 1.90$ m/s  and $\alpha(\omega) = 0.012\omega^{1.12}$ Np/m, respectively. Using PPM1D \cite{Tripathi2019_PPM1D_CT}, this gelatin was calibrated to be with the nonlinear parameter $\beta = 3.5 \pm 0.4$. The skull phantom was excited using a VTS-100 electromechanical shaker (Vibration Test Systems, Aurora, Ohio). A 75 Hz shear wave input with an 8-cycle -80 dB Chebychev window was used as an input to the shaker. The direction of gelatin motion is in the same direction as the ultrasound imaging wave i.e. along the $z$-axis and the direction of shear wave propagation is orthogonal to this axis, i.e. $xy$-plane. 
The resulting 2D shear wave propagation was observed by imaging throughout the gelatin surface using a Verasonics Vantage ultrasound scanner (Verasonics, Kirkland, WA, USA). The scanning of the gelatin surface was done by a 5.2 MHz ultrasound probe (ATL L7–4, Philips, Bothell, WA, USA) attached to a   six degree of freedom robotic arm (IRB 120, ABB Ltd, Zurich, Switzerland). It  had access to the surface at the top of the head phantom to obtain measurements at depth (up to 8 cm) within its entire volume. Custom high frame-rate (6000 images/second) imaging sequences were acquired and the beamformed RF data was processed with \cite{espindola2017High} adaptive and tracking algorithms \cite{pinton2014adaptive} to detect displacements smaller than 1 $\mu$m. By scanning the robot arm and subsequently stitching together the 2D movies,  the displacement estimates where obtained within the approximately 130 x 110 x 100 mm volume at 6000 volumes/second.

Shear wave focusing within the head, in the middle of anterior region, can be seen in the measured time-averaged RMS velocity averaged over a depth range of 60 to 100 mm (Fig. \ref{Fig:Skull-Exp}b). In addition to the prominent focal spot with a maximum RMS = 0.44 m/s at $x=1.04$ mm, $y=28.2$ mm, a number of other regions also give rise to local maxima in RMS velocity. The velocity measurements at the gelatin boundary, just inside the skull surface, and the linear, nonlinear, and viscous parameters of the brain-mimicking gelatin phantom were used an input to the simulation tool. With these inputs a close match between the experimental {(Fig. \ref{Fig:Skull-Exp}b)} and simulated {(Fig. \ref{Fig:Skull-Exp}c)} time-averaged RMS velocity was obtained. The local dynamics as a function of time are available throughout the volume. Four specific points in space were selected to illustrate the local time-dependent wave dynamics in terms of the velocity (Fig. \ref{Fig:Skull-Exp}d-g), acceleration {(Fig. \ref{Fig:Skull-Exp}h-k)}, and Lagrangian strain-rate {(Fig. \ref{Fig:Skull-Exp}l-o)}. At all positions there is a close match between experiments and simulations. For a low velocity, at location 1, the wave propagation is approximately linear and it retains the quasi-monochromatic sinusoidal shape that was originally generated by the shaker. As the particle velocity increases to its maximum, at location 4, the wave undergoes significant distortion that is well described by cubically nonlinear shear-stiffening elastodynamics \cite{Catheline2003,Tripathi2019_PPM2D_CT}. This behavior is observable in the waveform due to the high amplitude leading to stronger nonlinear effects thus generating the characteristic shark fin profile. Propagation movie in velocity for both experiment and simulation is provided in the supplementary material gives a better understanding of the focal effect.
The acceleration and strain-rate depend on the computation of a temporal derivative, which can be noise sensitive. These computations have been previously validated experimentally and numerically using simulations that model acoustic wave propagation in a medium undergoing shear wave deformation and they have been shown to be accurate to at least the 11th harmonic of the velocity signal, which coincides roughly to the sensitivity limit of the experimental ultrasound-based displacement estimates.\cite{Espindola2017} 
In the linear regime, the acceleration would retain a sinusoidal shape. However, any nonlinear distortion of the wave is magnified by the temporal derivative since it is rapidly amplified at the steep shock-front gradients. Thus, even the acceleration for the low-amplitude point {(Fig.~\ref{Fig:Skull-Exp}h)} does not have a purely sinusoidal shape. At large particle velocity amplitudes {(Fig.~\ref{Fig:Skull-Exp}k)} this effect is explosive and the local acceleration at the shock-front is dramatically amplified. The acceleration at the shaker surface, measured by an accelerometer (PCB Piezotronics, Inc., Depew, NY, USA) was 19$g$ which is 14 times smaller than the 266$g$ acceleration measured at the focal peak. Note that a 19$g$ impact is very rarely injurious and that the lower range of mild traumatic brain injuries occur for impacts above 35$g$ \cite{greenwald2008head,bayly2005deformation}. Estimates of the Lagrangian strain-rate (Fig.~\ref{Fig:Skull-Exp}l-o) exhibit trends that are similar to the acceleration. At the shock front the strain-rate increases dramatically and at the focus strain-rates up to 551 1/s were observed. Taken together these results thus demonstrate that shear shock waves are focused by the skull geometry and the cumulative nonlinear elastodynamic effects rapidly amplify the acceleration and strain-rates at the focus, though this is valid for this particular gelatin phantom.

\begin{figure}[htbp]
\vspace{10mm}
\hspace{-5mm}
\begin{picture}(500.8,0.75)(-0.05,0.25)
\setlength{\unitlength}{0.45\textwidth}
\put(0,-0.35){\includegraphics[trim=  5 1 18 20, clip, width=0.245\textwidth]{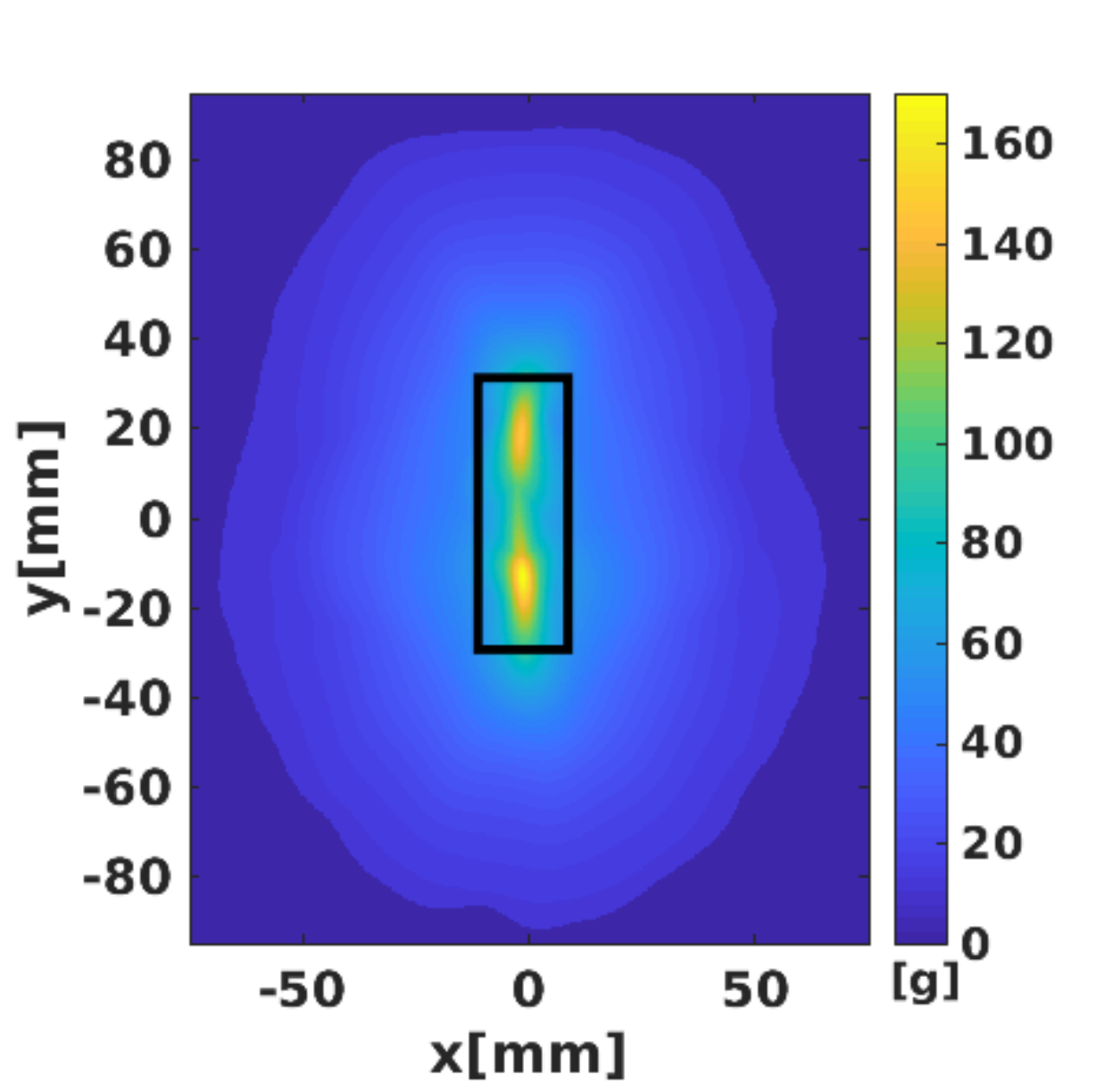}}
\put(0.357, -0.27){\includegraphics[trim=  55 55 200 35, clip, width=0.04\textwidth]{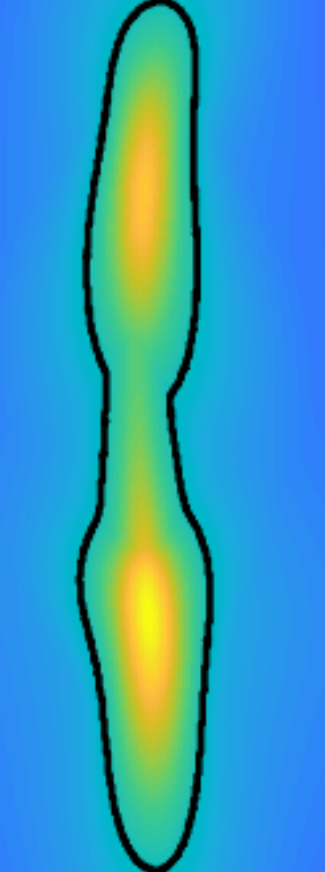}}
\put(0.09,0.14){\bf \color{white}\scriptsize{(a)}}
\put(0.58,-0.35){\includegraphics[trim= 5 1 18 20, clip, width=0.245\textwidth]{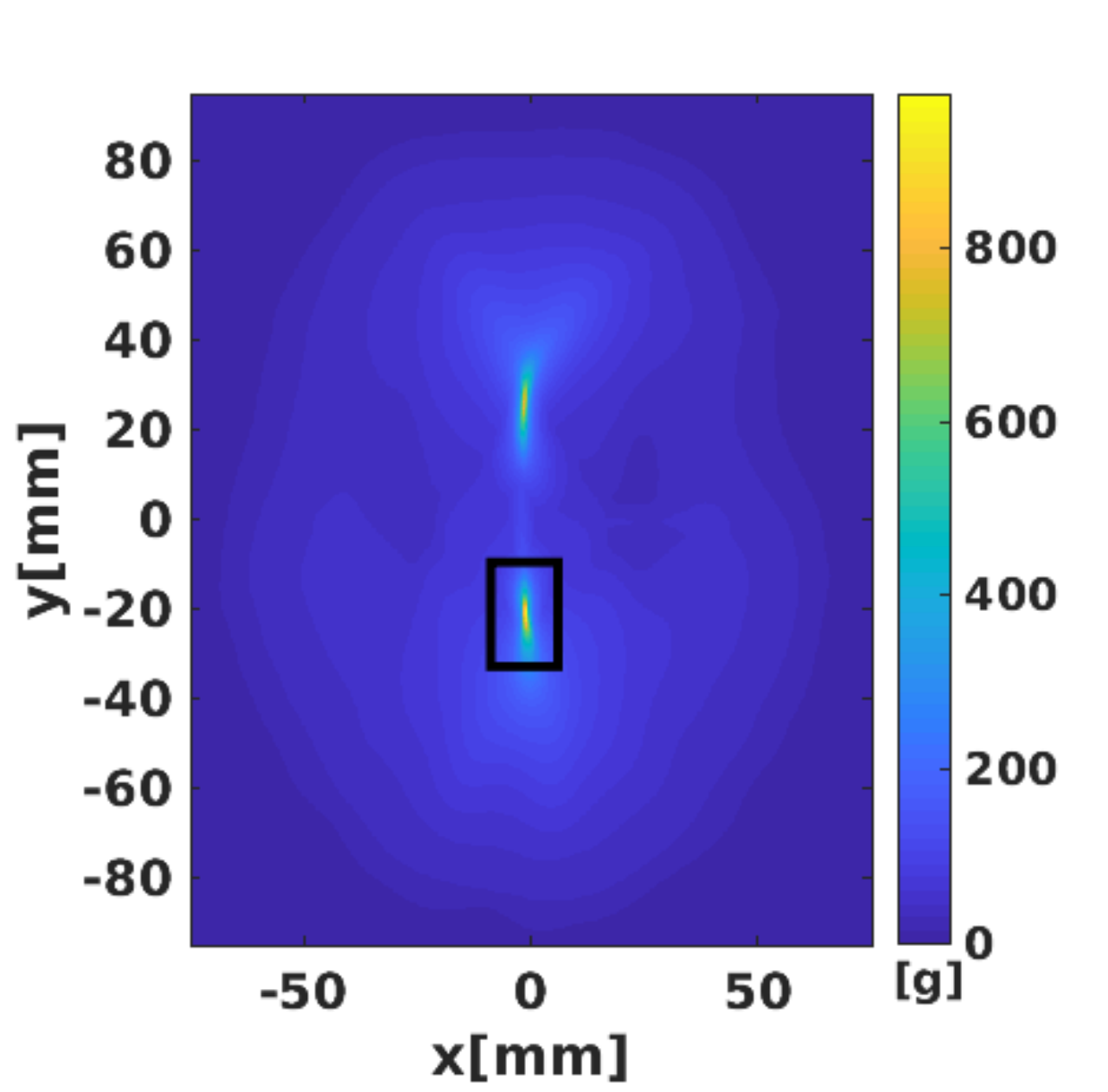}}
\put(0.91, -0.27){\includegraphics[trim=  55 55 170 40, clip, width=0.05\textwidth]{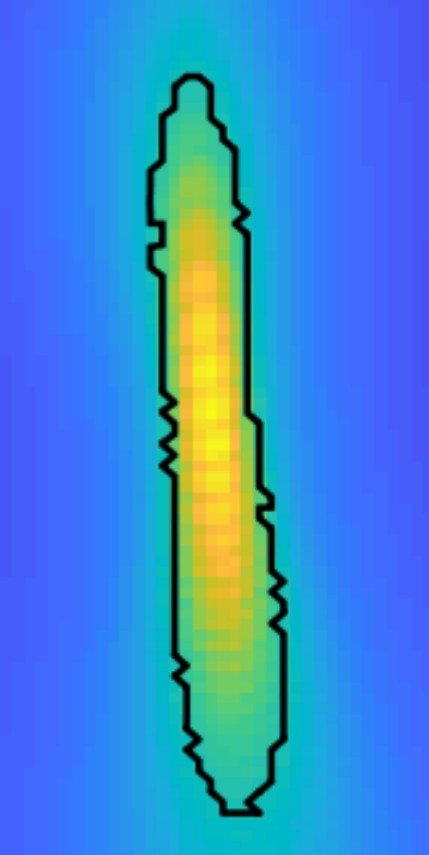}}
\put(0.68,0.14){\bf \color{white}\scriptsize{(b)}}
\put(1.16,-0.35){\includegraphics[trim= 5 1 18 20, clip, width=0.245\textwidth]{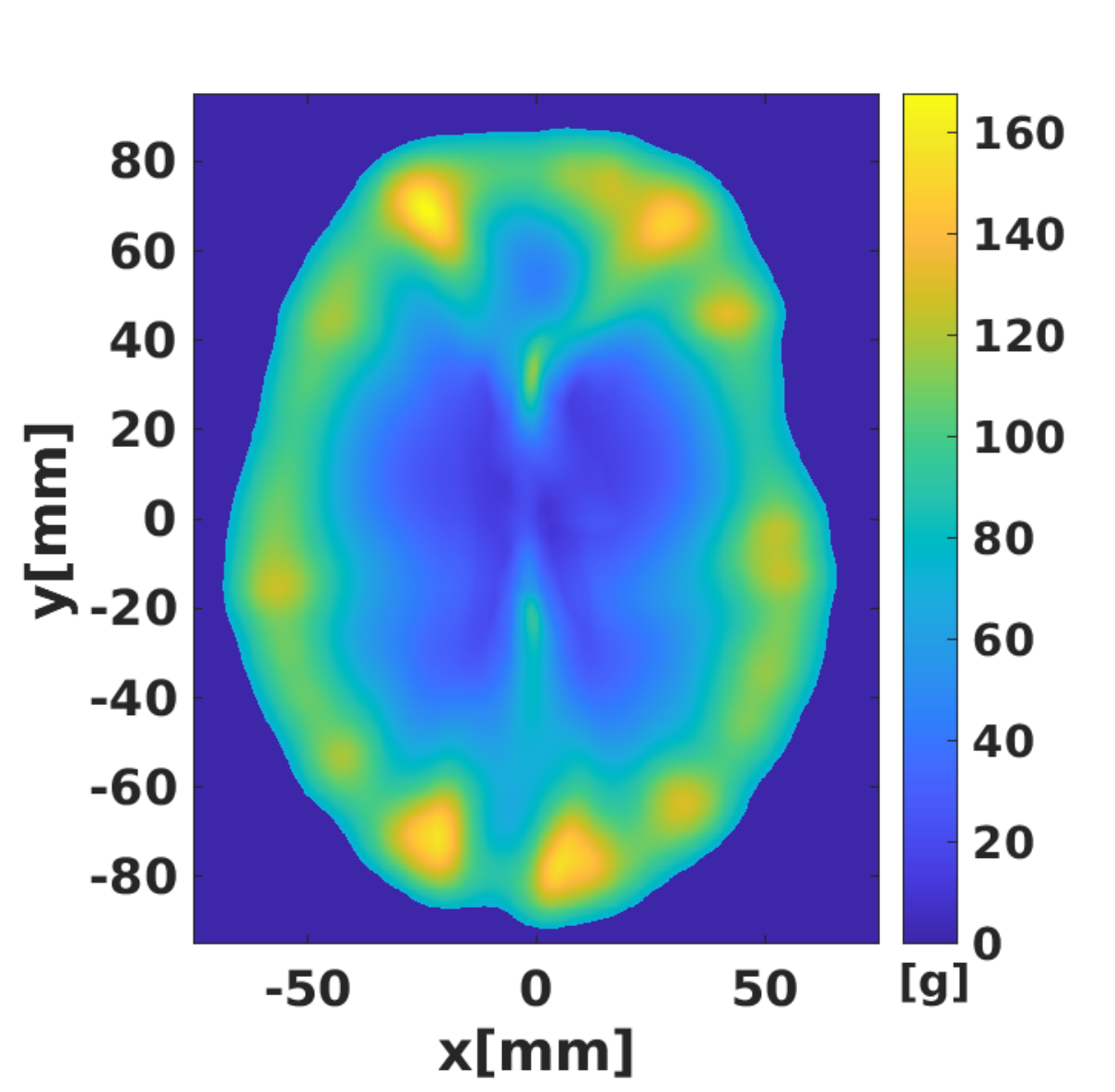}}
\put(1.25,0.14){\bf \color{white}\scriptsize{(c)}}
\put(1.735,-0.35){\includegraphics[trim= 5 1 18 20, clip, width=0.242\textwidth]{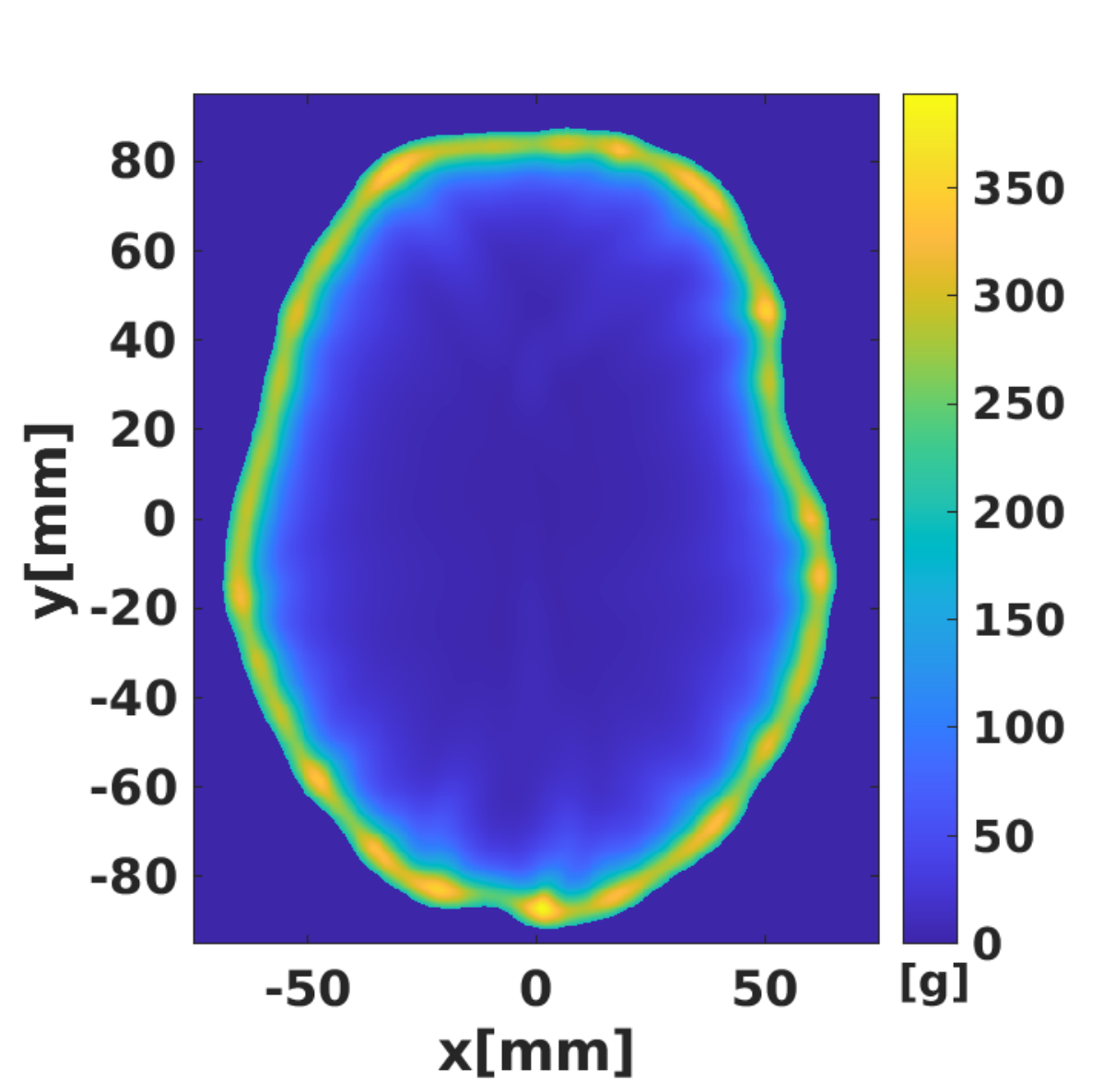}}
\put(1.83,0.14){\bf \color{white}\scriptsize{(d)}}
\end{picture}
\vspace{32mm}
\caption{Maximum acceleration ($g$) for frequencies 12.5, 25, 75, 200 Hz with amplitude 1.5 m/s inside human head is shown in subplots (a)-(d), respectively. Three focal regimes were observed: 1) focusing at the geometric foci (12.5, 25 Hz) 2) focusing at the geometric foci and at just under the surface as a ring (50, 75 Hz) 3) focusing only under the surface (75-200 Hz). Insets show the rice grain sized focal region inside the brain with the contour of the area corresponding to FWHM.  Propagation movies for 25, 75, 200 Hz  showing the three regimes are provided in the supplementary material.}
    \label{Fig:Skull_Print_MaxAcc_1p5}
\end{figure}
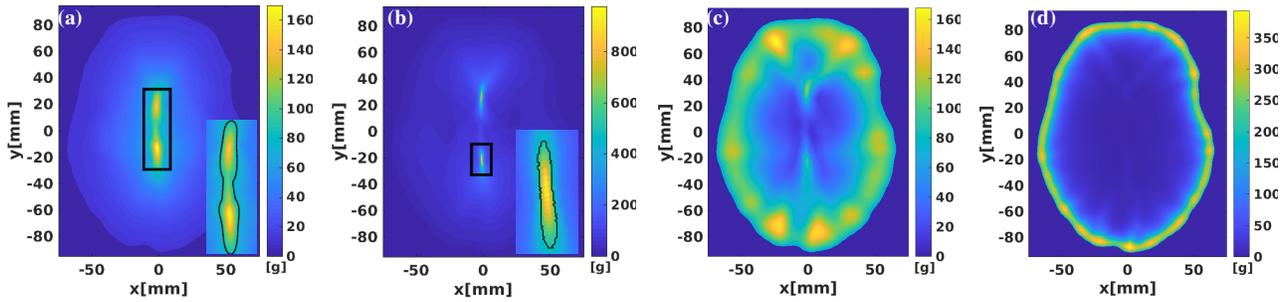

\begin{figure}[htbp]
    \centering
  \begin{subfigure}{0.45\textwidth}
    \centering
        \caption{}
    \includegraphics[trim= 2 2 25 10, clip, width=0.85\textwidth]{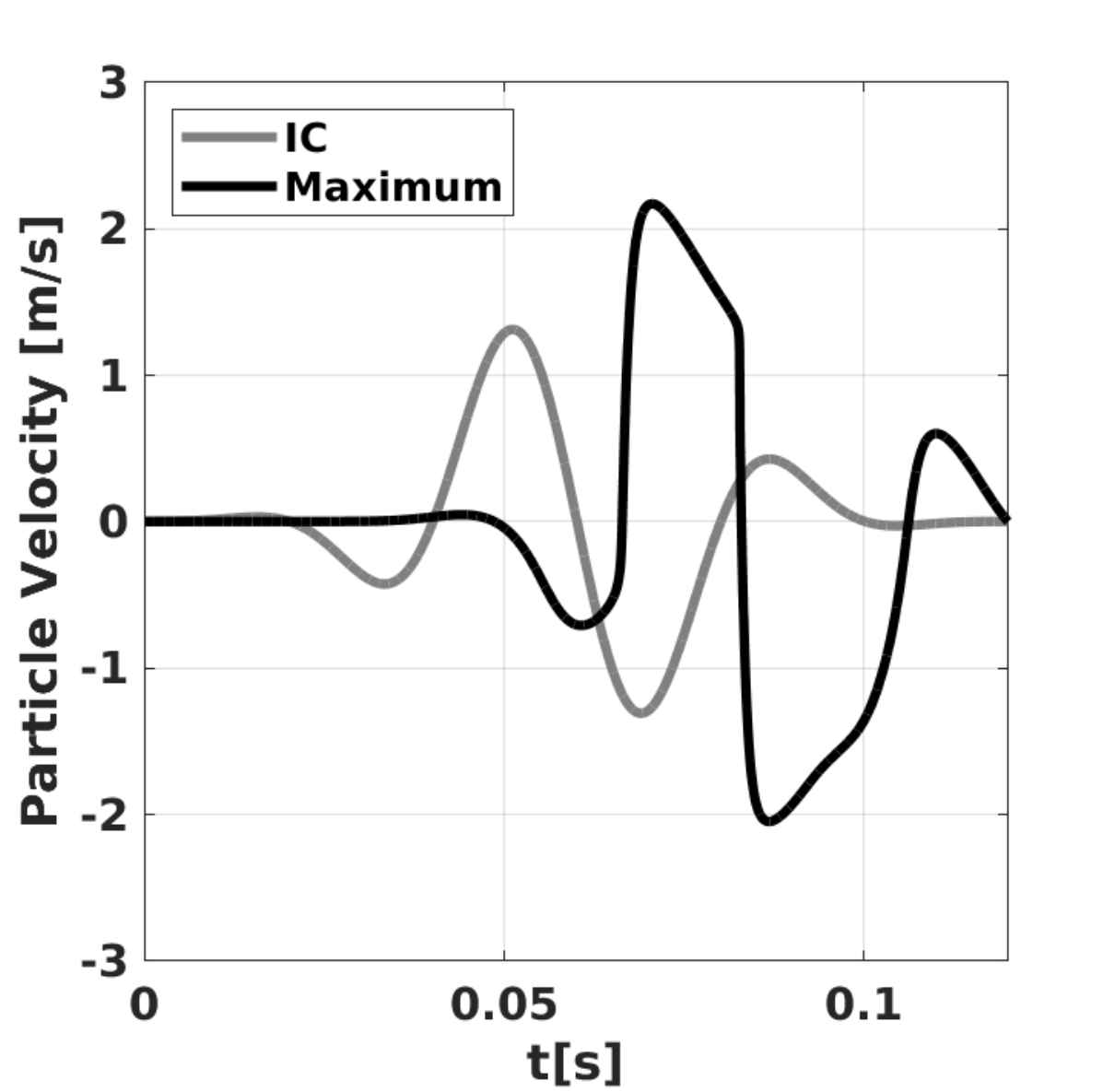}
        \label{Fig:Skull_Print_Max_Ratio-a}
    \end{subfigure}
     \begin{subfigure}{0.45\textwidth}
    \centering
        \caption{}
    \includegraphics[trim= 2 2 25 10, clip, width=0.85\textwidth]{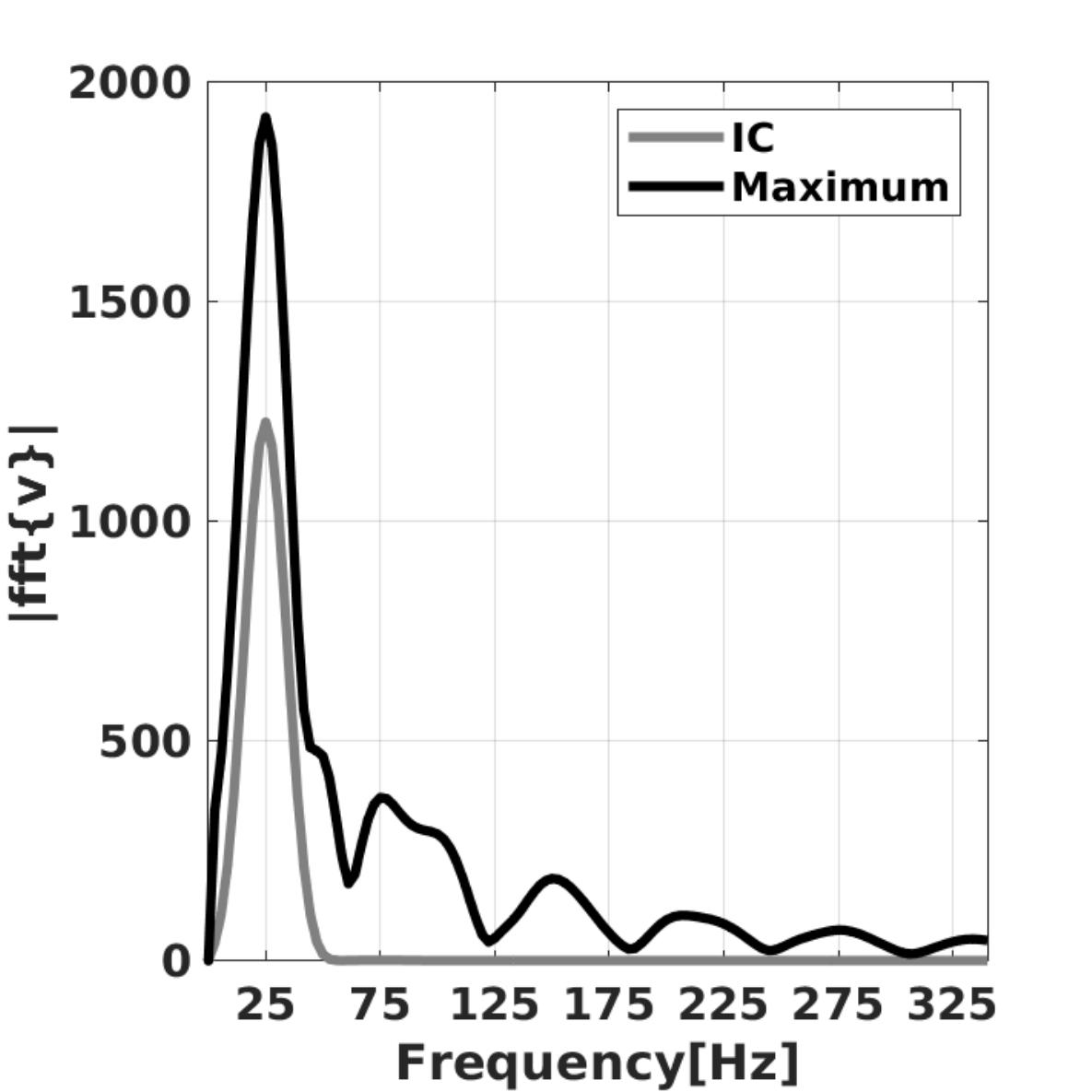}
        \label{Fig:Skull_Print_Max_Ratio-b}
    \end{subfigure}
    \caption{Particle velocity (a) and its spectrum (b) at the initial and the point of maximum acceleration for the amplitude 1.5 m/s and frequency 25 Hz. A strong shark-fin shaped shock is formed at the geometrical foci of the head. The cubic nonlinearity responsible for this peculiar shape is expressed by the generation of odd harmonics in the Fourier space.}
    \label{Fig:Shock_Spectrum}
\end{figure}

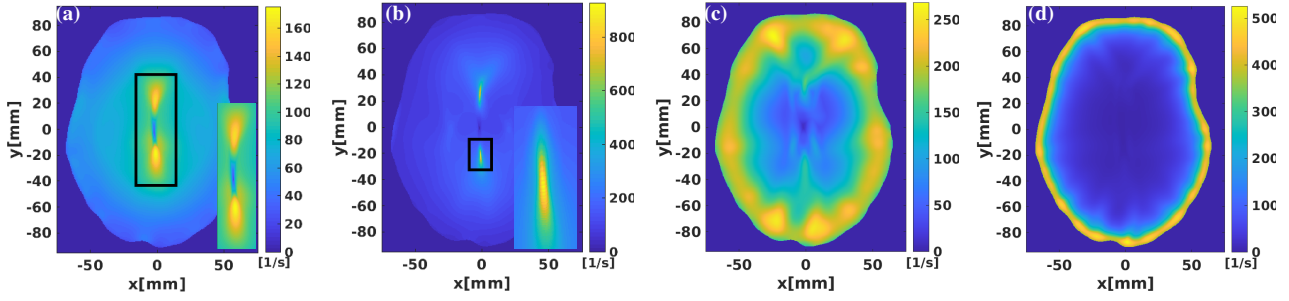
\begin{figure}[htbp]
\vspace{10mm}
\hspace{-5mm}
\begin{picture}(500.8,0.75)(-0.05,0.25)
\setlength{\unitlength}{0.45\textwidth}
\put(0,-0.35){\includegraphics[trim=  5 1 18 20, clip, width=0.245\textwidth]{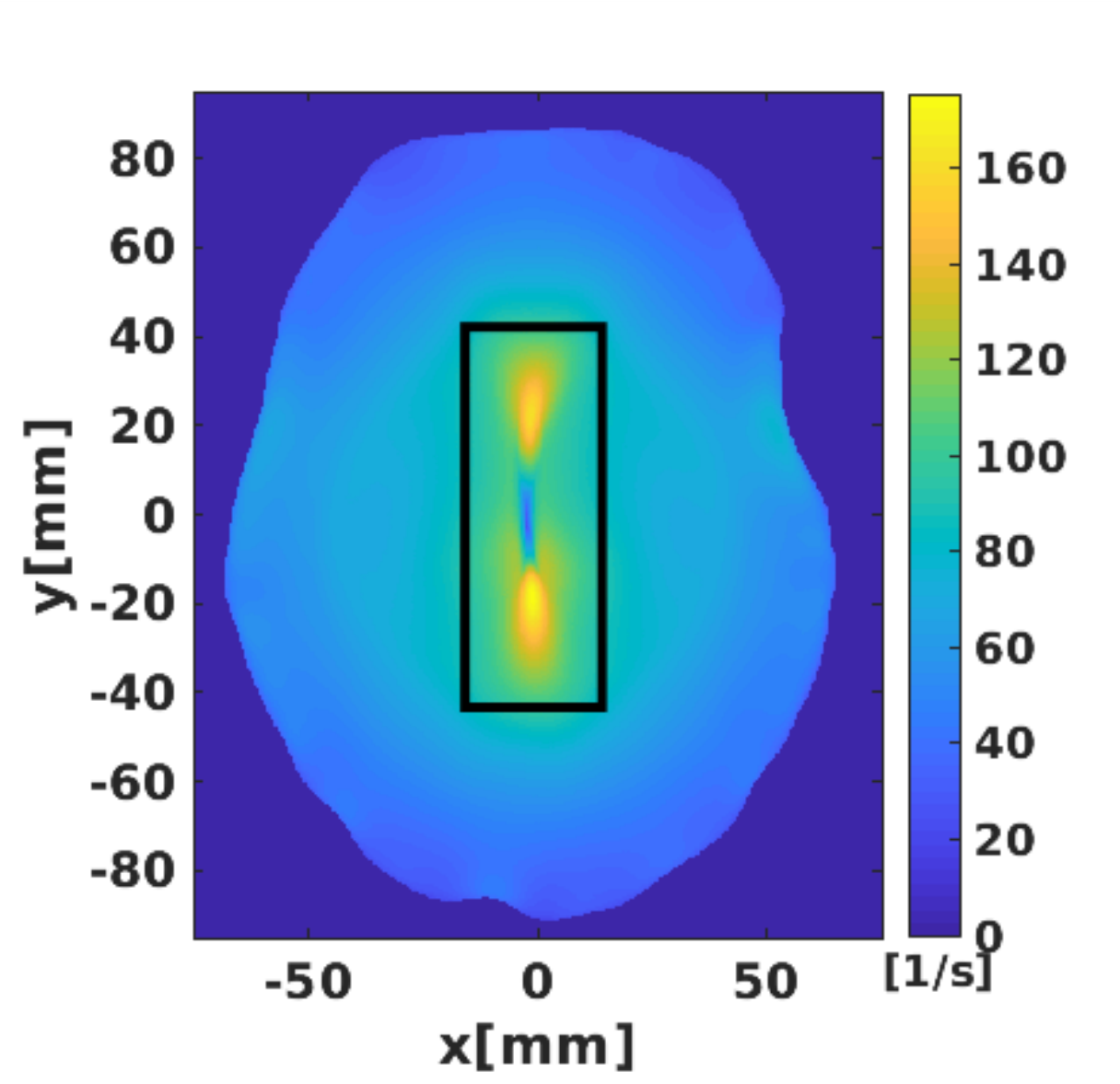}}
\put(0.38, -0.27){\includegraphics[trim=  61 55 220 35, clip, width=0.03\textwidth]{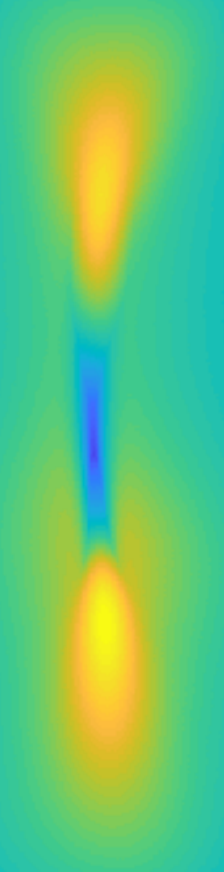}}
\put(0.09,0.14){\bf \color{white}\scriptsize{(a)}}
\put(0.58,-0.35){\includegraphics[trim= 5 1 18 20, clip, width=0.245\textwidth]{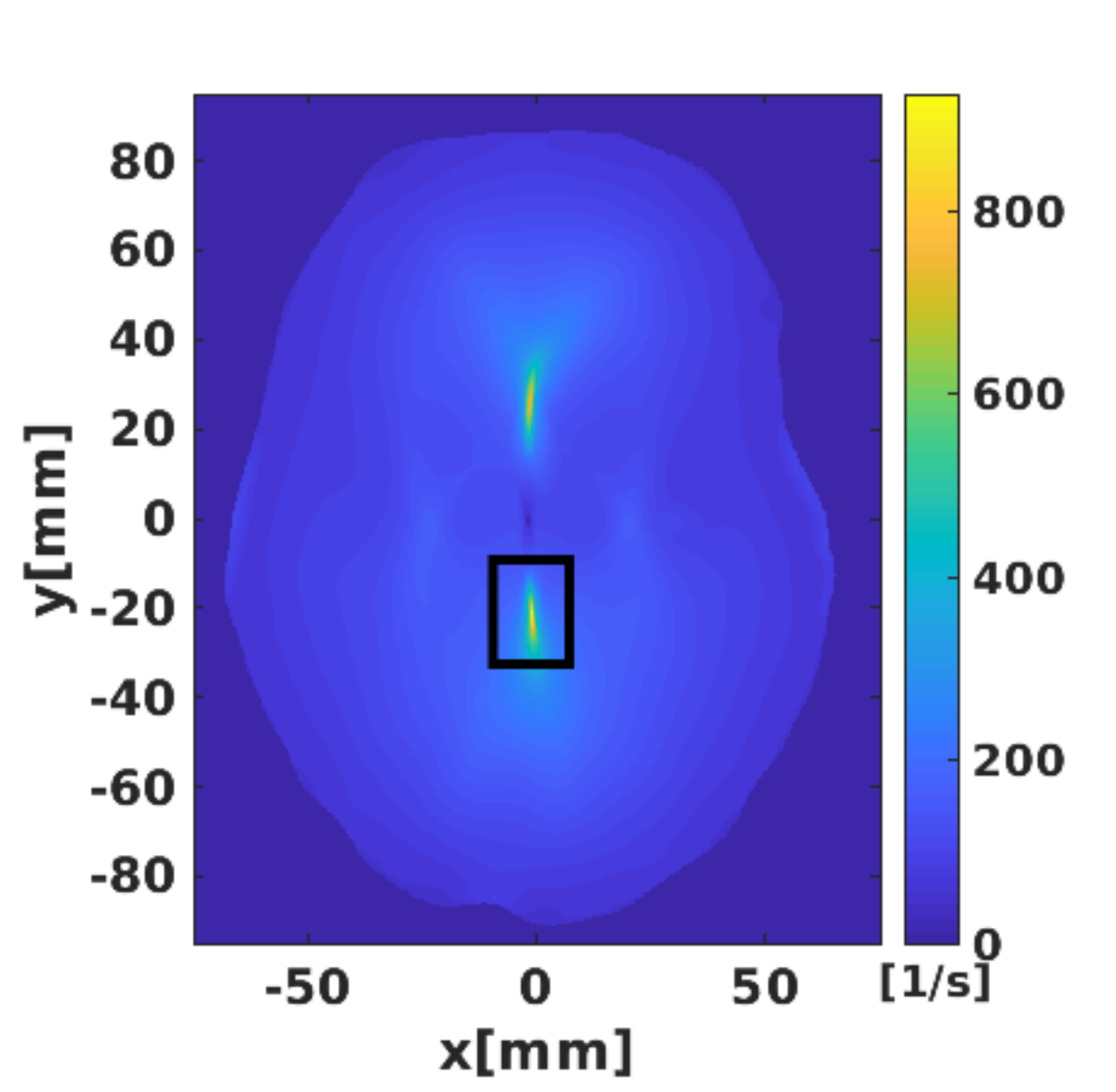}}
\put(0.91, -0.27){\includegraphics[trim=   70 90 185 40, clip, width=0.05\textwidth]{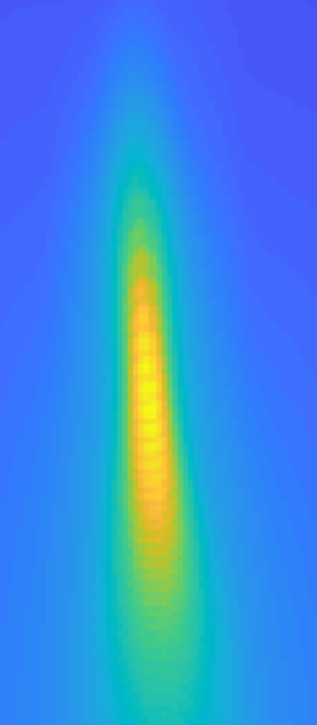}}
\put(0.68,0.14){\bf \color{white}\scriptsize{(b)}}
\put(1.16,-0.35){\includegraphics[trim= 5 1 18 20, clip, width=0.245\textwidth]{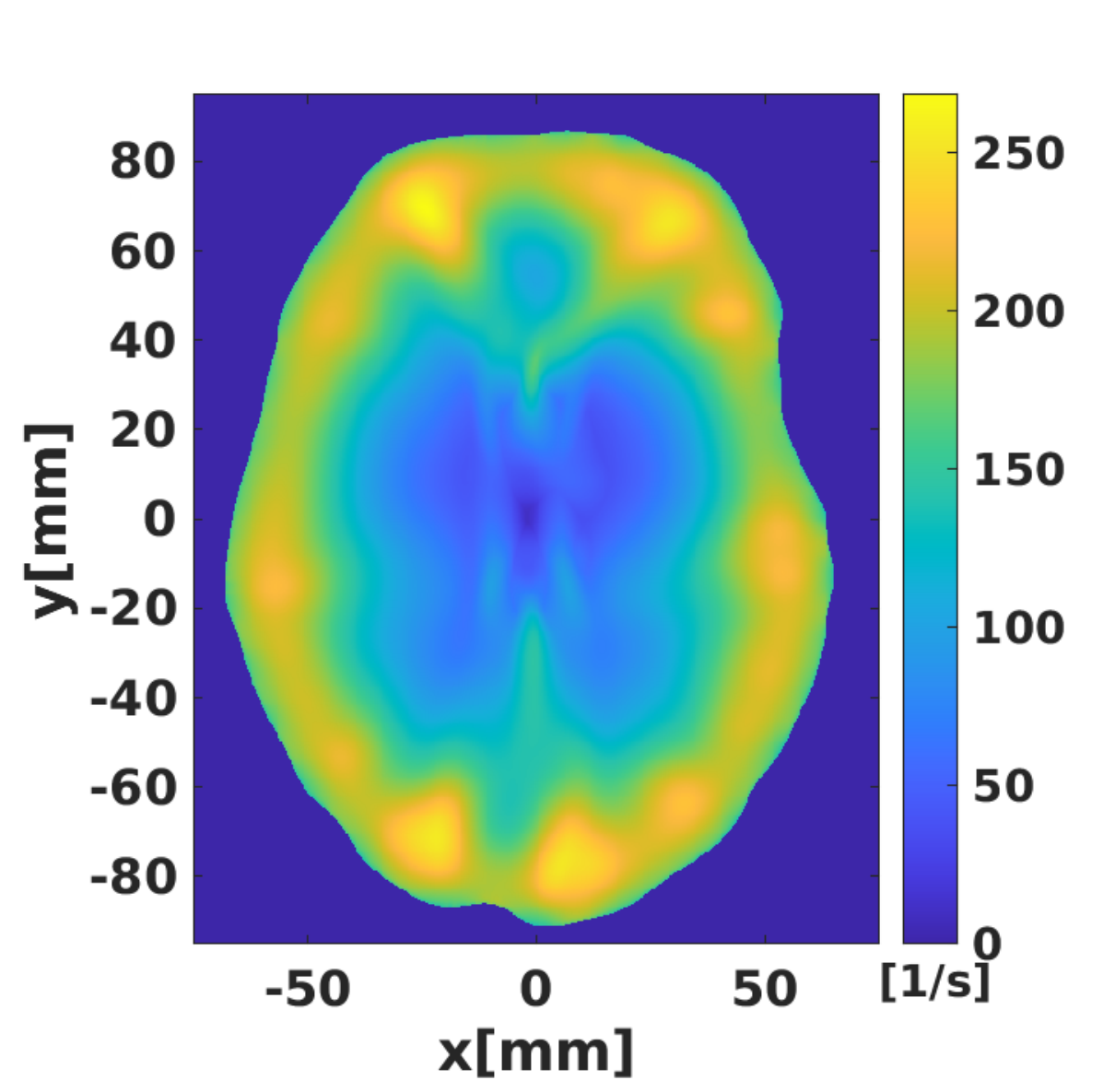}}
\put(1.25,0.14){\bf \color{white}\scriptsize{(c)}}
\put(1.735,-0.35){\includegraphics[trim= 5 1 18 20, clip, width=0.242\textwidth]{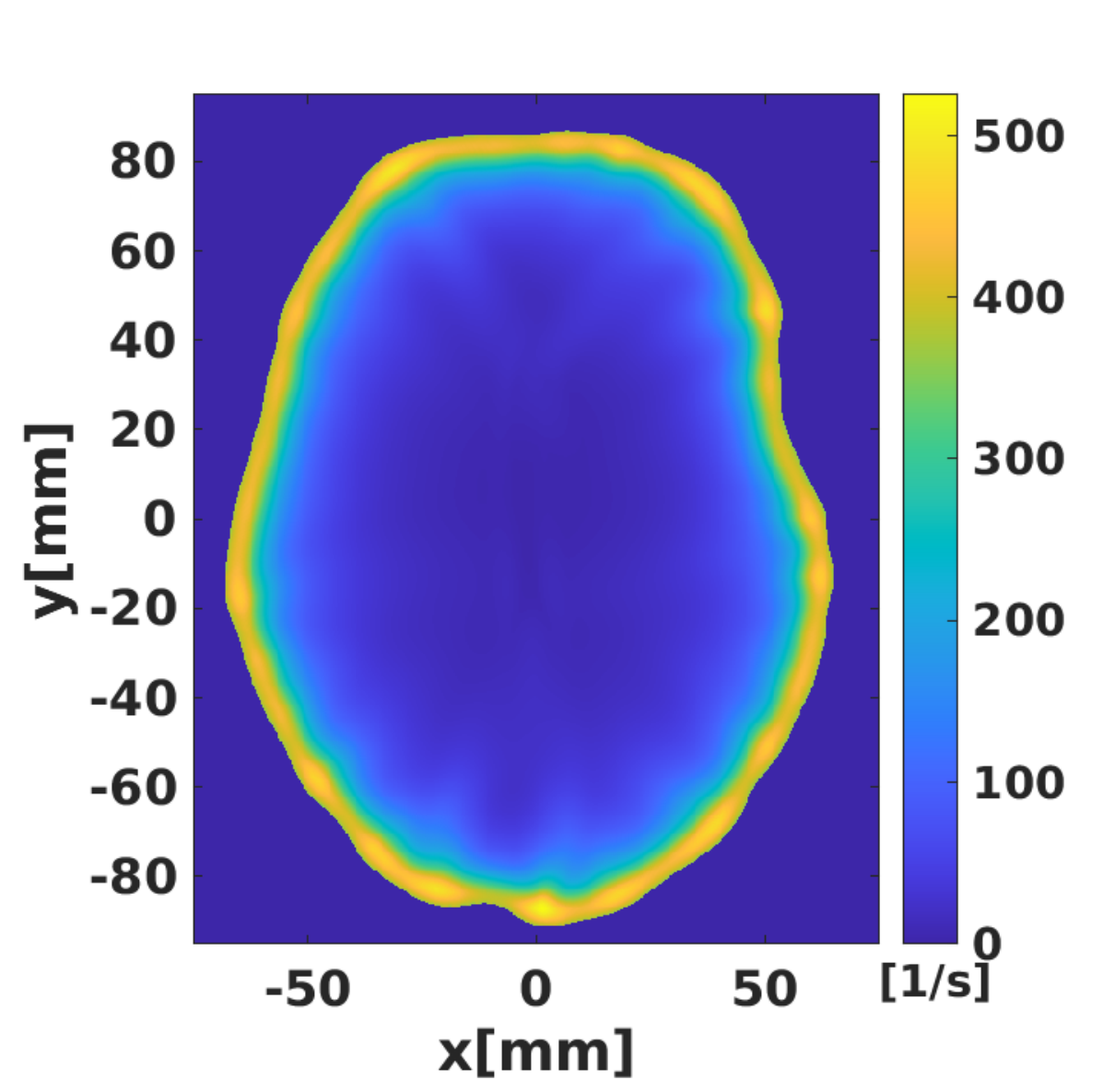}}
\put(1.83,0.14){\bf \color{white}\scriptsize{(d)}}
\end{picture}

\vspace{30mm}
\caption{Maximum strain-rate (1/s) for frequencies 12.5, 25, 75, 200 Hz with amplitude 1.5 m/s inside human head is shown in subplots (a)-(d), respectively. Like the maximum acceleration, here also three different regimes of shock focusing can be observed. A zoom of the focal region shows the minute region of peak strain-rate which could be damaging.}
\label{Fig:Skull_Print_StrainRate_1p5}
\end{figure}

\begin{figure}[t]
\vspace{20mm}
\hspace{-5mm}
\begin{picture}(500.8,0.75)(-0.05,0.25)
\setlength{\unitlength}{0.45\textwidth}
\put(0,-0.35){\includegraphics[trim=  5 1 15 20, clip, width=0.245\textwidth]{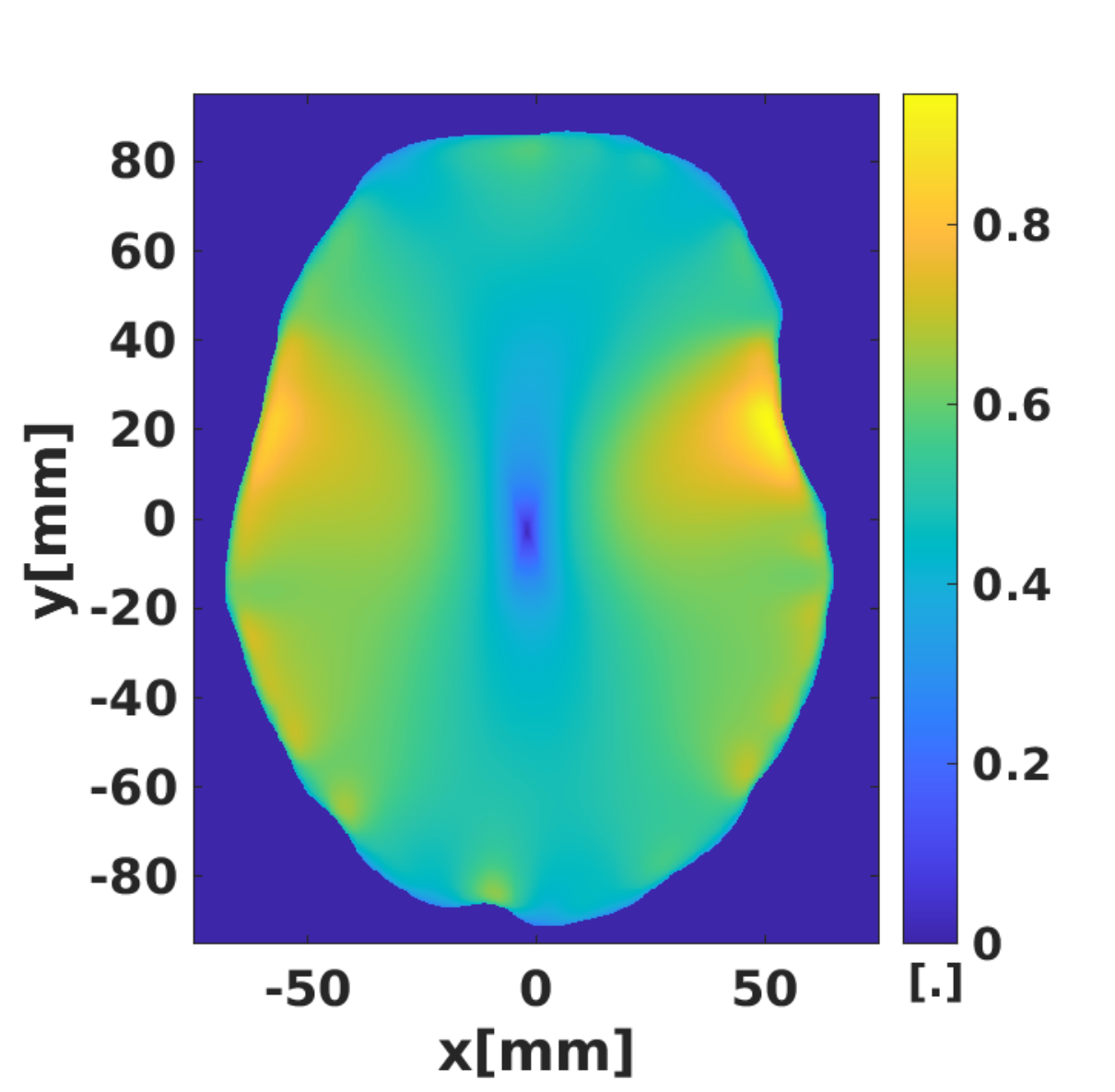}}
\put(0.09,0.14){\bf \color{white}\scriptsize{(a)}}
\put(0.58,-0.35){\includegraphics[trim= 5 1 15 20, clip, width=0.245\textwidth]{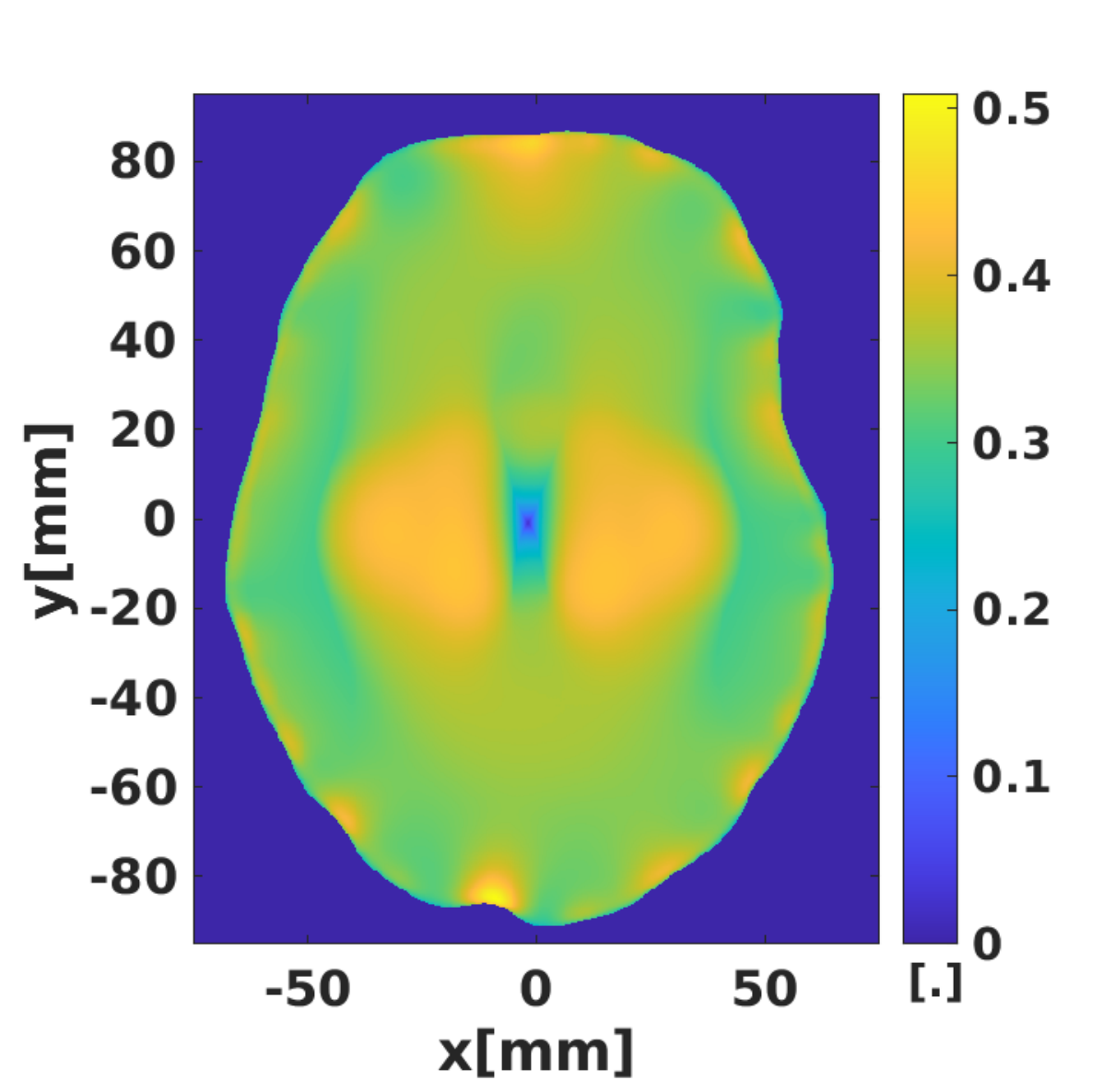}}
\put(0.68,0.14){\bf \color{white}\scriptsize{(b)}}
\put(1.16,-0.35){\includegraphics[trim= 5 1 15 20, clip, width=0.245\textwidth]{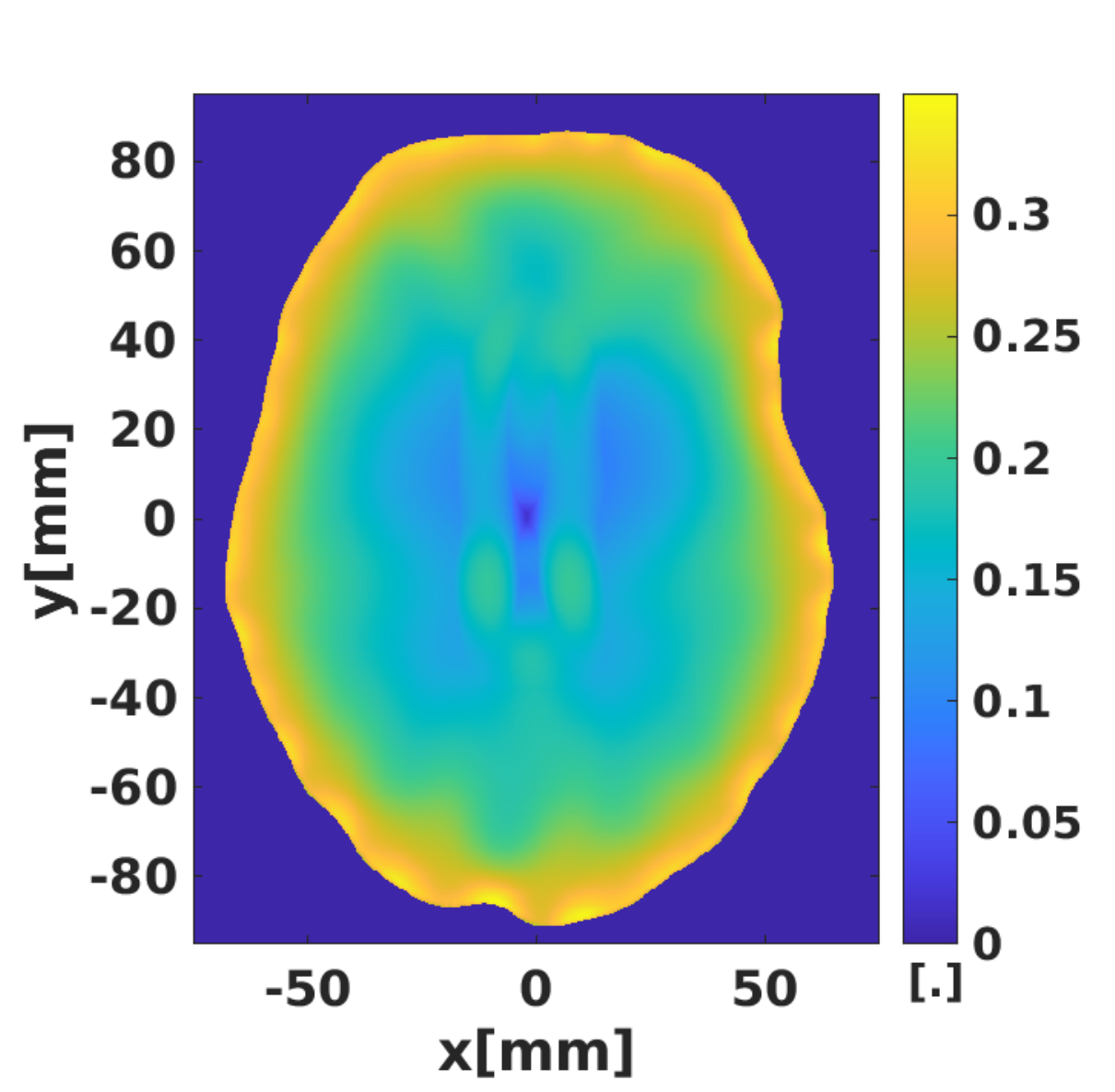}}
\put(1.25,0.14){\bf \color{white}\scriptsize{(c)}}
\put(1.735,-0.35){\includegraphics[trim= 5 1 15 20, clip, width=0.242\textwidth]{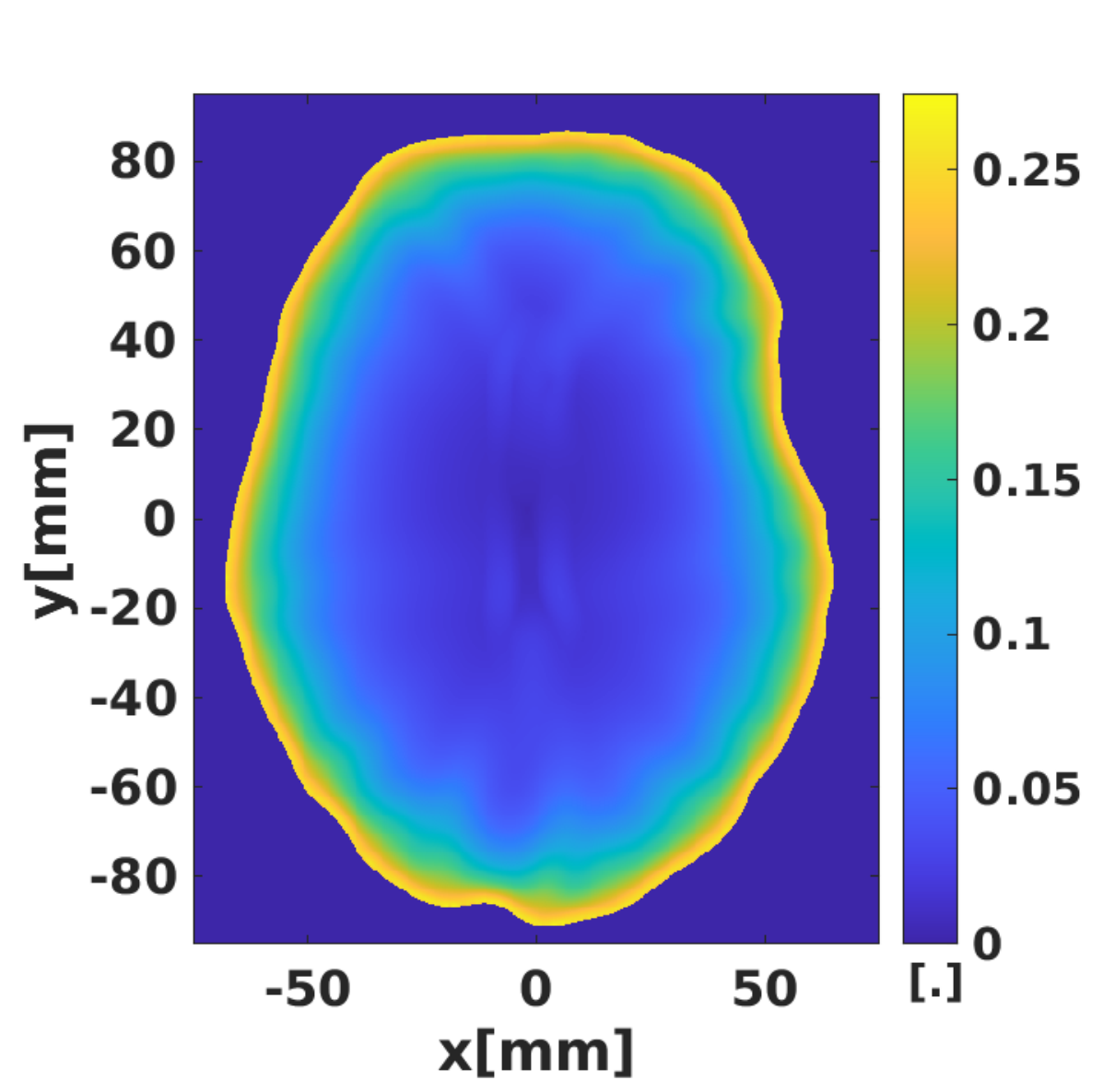}}
\put(1.83,0.14){\bf \color{white}\scriptsize{(d)}}
\end{picture}

\vspace{30mm}
\caption{Maximum strain for frequencies 12.5, 25, 75, 200 Hz with amplitude 1.5 m/s inside the human head is shown in subplots (a)-(d), respectively. Note the strain does not follow the same trend as that of the strain-rate. In fact, it is lower in regions with high strain-rate, this is probably due to the lower amplitude at the shock front (versus smooth regions) due to higher dissipation of higher frequencies responsible for the shock formation.}
    \label{Fig:Skull_Print_Strain_1p5}
\end{figure}
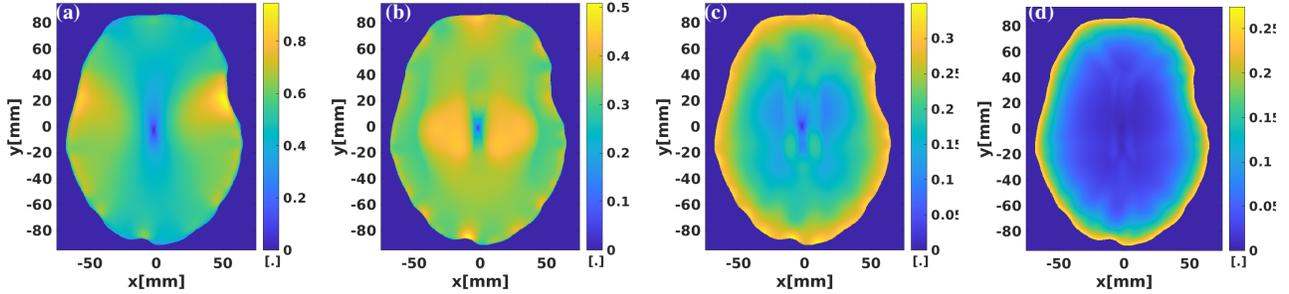

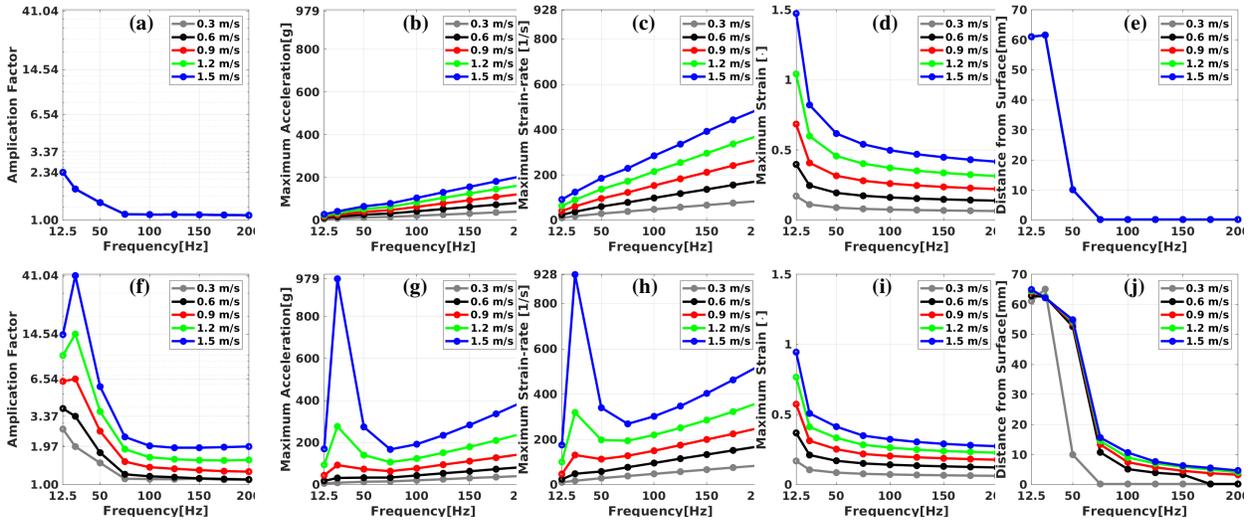
\begin{figure}
\begin{picture}(500.8,0.75)(-0.05,0.25)
\setlength{\unitlength}{0.45\textwidth}
\put(0,-0.35){\includegraphics[trim= 2 2 25 10, clip, width=0.2\textwidth]{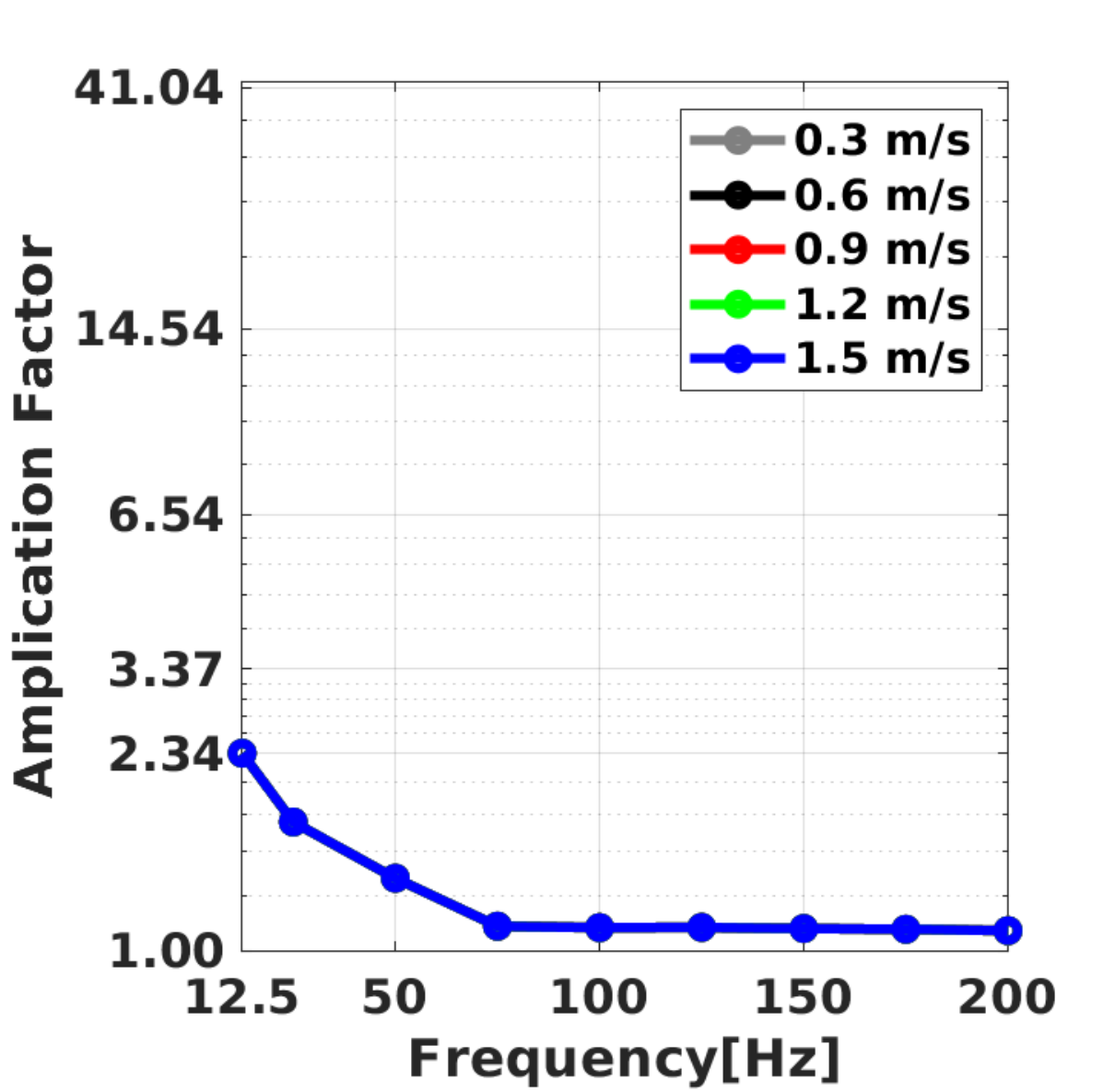}}
\put(0.22,0.05){\bf \color{black}\scriptsize{(a)}}

\put(0.49,-0.35){\includegraphics[trim= 2 2 25 10, clip, width=0.2\textwidth]{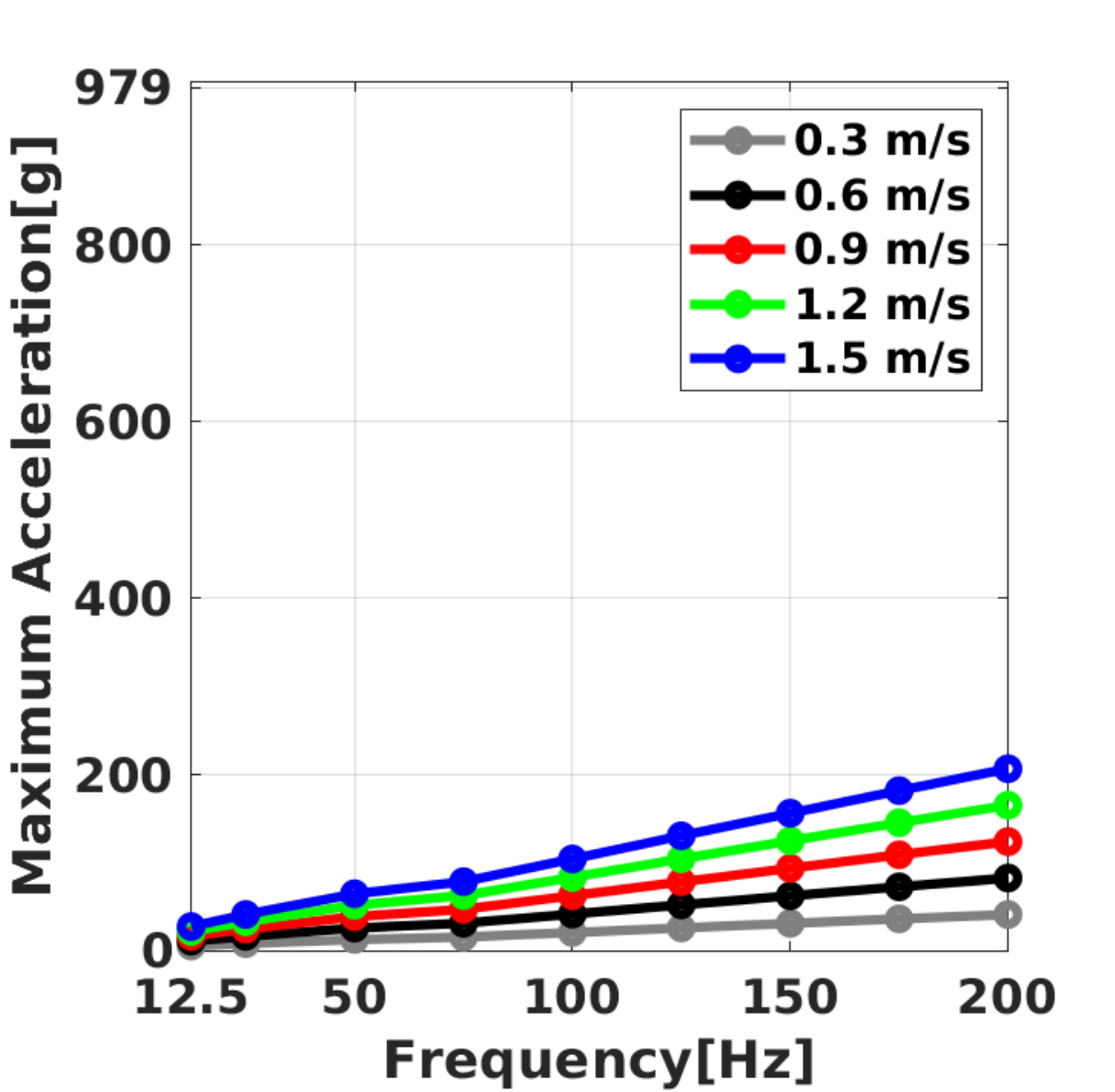}}
\put(0.71,0.05){\bf \color{black}\scriptsize{(b)}}

\put(0.915,-0.35){\includegraphics[trim= 2 2 25 10, clip, width=0.2\textwidth]{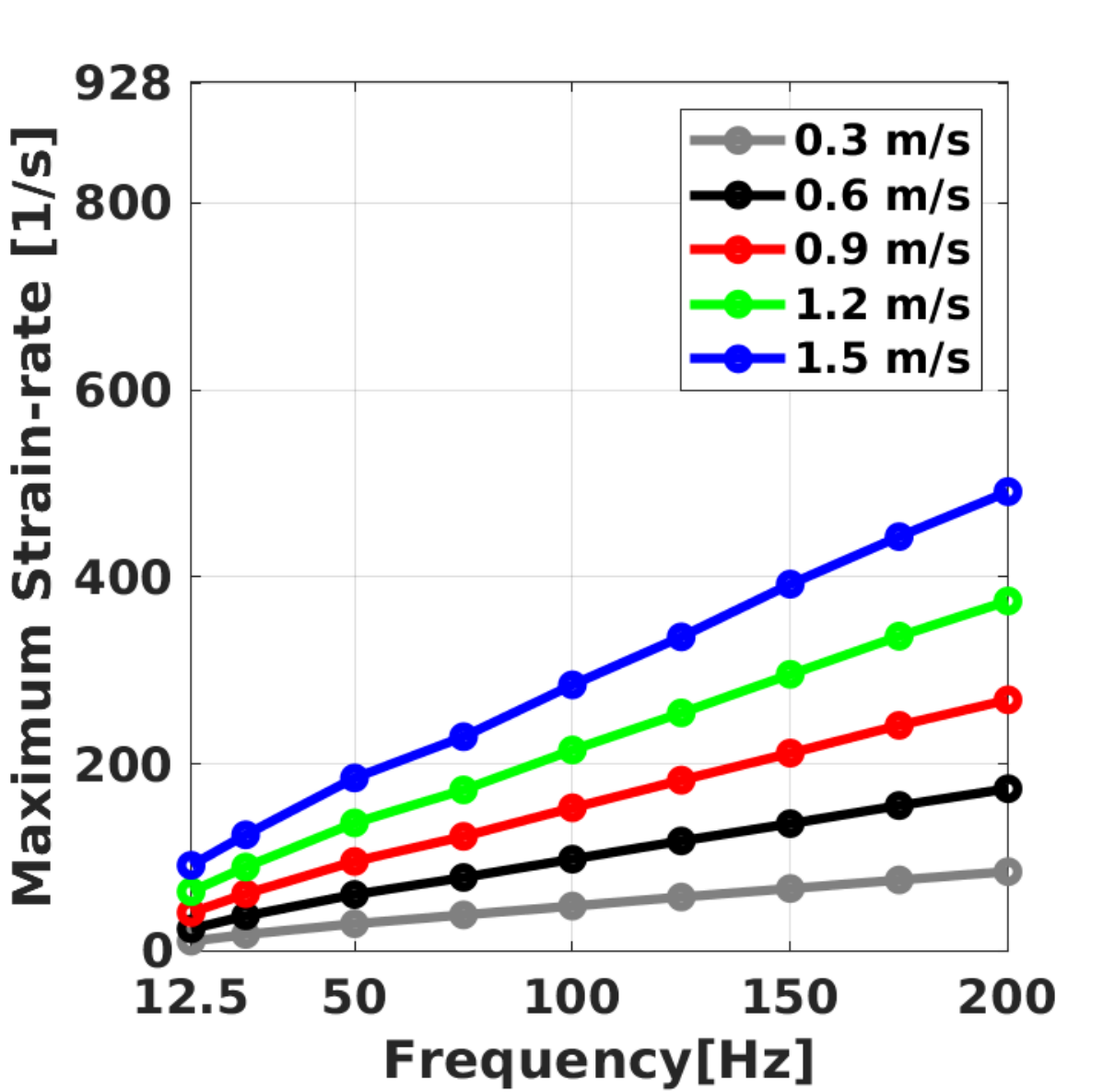}}
\put(1.12,0.05){\bf \color{black}\scriptsize{(c)}}

\put(1.34,-0.35){\includegraphics[trim= 2 2 25 10, clip, width=0.2\textwidth]{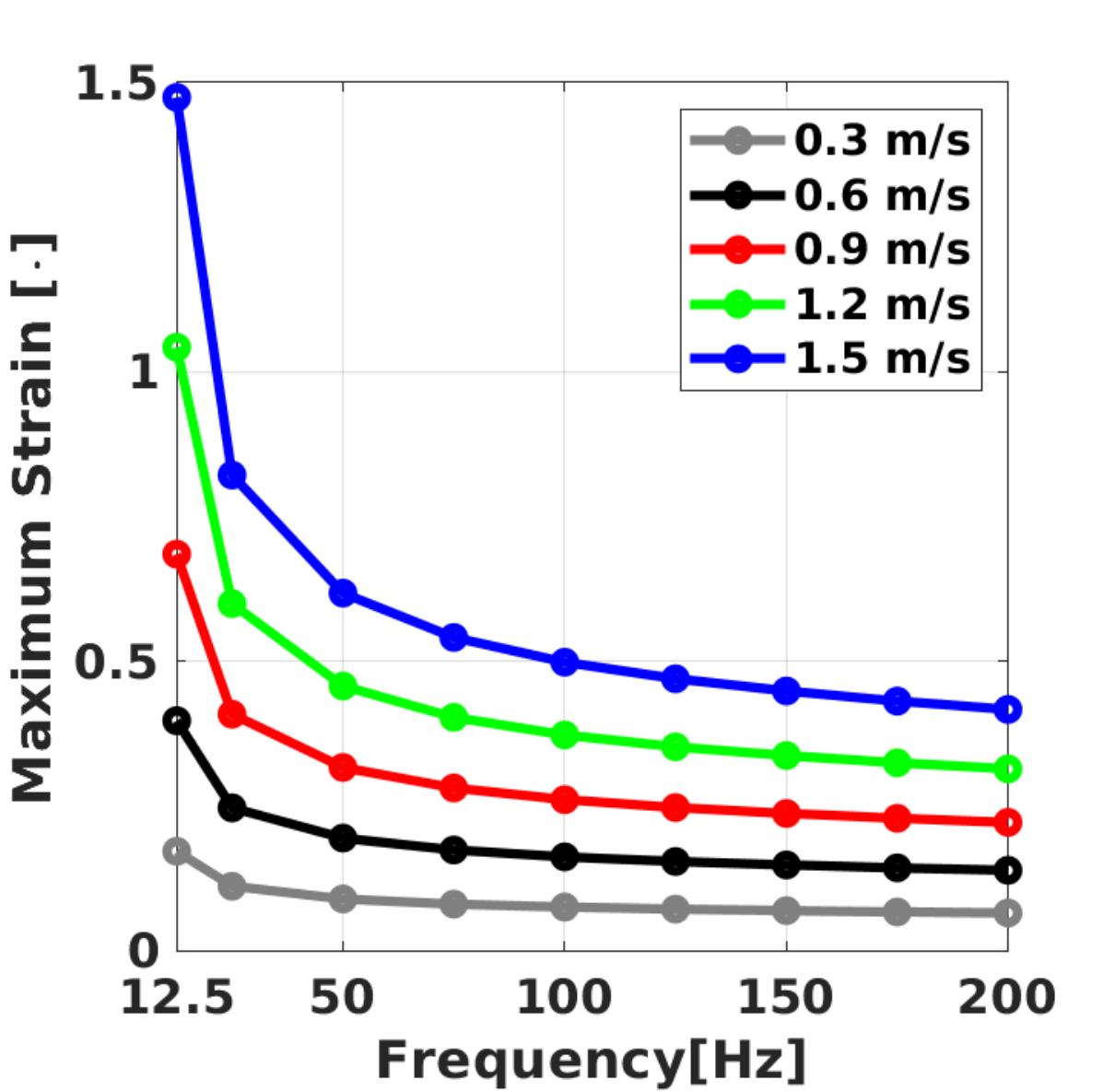}}
\put(1.55,0.05){\bf \color{black}\scriptsize{(d)}}

\put(1.77,-0.35){\includegraphics[trim= 2 2 25 10, clip, width=0.2\textwidth]{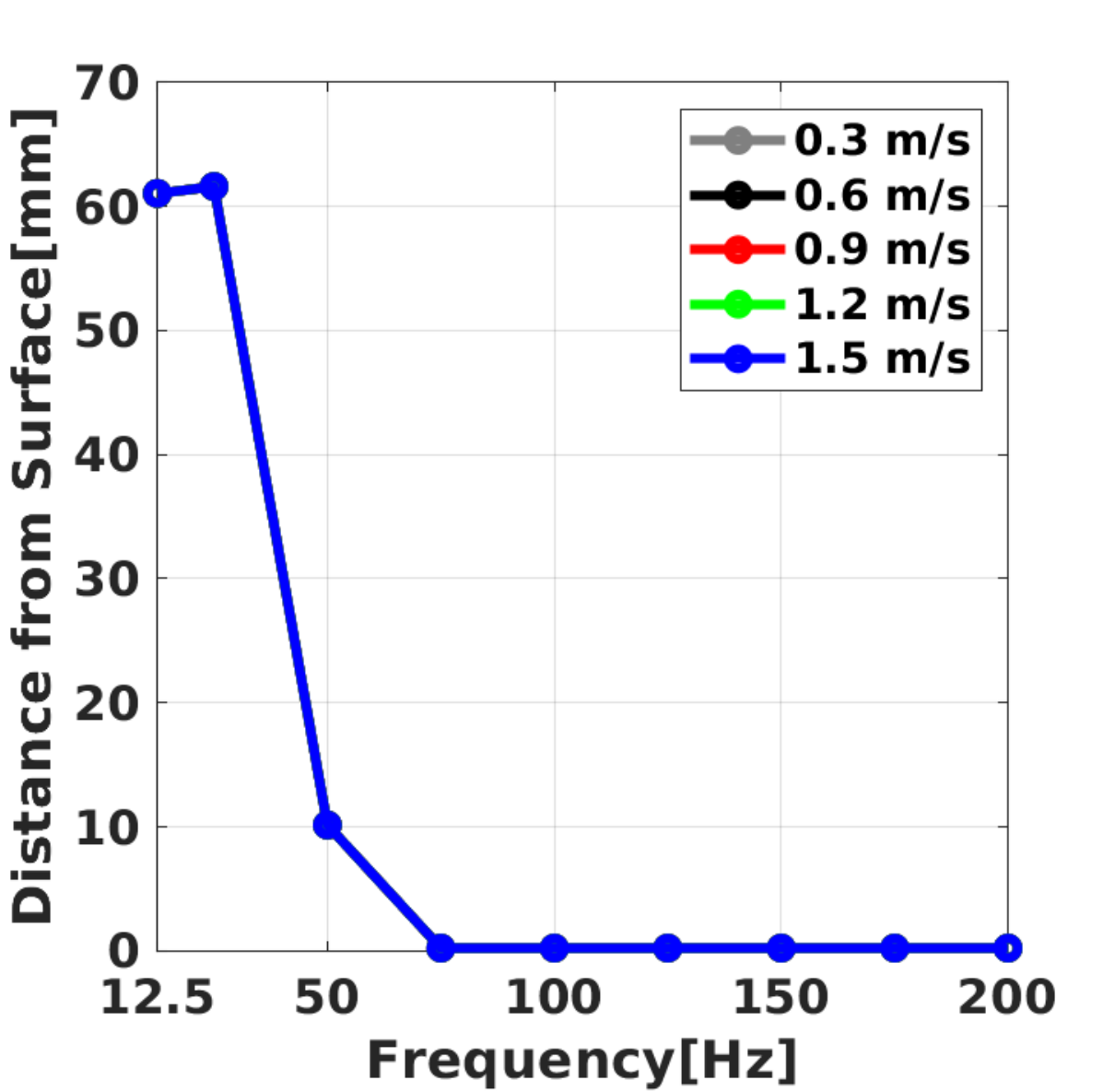}}
\put(2.0,0.05){\bf \color{black}\scriptsize{(e)}}
\end{picture}

\vspace{30mm}
\begin{picture}(500.8,0.75)(-0.05,0.25)
\setlength{\unitlength}{0.45\textwidth}
\put(0,-0.35){\includegraphics[trim= 2 2 25 10, clip, width=0.2\textwidth]{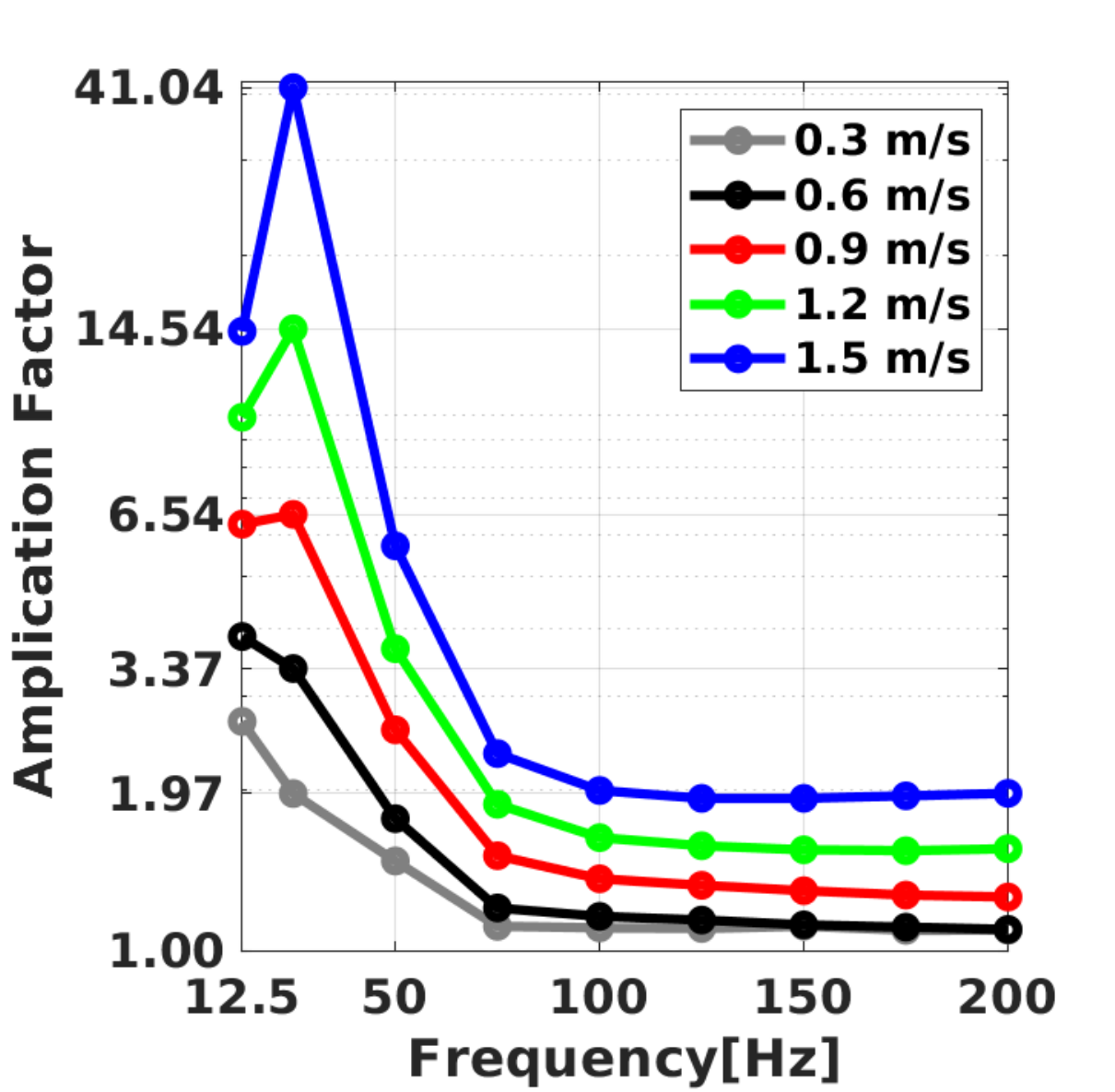}}
\put(0.22,0.05){\bf \color{black}\scriptsize{(f)}}

\put(0.49,-0.35){\includegraphics[trim= 2 2 25 10, clip, width=0.2\textwidth]{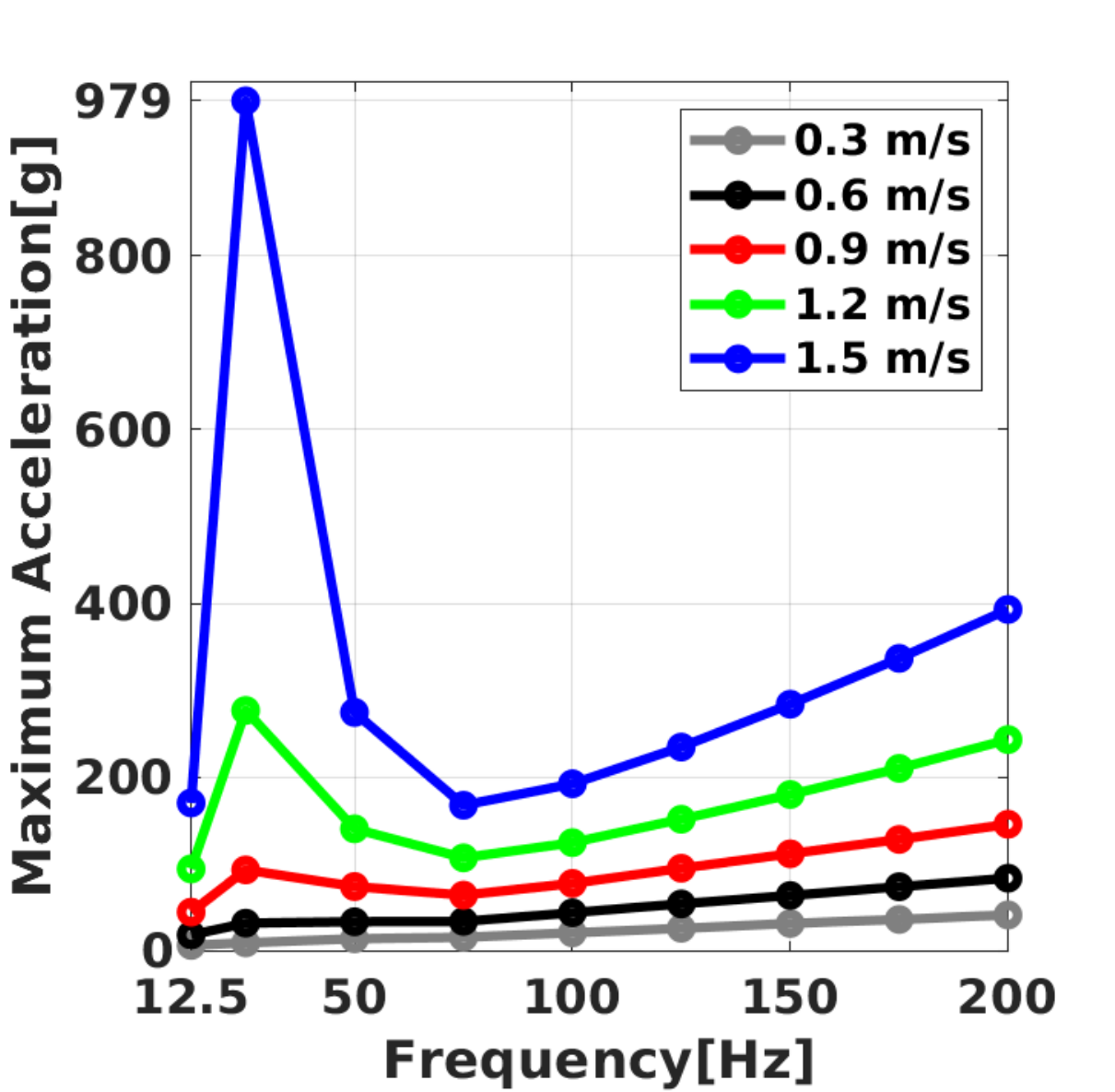}}
\put(0.71,0.05){\bf \color{black}\scriptsize{(g)}}

\put(0.915,-0.35){\includegraphics[trim= 2 2 25 10, clip, width=0.2\textwidth]{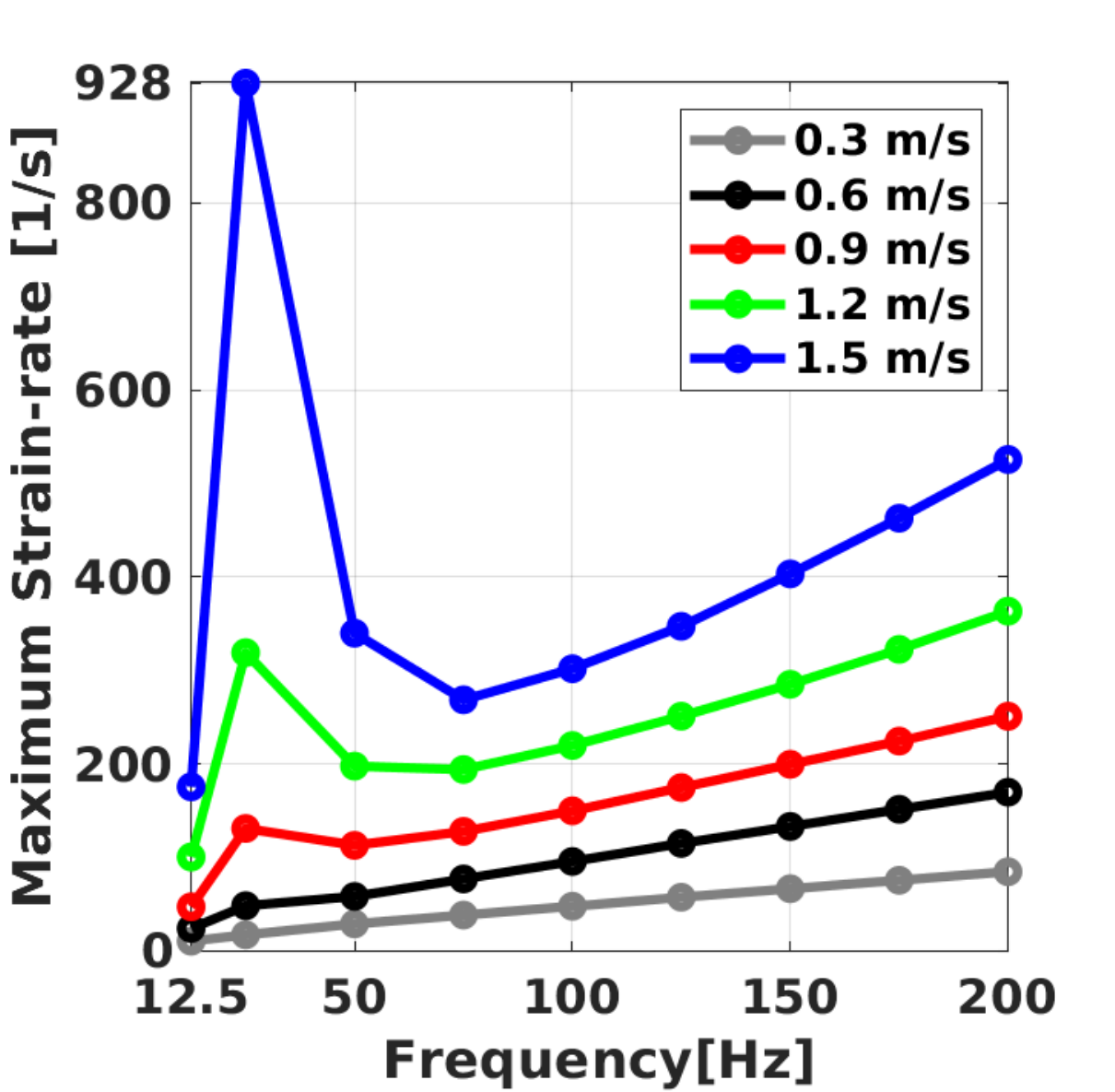}}
\put(1.12,0.05){\bf \color{black}\scriptsize{(h)}}

\put(1.34,-0.35){\includegraphics[trim= 2 2 25 10, clip, width=0.2\textwidth]{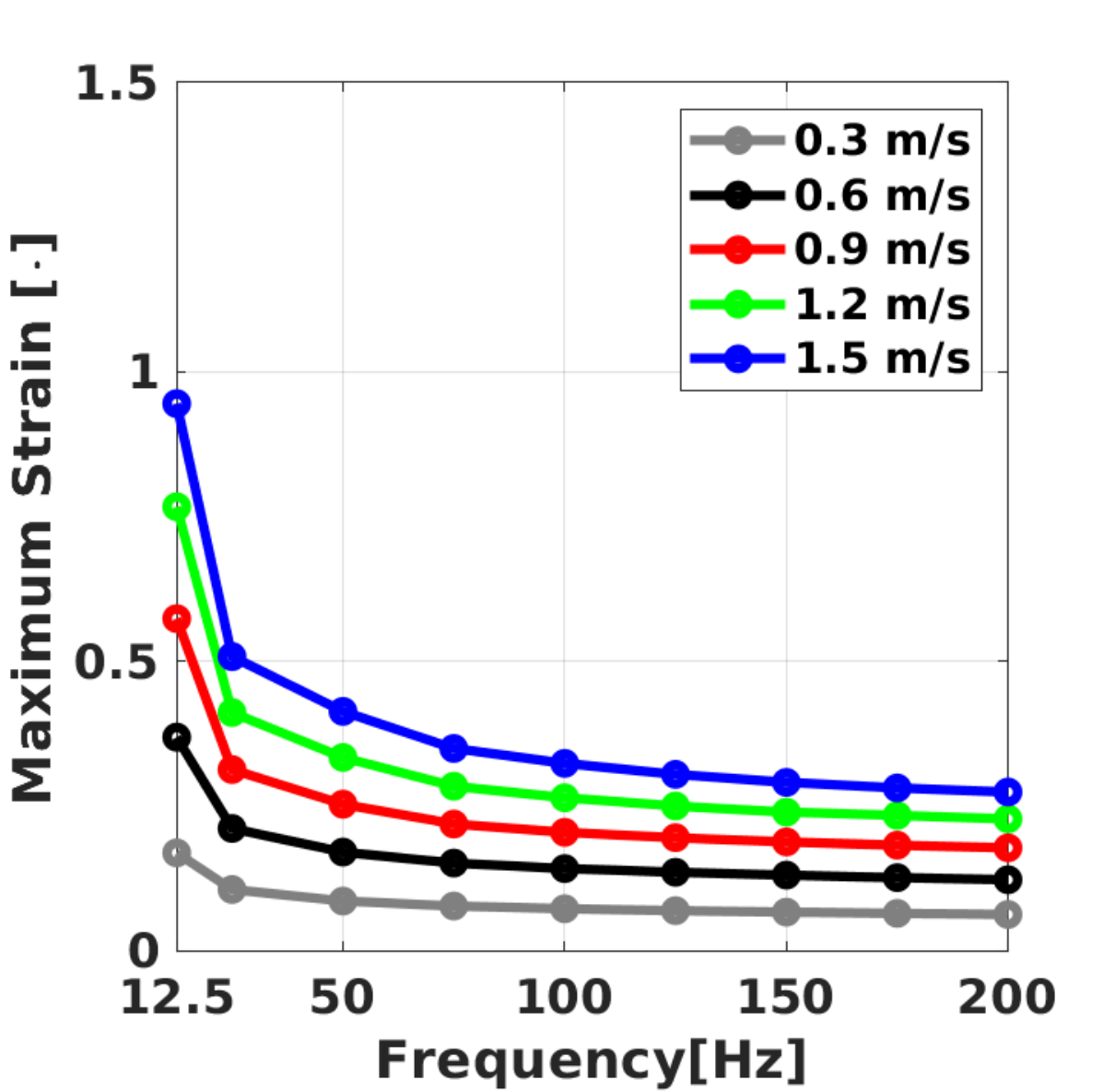}}
\put(1.55,0.05){\bf \color{black}\scriptsize{(i)}}

\put(1.77,-0.35){\includegraphics[trim= 2 2 25 10, clip, width=0.2\textwidth]{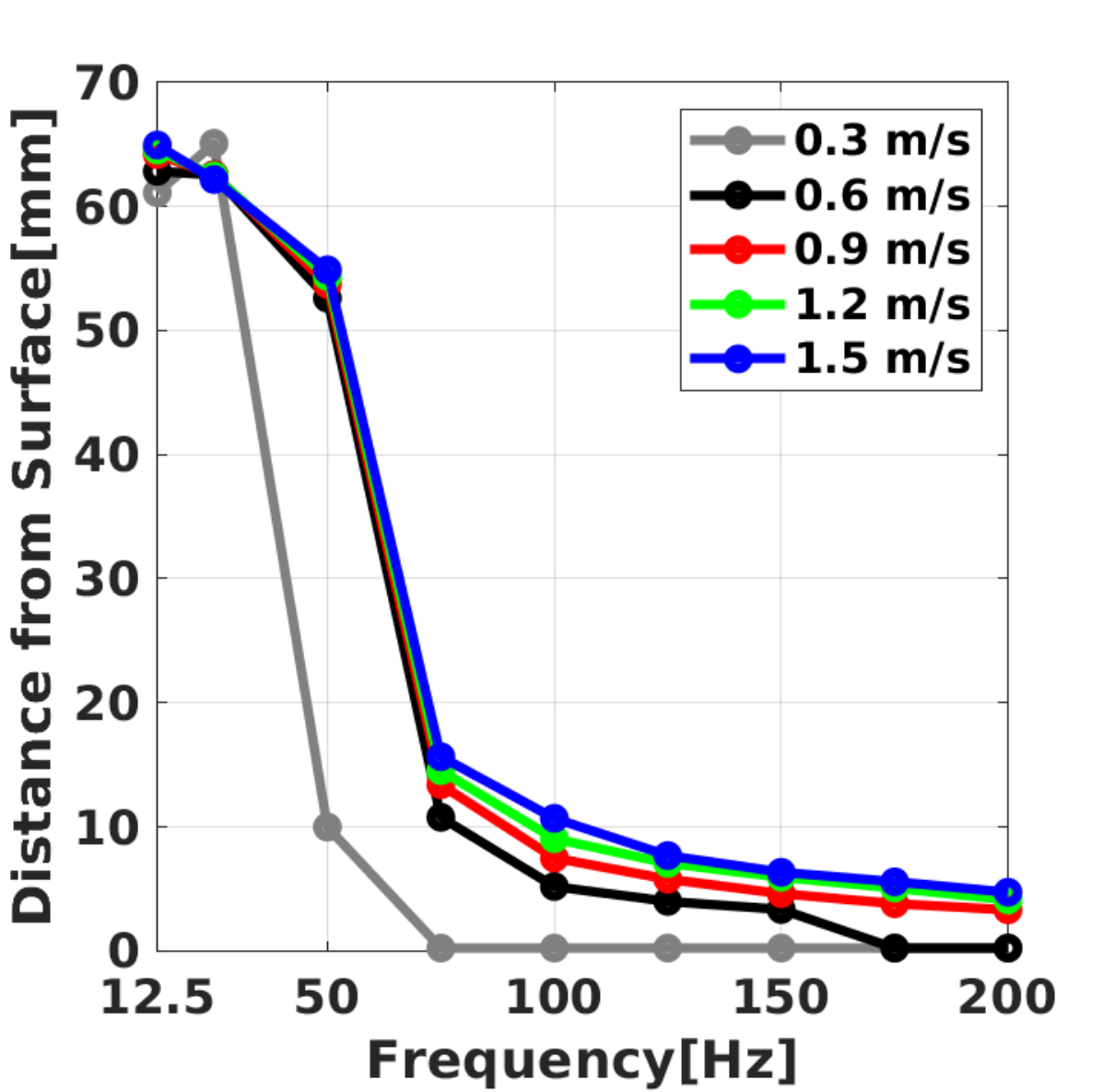}}
\put(2.0,0.05){\bf \color{black}\scriptsize{(j)}}

\end{picture}

\vspace{30mm}
\caption{Comparison of linear (top) versus nonlinear (bottom) simulations. 
    Subplot (a,f): Ratio of the maximum acceleration to the initial acceleration for each frequency-amplitude pair. 
    Subplot (b,g): Maximum of the magnitude of acceleration for each frequency-amplitude pair. 
    Subplot (c,h): Maximum of the magnitude strain-rate for each frequency-amplitude pair.
    Subplot (d,i): Maximum of the magnitude strain for each frequency-amplitude pair.
    Subplot (e,j): Shortest distance between the point of the maximum acceleration and the surface.}
    \label{Fig:Skull_Print_Max_Ratio}
\end{figure}

\begin{figure}[htbp]
    \centering
      \begin{subfigure}{0.35\textwidth}
    \centering
        \caption{}
    \includegraphics[trim= 2 2 25 10, clip, width=\textwidth]{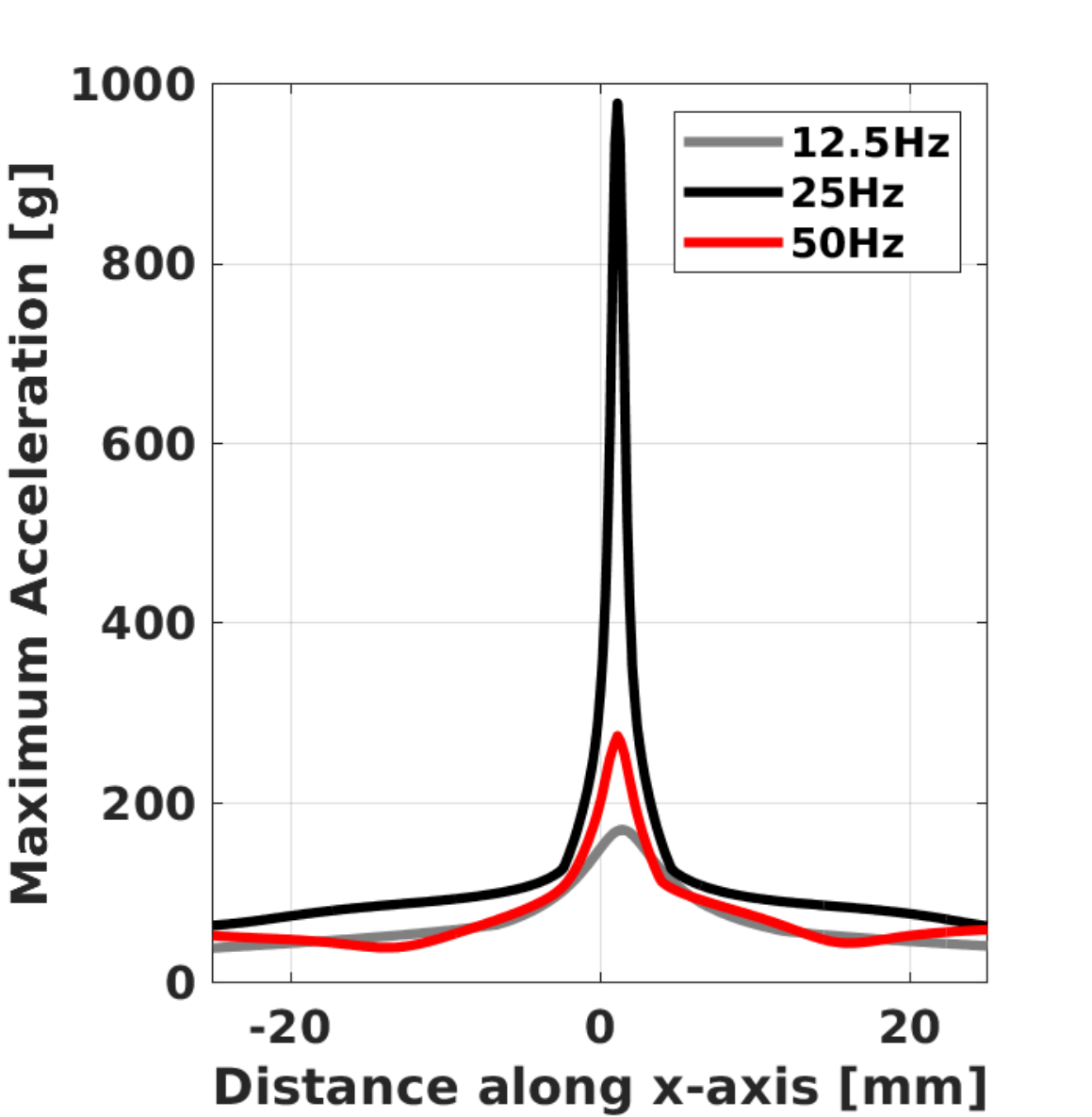}
        \label{Fig:Skull_Print_Max_Ratio-a}
    \end{subfigure}
     \begin{subfigure}{0.35\textwidth}
    \centering
        \caption{}
    \includegraphics[trim= 2 2 25 10, clip, width=\textwidth]{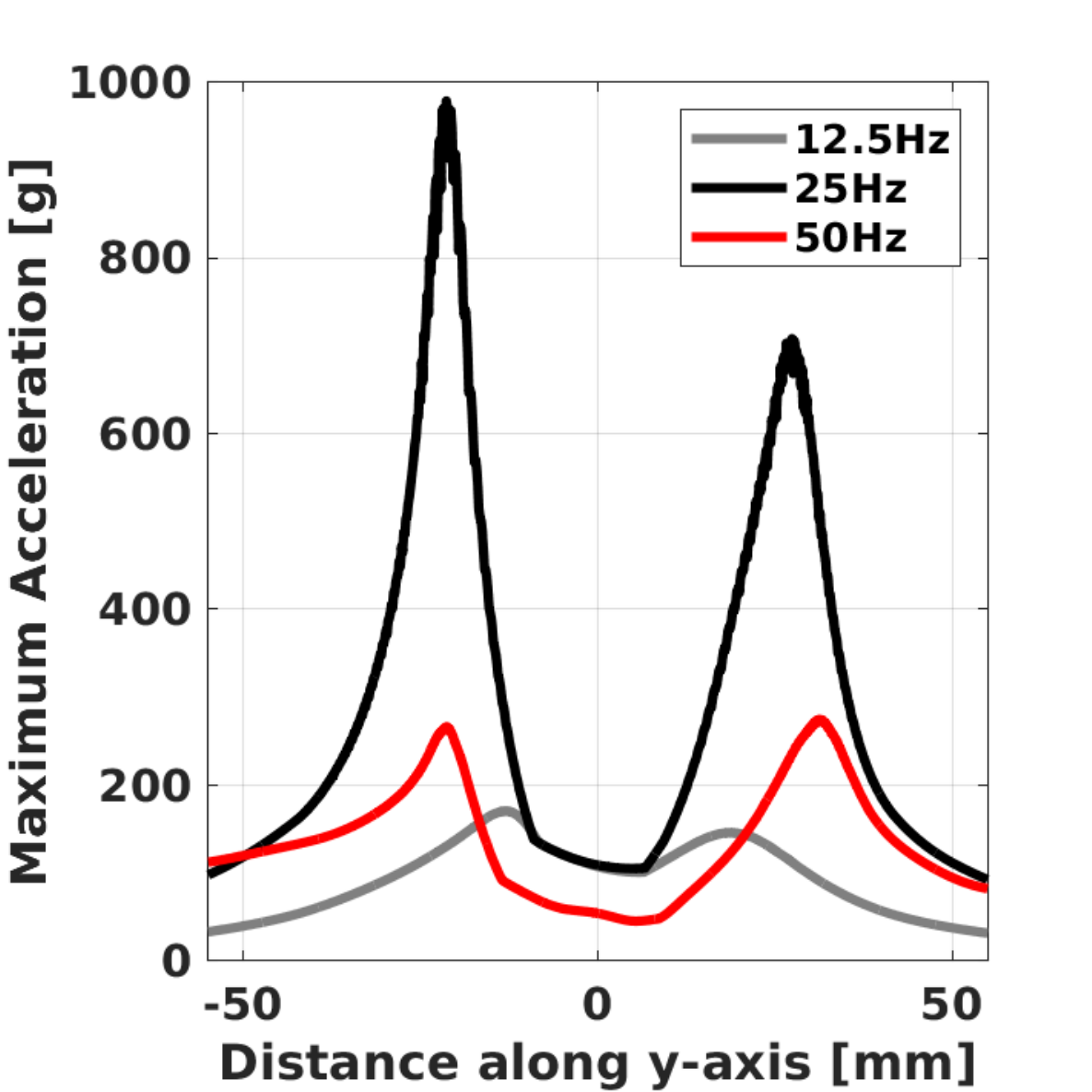}
        \label{Fig:Skull_Print_Max_Ratio-b}
    \end{subfigure}
    \caption{Beam plots of the maximum acceleration along $x$-axis and $y$-axis in subplots (a) and (b), respectively, for three different frequencies 12.5, 25, 50 Hz of 1.5 m/s. Higher the acceleration peak, smaller the FWHM, for instance, the FWHM for 12.5 Hz is 1.34 mm showing the super-resolved focusing due to the generation of higher harmonics.}
    \label{Fig:Beam_Plots}
\end{figure}

\section{Numerical Results}\label{Sec:Skull}



To understand the influence of shear shock formation in the brain the same head geometry with brain material properties was considered. The nonlinear elastic and attenuation parameters of fresh porcine brain were obtained through PPM1D\cite{Tripathi2019_PPM1D_CT} model calibration  based on  a previously published experimental data \cite{Espindola2017}. The inverse uncertainty quantification, using metropolis Markov chain Monte Carlo analysis along with the deterministic  1D piecewise parabolic method, was performed to estimate the nonlinear parameter $\beta = 44.24 \pm 14.77$. With the linear attenuation power-law $\alpha(\omega) = 0.06\omega^{1.05}$ Np/m, and the linear shear speed $c({\rm 75 Hz}) = 2.10$ m/s. Datasets from 3 different brains for 5 different amplitudes each was used in this estimation, a detailed paper is in preparation. 

Nonlinearity acts by transferring energy from low frequencies to higher harmonic frequencies and attenuation opposes this action by preferentially damping higher frequencies. It thus is not immediately obvious how raising the values of both nonlinearity and attenuation impacts the wave propagation dynamics, especially since the nonlinearity is amplitude-dependent whereas the attenuation is not. Simulations were performed for a range of frequencies ($12.5-200~{\rm Hz}$) and impact velocity amplitudes  ($0.3-1.5~ {\rm m/s}$). 
The domain was discretized with $\Delta x = 0.19$ mm, it ensured that the highest fundamental-frequency \ie 200 Hz had at least 50 points per wavelength. The time $t\in [0, 0.12]$ s was discretized using $\Delta t = \frac{C \Delta x }{c_0 + A} = 15.8~\mu$s with CFL number $C=0.3$, amplitude $A = 1.5$ m/s, and linear shear speed $c_0 = 2.1$ m/s. A monochromatic sine-pulse of its respective frequency and amplitude was taken with a 3-period Chebyshev window ($-80$ dB cut-off).
Additionally, reference linear visco-elastic simulations were also performed by setting the nonlinear parameter to zero ($\beta=0$). Within this parameter space three regimes emerged driven by this amplitude- and frequency-dependent interplay of nonlinearity and attenuation in the shock formation dynamics. 

First, for characteristic frequencies below 50 Hz, shear shocks formed deep inside the brain, at the natural geometric foci determined by the overall skull morphology {(Fig. \ref{Fig:Skull_Print_MaxAcc_1p5}a,b)}. In this shock focusing regime the local acceleration values can be enormous. For a 35$g$, 1.5 m/s, 25 Hz, impact at the surface of the brain, for example, the local acceleration at the focus, deep inside the brain, exceeds $900g$ {(Fig.~\ref{Fig:Skull_Print_Max_Ratio}g)}. This represents surface to focal amplification factor of 41 {(Fig.~\ref{Fig:Skull_Print_Max_Ratio}f)}. A linear wave of the same amplitude and frequency will only be amplified by a factor of 2.34 {(Fig.~\ref{Fig:Skull_Print_Max_Ratio}a)} with the absolute value of acceleration below $50g$ {(Fig.~\ref{Fig:Skull_Print_Max_Ratio}b)}, indicating that the majority of the amplification is driven by nonlinearity and not geometrical focusing. 
The temporal wave form at the surface and at the focus {(Fig \ref{Fig:Shock_Spectrum}a)} show how the shark-fin shape of the particle velocity produces a very high local acceleration i.e. the time derivative of the velocity at the nearly vertical shock front is large. In the linear regime the quasi-monochromatic wave retains its sinusoidal profile and does not undergo this shark-fin nonlinear distortion which is why its acceleration amplification is modest.
In the frequency domain {(Fig \ref{Fig:Shock_Spectrum}b)} the shape of the shear shock is supported by the odd harmonics, which is a specific feature of the cubic nonlinearity in the governing equations \cite{Catheline2003,Espindola2017}.

Second, between 50 Hz and 75 Hz, shear shocks also appear in a ring that is about a 15.6 mm under the brain surface {(Fig. \ref{Fig:Skull_Print_MaxAcc_1p5}c)}. The focal shock is also present, but its amplitude is reduced. This is due to attenuation which increases as a function of frequency and thus significantly reduces the shear wave amplitude over the long propagation lengths required to reach the focus. The appearance of the shock ring is due to a decrease in the shock formation distance as a function of frequency, i.e. the distance required for the peak of the wave to tip over and reach the trough decreases \cite{rothkopf1976shock}. 

In fact, as the frequency continues to increase above 75 Hz, the location of the shock ring migrates closer and closer to the surface {(Fig.~\ref{Fig:Skull_Print_Max_Ratio}j)}, whereas in the linear regime the maximum acceleration is always closer to the surface {(Fig.~\ref{Fig:Skull_Print_Max_Ratio}e)}, it is primarily governed by strong attenuation. In the third regime, the focal shocks are overwhelmed by attenuation and only the ring shocks occur {(Fig. \ref{Fig:Skull_Print_MaxAcc_1p5}d)}. At 200 Hz, for example, the maximum acceleration occurs in a shock ring at 4.7 mm from the surface where it is 393.2$g$. This is almost 2 times larger than the 199.5$g$ initial impact acceleration and it is a considerably smaller acceleration amplification factor than at 25 Hz. Propagation movies in acceleration for the 25, 75, 200 Hz provided in the supplementary material clearly differentiate the three different regimes.

A systematic analysis of the amplification factor as a function of frequency and amplitude  {(Fig.~\ref{Fig:Skull_Print_Max_Ratio}f)} shows that the acceleration amplification factor peaks at 25 Hz. As the frequency increases the amplification factor decreases due to the effects of attenuation. Note that it also decreases for frequencies below 25 Hz, however this requires a different explanation that will be discussed subsequently in the context of super-resolution. These acceleration amplification factors in the nonlinear regime are consistently larger than the amplification factors in the reference \textit{linear} visco-elastic regime {(Fig.~\ref{Fig:Skull_Print_Max_Ratio}a)} where there is no shock formation.
The overall maximum acceleration also occurs at 25 Hz {(Fig.~\ref{Fig:Skull_Print_Max_Ratio}f)} and the minimum occurs at 75 Hz. This suggests that the first regime, where the shocks occur only at the focus, is a particularly destructive, unlike the second regime where the shocks are distributed across the focal and ring regions.

The Lagrangian strain-rate distribution {(Fig.~\ref{Fig:Skull_Print_StrainRate_1p5}a,b,c,d)} and the maximum Lagrangian strain-rate {(Fig.~\ref{Fig:Skull_Print_Max_Ratio}c,h)} as a function of frequency and amplitude exhibit trends that closely match the acceleration {(Fig.~\ref{Fig:Skull_Print_Max_Ratio}b,g)}. 
The lowest strain rates (175 1/s) occur at 12.5 Hz and the highest strain-rates (928 1/s) occur 25 Hz. At  75 Hz i.e. in the second regime there is a local strain rate minimum (268 1/s). In the third regime, the strain-rates increase with frequency to 525 1/s at 200 Hz. These simulated estimates of the strain-rate are consistent with experimental measurements of the strain-rate at shear shock fronts imaged in fresh porcine brain, where strain-rates as high as 600 1/s were observed at 75 Hz \cite{Espindola2017}.

The Lagrangian strain distribution {(Fig.~\ref{Fig:Skull_Print_Strain_1p5}a,b,c,d)} and
the maximum Lagrangian strain as a function of frequency and amplitude {(Fig. \ref{Fig:Skull_Print_Max_Ratio}d,i)} behaves in a somewhat counter-intuitive fashion. The strain is often \textit{lower} in regions of high strain-rate especially in the focal regions. 
Furthermore, the maximum strain in the linear regime {(Fig.~\ref{Fig:Skull_Print_Max_Ratio}d)} is about 50\% \textit{larger} than in the nonlinear regime. Nonlinearity generates higher harmonics which are in turn more strongly attenuated thus reducing the overall particle velocity amplitude and strain estimates in comparison to the linear case. Therefore this indicates it is the rate or time-derivative-dependent behavior that is most strongly affected by the shear shock wave physics rather than strain-dependent estimates directly. 

The focal region for the acceleration  {(Fig. \ref{Fig:Skull_Print_MaxAcc_1p5}a,b)} 
and strain-rate {(Fig.~\ref{Fig:Skull_Print_StrainRate_1p5}a,b)} 
is much smaller than the wavelength of the initial impact. At 25 Hz, for example, the wavelength is 7.52 cm and the full-width half-max of the acceleration focal zone along the x-axis is 1.34 mm {(Fig. \ref{Fig:Beam_Plots}a,b)} i.e. 56 times smaller than the impact wavelength. The strain-rate focal zone is similarly super-resolved by a factor of 43 at 25 Hz. The ability to super-resolve by over an order of magnitude is due to the broad spectral content of the shock wave and the substantial nonlinear energy transfer to frequencies over ten times higher than the initial impact.
{(Fig.~\ref{Fig:Shock_Spectrum}b)}.
At 12.5 Hz the wave is super-resolved by a factor of 13 but the large 14.08 cm wavelength results in a larger focal zone and thus an overall acceleration or strain-rate amplification is smaller than at 25 Hz. 

\section{Discussion and Conclusions}\label{Sec:Conclusions}
Previously unappreciated shear shock wave physics has been shown to play a determining role in estimates of gelatin/brain motion during an impacts. Estimates of the acceleration and strain-rate, critical parameters in brain injury biomechanics, are over an order of magnitude larger when taking into account shear shock formation. The use of high frame-rate and high motion sensitivity quantitative ultrasound imaging that can directly observe this behavior is a crucial component that is required to inform the local viscoelastodynamics. 

This study is confined to the propagation of linearly polarized shear shock waves in a homogeneous, isotropic, relaxing soft solid without considering other important physical effects like heterogeneous composition of the brain, including ventricles, fluid-solid interfaces, and non-polarized 3D propagation. Therefore this work cannot be directly linked to the real world injury scenarios, nonetheless it is the first work which demonstrates the formation of shear shock waves in realistic  morphology of human head in the event of an impact and opens a paradigm for further research. 
This simulation tool was validated with direct ultrasound-based  quantitative imaging of shear shock wave formation at depth within a human head phantom filled with gelatin thus establishing a high level of confidence in its ability to model the relevant nonlinear viscoelastodynamics. The extremely nonlinear shear behavior observed here easily yields Mach numbers that are greater than one deep inside the gelatin for relatively mild impacts. Strong attenuation contributes significantly to the richness of the observed behavior and its frequency-dependence. 
Based on an analysis of the local velocity, acceleration, strain, and strain-rate three distinct regimes emerged depending on the frequency and amplitude of the impact.

At lower frequencies and higher amplitudes, the long propagation lengths and high nonlinearity work together to generate extremely nonlinear focal shocks. For example, at 25 Hz for a mild 1.5 m/s impact the local particle velocity at the focus is 2.16 m/s which corresponds to a Mach number of 1.14 (=2.16/1.88). The characteristic attenuation length scale (78 mm) is slightly larger than the wavelength ($\lambda = 75$ mm) and the shear wave easily propagate to the middle of the human brain, which has a typical diameter of 12 cm. Consequently there is a significant focal gain. At these Mach numbers shocks form very quickly, and the explosive gradients at the shock send the local acceleration to 979$g$  and the strain rate to 927 1/s. Together, focusing and nonlinearity amplify the local acceleration by 41 times compared to the impact acceleration, with nonlinearity accounting for the majority of that amplification and focusing accounting for a factor of 1.74. In fact, the nonlinearity acts on such short length scales that the focal spot size
is super-resolved to a FWHM of 1.34 mm which is over 56 times  smaller than the 75 mm impact wavelength. The ability to super-resolve beyond linear diffraction limits theory is due to harmonic generation. As the wave develops into a shock higher frequencies or smaller wavelengths are required to support the sharp features. Thus the local frequency content at the focal shock has a substantial amount of energy at frequencies that are over ten times higher then the fundamental. The impact initial conditions that generate this highly nonlinear behavior correspond to a low scale of what is observed in traumatic brain injuries. In NFL players with head injuries the average head impact velocity is 9.3 m/s acceleration is 98$g$ and the corresponding characteristic average impact frequency is 10.6 Hz \cite{Pellman2003,Sanchez2019}. 

At high frequencies, the attenuation is strong and shocks appear in a ring close to the skull surface and not at all at the geometric focus. At 200 Hz, for example the characteristic attenuation length 9.0 mm is short compared to the size of the brain.
At the location of peak acceleration, which is 4.7 mm from the skull surface, the local velocity is 1.11 which corresponds to a Mach number of 0.84, the peak acceleration is 393.2$g$ and the peak strain-rate is 525.5 1/s.  Beyond the this distance the attenuation dominates and is insufficient to overcome the focal gain thus sparing deep parts of the brain. However a shock ring forms just under the surface of the brain potentially causing injury in a wide range of superficial regions. 

At intermediate frequencies, shocks appear at both the focus deep inside the brain and in the ring just under the brain surface.  This regime corresponds to a local minimum in acceleration and strain-rates because the energy is distributed between the two regions. For 75 Hz case, for example, the peak acceleration is 167.7$g$ and the peak strain-rate is 268.6 1/s. 
The location of peak acceleration in the band region, where propagation is still quasi-planar, is 15.65 mm from the skull surface, the local velocity is 1.19 m/s which corresponds to a Mach number of 0.91. 

Thus even a small change in the characteristic frequency of the impact can have a large effect on the local acceleration, strain-rate and their distribution within the brain. This also suggests that there is an optimum impact frequency, around 75 Hz, that can minimize local biomechanical injury metrics. Counter-intuitively this also indicates that damping out high frequencies while preserving low frequencies may be entirely counterproductive. Wearing protective equipment, for example, encourages highly competitive athletes to take more risks and absorb larger impacts~\cite{chen2019winning}. This is due to the fact that protective equipment will dampen the high frequencies that trigger superficial pain receptors. A boxing glove, for example, will damp out the high frequencies that would otherwise be present in a hard and painful fist-to-face contact. However, the low frequency component of the impact, which is less painful to superficial receptors, is readily transmitted to the brain, where there are no pain receptors. At these low frequencies, even for mild impacts, the local acceleration can be focused into highly destructive and highly localized super-resolved shear shocks that tear and damage tissue. Above local acceleration measurements of 266$g$ in our head phantom, for example, the tissue-mimicking gelatin completely fractured at the focus. However the acceleration at the brain phantom surface was measured to be only 19$g$ i.e. 5 times lower than the average injurious acceleration in the NFL~\cite{Pellman2003,Sanchez2019}. The size of this high acceleration region in the focal regime is small. The FWHM extends over an area of just 14.66 mm$^2$, which is about the size of a grain of rice. Thus a single mild impact may incur devastating damage but only to a small region. However, over the course of an athletic career the accumulation of many tiny mm-scale injuries could explain why repeated exposure to mild events can lead to staggering rates of CTE, such as 99\% observed in the NFL \cite{mez2017clinicopathological}. The focal location, which depends on an specific impact, may also explain the wide variety of neurological symptoms that follow a TBI.

In conclusion, the evidence that shear shock wave physics is a necessary and primary component of brain biomechanics and, we hypothesize, brain injury is overwhelming. Local measurements and simulations of this shock wave behavior, which are absent from current biomechanical models of the brain,  may fundamentally change the way we approach the design of protective equipment in transportation, sports, playground safety, falls and our understanding of the extreme biomechanical environment to which our brains can be subjected.

\section{Acknowledgement}
We would like to acknowledge funding from the NIH (R01 NS091195).

\bibliographystyle{elsarticle-num}





\begin{thebibliography}{10}
\expandafter\ifx\csname url\endcsname\relax
  \def\url#1{\texttt{#1}}\fi
\expandafter\ifx\csname urlprefix\endcsname\relax\def\urlprefix{URL }\fi
\expandafter\ifx\csname href\endcsname\relax
  \def\href#1#2{#2} \def\path#1{#1}\fi

\bibitem{corrigan2010epidemiology}
J.~D. Corrigan, A.~W. Selassie, J.~A.~L. Orman, The epidemiology of traumatic
  brain injury, The Journal of head trauma rehabilitation 25~(2) (2010) 72--80.

\bibitem{guskiewicz2000epidemiology}
K.~M. Guskiewicz, N.~L. Weaver, D.~A. Padua, W.~E. Garrett, Epidemiology of
  concussion in collegiate and high school football players, The American
  Journal of Sports Medicine 28~(5) (2000) 643--650.

\bibitem{meythaler2001current}
J.~Meythaler, J.~Peduzzi, E.~Eleftheriou, T.~Novack, {Current concepts: Diffuse
  axonal injury-- associated traumatic brain injury}, Archives of physical
  medicine and rehabilitation 82~(10) (2001) 1461--1471.

\bibitem{tagliaferri2006systematic}
F.~Tagliaferri, C.~Compagnone, M.~Korsic, F.~Servadei, J.~Kraus, A systematic
  review of brain injury epidemiology in europe, Acta neurochirurgica 148~(3)
  (2006) 255--268.

\bibitem{maas2008moderate}
A.~Maas, N.~Stocchetti, R.~Bullock, {Moderate and severe traumatic brain injury
  in adults}, The Lancet Neurology 7~(8) (2008) 728--741.

\bibitem{chen2004long}
X.-H. Chen, R.~Siman, A.~Iwata, D.~F. Meaney, J.~Q. Trojanowski, D.~H. Smith,
  Long-term accumulation of amyloid-$\beta$, $\beta$-secretase, presenilin-1,
  and caspase-3 in damaged axons following brain trauma, The American journal
  of pathology 165~(2) (2004) 357--371.

\bibitem{mckee2009chronic}
A.~C. McKee, R.~C. Cantu, C.~J. Nowinski, E.~T. Hedley-Whyte, B.~E. Gavett,
  A.~E. Budson, V.~E. Santini, H.-S. Lee, C.~A. Kubilus, R.~A. Stern, Chronic
  traumatic encephalopathy in athletes: progressive tauopathy after repetitive
  head injury, Journal of Neuropathology \& Experimental Neurology 68~(7)
  (2009) 709--735.

\bibitem{johnson2010traumatic}
V.~E. Johnson, W.~Stewart, D.~H. Smith, Traumatic brain injury and
  amyloid-$\beta$ pathology: a link to alzheimer's disease?, Nature Reviews
  Neuroscience 11~(5) (2010) 361--370.

\bibitem{Beckwith2012}
J.~G. Beckwith, R.~M. Greenwald, J.~J. Chu, Measuring head kinematics in
  football: correlation between the head impact telemetry system and hybrid iii
  headform, Annals of biomedical engineering 40~(1) (2012) 237--248.

\bibitem{Camarillo2013}
D.~B. Camarillo, P.~B. Shull, J.~Mattson, R.~Shultz, D.~Garza, An instrumented
  mouthguard for measuring linear and angular head impact kinematics in
  american football, Annals of biomedical engineering 41~(9) (2013) 1939--1949.

\bibitem{Salzar2008_Bass}
R.~S. Salzar, R.~Cameron, J.~A. Pellettiere, Improving earpiece accelerometer
  coupling to the head, SAE International Journal of Passenger Cars-Mechanical
  Systems 1~(2008-01-2978) (2008) 1367--1381.

\bibitem{Kim1993}
W.~Kim, A.~Voloshin, S.~Johnson, A.~Simkin, Measurement of the impulsive bone
  motion by skin-mounted accelerometers, J Biomech Eng. 115~(1) (1993) 47--52.

\bibitem{rimel1981disability}
R.~W. Rimel, B.~Giordani, J.~T. Barth, T.~J. Boll, J.~A. Jane, Disability
  caused by minor head injury, Neurosurgery 9~(3) (1981) 221--228.

\bibitem{greenwald2008head}
R.~M. Greenwald, J.~T. Gwin, J.~J. Chu, J.~J. Crisco, Head impact severity
  measures for evaluating mild traumatic brain injury risk exposure,
  Neurosurgery 62~(4) (2008) 789.

\bibitem{Wu2016_Camarillo}
L.~C. Wu, V.~Nangia, K.~Bui, B.~Hammoor, M.~Kurt, F.~Hernandez, C.~Kuo, D.~B.
  Camarillo, In vivo evaluation of wearable head impact sensors, Annals of
  biomedical engineering 44~(4) (2016) 1234--1245.

\bibitem{guskiewicz2007measurement}
K.~M. Guskiewicz, J.~P. Mihalik, V.~Shankar, S.~W. Marshall, D.~H. Crowell,
  S.~M. Oliaro, M.~F. Ciocca, D.~N. Hooker, Measurement of head impacts in
  collegiate football players: relationship between head impact biomechanics
  and acute clinical outcome after concussion, Neurosurgery 61~(6) (2007)
  1244--1253.

\bibitem{margulies1990physical}
S.~S. Margulies, L.~E. Thibault, T.~A. Gennarelli, Physical model simulations
  of brain injury in the primate, Journal of biomechanics 23~(8) (1990)
  823--836.

\bibitem{meaney1995biomechanical}
D.~F. Meaney, D.~H. Smith, D.~I. Shreiber, A.~C. Bain, R.~T. Miller, D.~T.
  Ross, T.~A. Gennarelli, Biomechanical analysis of experimental diffuse axonal
  injury, Journal of neurotrauma 12~(4) (1995) 689--694.

\bibitem{hardy2001investigation}
W.~N. Hardy, C.~D. Foster, M.~J. Mason, K.~H. Yang, A.~I. King, S.~Tashman,
  Investigation of head injury mechanisms using neutral density technology and
  high-speed biplanar x-ray, Tech. rep., SAE Technical Paper (2001).

\bibitem{bayly2005deformation}
P.~Bayly, T.~Cohen, E.~Leister, D.~Ajo, E.~Leuthardt, G.~Genin, Deformation of
  the human brain induced by mild acceleration, Journal of neurotrauma 22~(8)
  (2005) 845--856.

\bibitem{clayton2012transmission}
E.~H. Clayton, G.~M. Genin, P.~V. Bayly, Transmission, attenuation and
  reflection of shear waves in the human brain, Journal of The Royal Society
  Interface 9~(76) (2012) 2899--2910.

\bibitem{Macmanus2018}
D.~B. MacManus, J.~G. Murphy, M.~D. Gilchrist, Mechanical characterisation of
  brain tissue up to 35\% strain at 1, 10, and 100/s using a custom-built
  micro-indentation apparatus, Journal of the mechanical behavior of biomedical
  materials 87 (2018) 256--266.

\bibitem{Ahmadzadeh2014}
H.~Ahmadzadeh, D.~H. Smith, V.~B. Shenoy, Viscoelasticity of tau proteins leads
  to strain rate-dependent breaking of microtubules during axonal stretch
  injury: predictions from a mathematical model, Biophysical journal 106~(5)
  (2014) 1123--1133.

\bibitem{Ghajari2017}
M.~Ghajari, P.~J. Hellyer, D.~J. Sharp, Computational modelling of traumatic
  brain injury predicts the location of chronic traumatic encephalopathy
  pathology, Brain 140~(2) (2017) 333--343.

\bibitem{Espindola2017}
D.~Esp{\'\i}ndola, S.~Lee, G.~Pinton, Shear shock waves observed in the brain,
  Physical Review Applied 8~(4) (2017) 044024.

\bibitem{Tripathi2019_PPM1D_CT}
B.~B. Tripathi, D.~Esp{\'{i}}ndola, G.~F. Pinton, {Piecewise Parabolic Method
  for Propagation of Shear Shock Waves in Relaxing Soft Solids: One Dimensional
  Case}, Int. J. Num. Meth. Bio. Med. Eng. 35~(5) (2019) e3187.

\bibitem{Tripathi2019_PPM2D_CT}
B.~B. Tripathi, D.~Esp{\'{i}}ndola, G.~F. Pinton, {Modeling and Simulations of
  Two Dimensional Propagation of Shear Shock Waves in Relaxing Soft Solids}, J.
  Comput. Phys. 395 (2019) 205--222.

\bibitem{Pinton2010}
G.~Pinton, F.~Coulouvrat, J.-L. Gennisson, M.~Tanter, {Nonlinear reflection of
  shock shear waves in soft elastic media.}, The Journal of the Acoustical
  Society of America 127~(2) (2010) 683--91.

\bibitem{Landau1986}
L.~D. Landau, E.~M. Lifshitz, {Theory of Elasticity, Vol. 7}, 3rd Edition,
  Elsevier, 1986.

\bibitem{Zabolotskaya2004}
E.~A. Zabolotskaya, M.~Hamilton, Y.~A. Ilinskii, G.~D. Meegan, {Modeling of
  nonlinear shear waves in soft solids}, The Journal of the Acoustical Society
  of America 116~(5) (2004) 2807.

\bibitem{Destrade2010}
M.~Destrade, M.~D. Gilchrist, G.~Saccomandi, Third-and fourth-order constants
  of incompressible soft solids and the acousto-elastic effect, The Journal of
  the Acoustical Society of America 127~(5) (2010) 2759--2763.

\bibitem{Chockalingam2020}
S.~Chockalingam, T.~Cohen, Shear shock evolution in incompressible soft solids,
  Journal of the Mechanics and Physics of Solids 134 (2020) 103746.

\bibitem{Ziv2019}
R.~Ziv, G.~Shmuel, Smooth waves and shocks of finite amplitude in soft
  materials, Mechanics of Materials 135 (2019) 67--76.

\bibitem{Giammarinaro2016}
B.~Giammarinaro, F.~Coulouvrat, G.~Pinton, Numerical simulation of focused
  shock shear waves in soft solids and a two-dimensional nonlinear homogeneous
  model of the brain, Journal of biomechanical engineering 138~(4) (2016)
  041003.

\bibitem{Tripathi2017}
B.~B. Tripathi, D.~Esp{\'{i}}ndola, G.~F. Pinton, {Piecewise parabolic method
  for simulating one-dimensional shear shock wave propagation in
  tissue-mimicking phantoms}, Shock Waves 27~(6) (2017) 879--888.

\bibitem{chatelin2010fifty}
S.~Chatelin, A.~Constantinesco, R.~Willinger, Fifty years of brain tissue
  mechanical testing: from in vitro to in vivo investigations, Biorheology
  47~(5-6) (2010) 255--276.

\bibitem{Dixit2017}
P.~Dixit, G.~Liu, A review on recent development of finite element models for
  head injury simulations, Archives of Computational Methods in Engineering
  24~(4) (2017) 979--1031.

\bibitem{horgan2003creation}
T.~J. Horgan, M.~D. Gilchrist, The creation of three-dimensional finite element
  models for simulating head impact biomechanics, International Journal of
  Crashworthiness 8~(4) (2003) 353--366.

\bibitem{horgan2004influence}
T.~J. Horgan, M.~D. Gilchrist, Influence of fe model variability in predicting
  brain motion and intracranial pressure changes in head impact simulations,
  International Journal of Crashworthiness 9~(4) (2004) 401--418.

\bibitem{Taylor2009}
P.~A. Taylor, C.~C. Ford, Simulation of blast-induced early-time intracranial
  wave physics leading to traumatic brain injury, Journal of biomechanical
  engineering 131~(6) (2009) 061007.

\bibitem{wittek2011algorithms}
A.~Wittek, G.~Joldes, K.~Miller, Algorithms for computational biomechanics of
  the brain, in: Biomechanics of the Brain, Springer, 2011, pp. 189--219.

\bibitem{yang2019modelling}
K.~H. Yang, H.~Mao, Modelling of the brain for injury simulation and
  prevention, in: Biomechanics of the Brain, Springer, 2019, pp. 97--133.

\bibitem{Zienkiewicz2005}
O.~C. Zienkiewicz, R.~L. Taylor, The finite element method for solid and
  structural mechanics, Elsevier, 2005.

\bibitem{Ye2017b}
W.~Ye, A.~Bel-Brunon, S.~Catheline, A.~Combescure, M.~Rochette, Simulation of
  non-linear transient elastography: finite element model for the propagation
  of shear waves in homogeneous soft tissues, International journal for
  numerical methods in biomedical engineering (2017).

\bibitem{Colella1984}
P.~Colella, P.~R. Woodward, {The Piecewise Parabolic Method (PPM) for
  gas-dynamical simulations}, Journal of Computational Physics 54~(1) (1984)
  174--201.
\newblock \href {https://doi.org/10.1016/0021-9991(84)90143-8}
  {\path{doi:10.1016/0021-9991(84)90143-8}}.

\bibitem{Miller2002}
G.~Miller, P.~Colella, {A Conservative Three--Dimensional Eulerian Method for
  Coupled Solid--Fluid Shock Capturing}, Journal of Computational Physics
  183~(1) (2002) 26--82.
\newblock \href {https://doi.org/10.1006/jcph.2002.7158}
  {\path{doi:10.1006/jcph.2002.7158}}.

\bibitem{Smoller2012}
J.~Smoller, Shock waves and reaction diffusion equations, Vol. 258, Springer
  Science \& Business Media, 2012.

\bibitem{espindola2017High}
D.~Espindola, G.~Pinton, High frame-rate imaging and adaptive tracking of shear
  shock wave formation in the brain: A fullwave and experimental study, in:
  2017 IEEE International Ultrasonics Symposium (IUS), 2017, pp. 1--1.

\bibitem{pinton2014adaptive}
G.~Pinton, J.-L. Gennisson, M.~Tanter, F.~Coulouvrat, Adaptive motion
  estimation of shear shock waves in soft solids and tissue with ultrasound,
  IEEE transactions on ultrasonics, ferroelectrics, and frequency control
  61~(9) (2014) 1489--1503.

\bibitem{Catheline2003}
S.~Catheline, J.-L. Gennisson, M.~Tanter, M.~Fink, {Observation of shock
  transverse waves in elastic media.}, Physical review letters 91~(16) (2003)
  164301.

\bibitem{rothkopf1976shock}
E.~Rothkopf, W.~Low, Shock formation distance in a pressure driven shock tube,
  The Physics of Fluids 19~(12) (1976) 1885--1888.

\bibitem{Pellman2003}
E.~J. Pellman, D.~C. Viano, A.~M. Tucker, I.~R. Casson, Concussion in
  professional football: Location and direction of helmet impacts—part 2,
  Neurosurgery 53~(6) (2003) 1328--1341.

\bibitem{Sanchez2019}
E.~J. Sanchez, L.~F. Gabler, A.~B. Good, J.~R. Funk, J.~R. Crandall, M.~B.
  Panzer, A reanalysis of football impact reconstructions for head kinematics
  and finite element modeling, Clinical biomechanics 64 (2019) 82--89.

\bibitem{chen2019winning}
Y.~Chen, C.~Buggy, S.~Kelly, Winning at all costs: a review of risk-taking
  behaviour and sporting injury from an occupational safety and health
  perspective, Sports medicine-open 5~(1) (2019) 15.

\bibitem{mez2017clinicopathological}
J.~Mez, D.~H. Daneshvar, P.~T. Kiernan, B.~Abdolmohammadi, V.~E. Alvarez, B.~R.
  Huber, M.~L. Alosco, T.~M. Solomon, C.~J. Nowinski, L.~McHale, et~al.,
  Clinicopathological evaluation of chronic traumatic encephalopathy in players
  of american football, Jama 318~(4) (2017) 360--370.

\end{thebibliography}





\end{document}